\documentclass[a4paper,11pt]{article}
\usepackage{jheppub}
\usepackage{amsfonts,amsmath,amssymb,mathrsfs,textcomp,dsfont,esint,array,multirow,url,color,bm}
\usepackage[mathscr]{euscript}
\usepackage{tikz,epsfig,graphicx}

\title{Tetrahedron instantons}

\author[1]{Elli Pomoni,}
\author[2]{Wenbin Yan,}
\author[1,3]{Xinyu Zhang}
\affiliation[1]{Deutsches Elektronen-Synchrotron DESY, Notkestr. 85, 22607 Hamburg, Germany}
\affiliation[2]{Yau Mathematical Sciences Center, Tsinghua University, Beijing, 10084, China}
\affiliation[3]{New High Energy Theory Center and Department of Physics and Astronomy, Rutgers University, Piscataway, New Jersey 08854, USA}
\emailAdd{elli.pomoni@desy.de}
\emailAdd{wbyan@tsinghua.edu.cn}
\emailAdd{zhangxinyuphysics@gmail.com}

\preprint{DESY 21-087}

\abstract{
We introduce and study tetrahedron instantons, which can be realized
in string theory by D$1$-branes probing a configuration of intersecting
D$7$-branes in flat spacetime with a proper constant $B$-field.
Physically they capture instantons on $\mathbb{C}^{3}$ in the presence
of the most general intersecting real codimension-two supersymmetric
defects. Moreover, we construct the tetrahedron instantons as particular
solutions of general instanton equations in noncommutative field theory.
We analyze the moduli space of tetrahedron instantons and discuss
the geometric interpretations. We compute the instanton partition
function both via the equivariant localization on the moduli space
of tetrahedron instantons and via the elliptic genus of the worldvolume
theory on the D$1$-branes probing the intersecting D$7$-branes,
obtaining the same result. The instanton partition function of the
tetrahedron instantons lies between the higher-rank Donaldson-Thomas
invariants on $\mathbb{C}^{3}$ and the partition function of the
magnificent four model, which is conjectured to be the mother of all
instanton partition functions. Finally, we show that the instanton
partition function admits a free field representation, suggesting
the existence of a novel kind of symmetry which acts on the cohomology
of the moduli spaces of tetrahedron instantons.
}

\begin{document}
\maketitle


\section{Introduction}

Since the discovery of Yang-Mills instantons as topologically nontrivial
field configurations that minimize the Yang-Mills action in four-dimensional
Euclidean spacetime \cite{Belavin:1975fg}, many important developments
on the applications of instantons arose in both physics \cite{Coleman:1978ae,Vainshtein:1981wh,Schafer:1996wv}
and mathematics \cite{Freed:1984xe,DonaldsonKronheimer}. In the Atiyah-Drinfield-Hitchin-Manin
(ADHM) construction \cite{Atiyah:1978ri}, the moduli space of Yang-Mills
instantons on $\mathbb{R}^{4}$ is given as a hyperkahler quotient.
In addition, the ADHM construction can be derived in a physically
intuitive way using string theory \cite{Witten:1994tz,Douglas:1995bn,Douglas:1996uz}.
For example, the moduli space $\mathcal{M}_{n,k}$ of $\mathrm{SU}(n)$
instantons of charge $k$ is given by the Higgs branch of the supersymmetric
gauge theory living on $k$ D$1$-branes probing a stack of $n$ coincident
D$5$-branes in type IIB superstring theory. To avoid the noncompactness
of $\mathcal{M}_{n,k}$ due to small instantons, Nakajima introduced
a smooth manifold $\widetilde{\mathcal{M}}_{n,k}$, which can be obtained
from the Uhlenbeck compactification of $\mathcal{M}_{n,k}$ by resolving
the singularities \cite{nakajima1994resolutions}. Thereafter Nekrasov
and Schwarz interpreted $\widetilde{\mathcal{M}}_{n,k}$ as the moduli
space of $\mathrm{U}(n)$ instantons on noncommutative $\mathbb{R}^{4}$
\cite{Nekrasov:1998ss}, where the noncommutativity of the spacetime
coordinates can be produced in string theory by turning on a proper
constant $B$-field \cite{Seiberg:1999vs}. 

The moduli space $\widetilde{\mathcal{M}}_{n,k}$ admits a $\mathrm{U}(1)^{2}$
action which stems from the rotation symmetry of the spacetime $\mathbb{R}^{4}$,
and a $\mathrm{U}(n)$ action which rotates the gauge orientation
at infinity. Although $\widetilde{\mathcal{M}}_{n,k}$ is still noncompact
because the instantons can run away to infinity of the spacetime $\mathbb{R}^{4}$,
the $\mathbf{T}$-equivariant symplectic volume $\mathcal{Z}_{k}$
of $\widetilde{\mathcal{M}}_{n,k}$ is well-defined \cite{Moore:1997dj},
with $\mathbf{T}$ being the maximal torus of $\mathrm{U}(1)^{2}\times\mathrm{U}(n)$.
Using the equivariant localization theorem \cite{Atiyah:1984px},
$\mathcal{Z}_{k}$ can be evaluated exactly and is given by a sum
over a collection of random partitions. Assembling $\mathcal{Z}_{k}$
with all $k\geq0$ into a generating function, Nekrasov obtained the
instanton partition function $\mathcal{Z}=\sum_{k\ge0}\mathtt{q}^{k}\mathcal{Z}_{k}$
of four-dimensional $\mathcal{N}=2$ $\mathrm{SU}(n)$ supersymmetric
Yang-Mills theory in the $\Omega$-background \cite{Nekrasov:2002qd}.
It turns out that both the Seiberg-Witten effective prepotential \cite{Seiberg:1994rs,Seiberg:1994aj}
and the couplings to the background gravitational fields \cite{Bershadsky:1993cx,Antoniadis:1993ze}
can be derived rigorously from $\mathcal{Z}$ \cite{Nekrasov:2003rj,Nakajima:2003uh,Nekrasov:2004vw,Nekrasov:2012xe,Zhang:2019msw,Manschot:2019pog}.
The instanton partition function is also related to the A-model topological
strings on two-dimensional Riemann surfaces \cite{Losev:2003py,Marshakov:2006ii,Nekrasov:2009zza,Zhang:2016xqi},
the Virasoro/W-algebra conformal blocks \cite{Alday:2009aq,Wyllard:2009hg},
and quantum integrable systems \cite{Nekrasov:2009rc,Nekrasov:2011bc,Nekrasov:2013xda}.
Recently, based on the computation of the elliptic genus using supersymmetric
localization techniques \cite{Benini:2013nda,Benini:2013xpa}, an
alternative general approach to computing $\mathcal{Z}$ was provided
in \cite{Hwang:2014uwa}. The major advantage of this approach is
that we no longer need to know $\widetilde{\mathcal{M}}_{n,k}$ explicitly,
and $\mathcal{Z}$ is given in terms of contour integrals with Jeffrey-Kirwan
(JK) residue prescription \cite{jeffrey1995localization}.

Over the past few years, there have been several fascinating generalizations
of the Yang-Mills instantons on $\mathbb{R}^{4}$. The ADHM-type constructions
of Yang-Mills instantons on some other four-manifolds have been found
\cite{king1989instantons,Kronheimer1990,Douglas:1996sw,Cherkis:2008ip,Witten:2009xu}.
Instantons also appear in higher-dimensional gauge theories \cite{Donaldson:1996kp,Acharya:1997gp,Baulieu:1997jx},
and we can get an ADHM-type construction of instanon moduli spaces
from the low-energy worldvolume theory on D$1$-branes probing D$(2p+1)$-branes
with $p=3,4$ \cite{Moore:1998et}. The instanton partition function
$\mathcal{Z}$ is given by a statistical sum over random plane partitions
($p=3$) or solid partitions ($p=4$), see \cite{Kanno:2020ybd} for
a recent review. The $p=3$ case gives the equivariant Donaldson-Thomas
invariants of toric Calabi-Yau threefolds \cite{maulik2006gromov1,maulik2006gromov2,Awata:2009dd,maulik2011gromov,Cirafici:2008sn,Cirafici:2010bd,Benini:2018hjy,Fasola:2020hqa},
while the $p=4$ case defines the magnificent four model \cite{Nekrasov:2017cih,Nekrasov:2018xsb},
and can be interpreted in terms of equivariant Donaldson-Thomas invariants
of toric Calabi-Yau fourfolds \cite{Cao:2017swr,Cao:2018rbp,Cao:2020otr}.
The partition function of the magnificent four model is envisioned
to be the mother of all instanton partition functions \cite{Nekrasov:2017cih}.

In yet a different line of research, the concept of generalized field
theory, which is constructed by merging several ordinary field theories
across defects, has been emerging in recent years. The spacetime $X$
of a generalized gauge theory contains several intersecting components,
$X=\bigcup_{A}X_{A}$. The fields and the gauge groups $G_{A}=\left.G\right|_{A}$
on different components can be different, and the matter fields living
on the intersection $X_{A}\bigcap X_{B}$ transform in the bifundamental
representation of the product group $G_{A}\times G_{B}$. For instance,
D$1$-branes probing a configuration of intersecting (anti-)D$5$-branes,
in the presence of a proper $B$-field, give rise to the spiked instantons
in a generalized gauge theory \cite{Nekrasov:2015wsu,Nekrasov:2016qym,Nekrasov:2016gud,Prabhakar:2017oyk}.
When each component $X_{A}$ of the spacetime is a noncompact toric
surface, the generating function of equivariant symplectic volumes
of the instanton moduli spaces can be similarly defined and is called
the gauge origami partition function \cite{Nekrasov:2016ydq}. Applying
the equivariant localization theorem, the gauge origami partition
function can be expressed as a statistical sum over collections of
random partitions, and provides a unified treatment of instanton partition
functions of four-dimensional $\mathcal{N}=2$ supersymmetric gauge
theories \cite{Nekrasov:2012xe}, possibly with local or surface defects
\cite{Alday:2009fs,Gaiotto:2009fs,Alday:2010vg,Awata:2010bz,Kozcaz:2010yp,Kanno:2011fw,Pan:2016fbl,Nekrasov:2017rqy}.
Nekrasov also derived an infinite set of nonperturbative Dyson-Schwinger
equations from the gauge origami partition function \cite{Nekrasov:2015wsu},
leading to a number of interesting applications \cite{Jeong:2017pai,Jeong:2017mfh,Nekrasov:2017gzb,Jeong:2018qpc,Jeong:2019fgx,Lee:2020hfu,Nekrasov:2021tik}.

It is the goal of the present work to piece together the jigsaw puzzle
of instantons by studying D$1$-branes probing a configuration of
intersecting D$7$-branes. With a proper $B$-field, the ground state
of the D-brane system is supersymmetric, and the low-energy theory
on D$1$-branes preserves $\mathcal{N}=\left(0,2\right)$ supersymmetry
in two dimensions. Since the arrangement of various D-branes and open
strings attached to them can be naturally associated with the vertices,
edges and faces of a tetrahedron, we will call these instantons the
\emph{tetrahedron instantons}. We carefully work out the moduli space
of tetrahedron instantons, which can be viewed as an interpolation
between other instanton moduli spaces that have been explored extensively.
It is also a generalization of the moduli space of solutions to the
Donaldson-Uhlenbeck-Yau equations \cite{donaldson1985anti,uhlenbeck1986existence},
which describe the BPS configurations in higher dimensional super-Yang-Mills
theory \cite{Donaldson:1996kp,Acharya:1997gp,Baulieu:1997jx}. We
compute the instanton partition function $\mathcal{Z}$ using two
approaches. Furthermore, we show that $\mathcal{Z}$ admits a free
field representation, suggesting the existence of a novel kind of
symmetry which acts on the cohomology of the moduli spaces of tetrahedron
instantons.

The paper is organized as follows. In section \ref{sec:string} we
provide a string theory construction of tetrahedron instantons and
work out the instanton moduli spaces. In section \ref{sec:Noncommutative},
we describe the tetrahedron instantons in the framework of noncommutative
field theory. In section \ref{sec:moduli} we analyze the moduli space
of tetrahedron instantons. In section \ref{sec:Zloc}, we compute
the instanton partition function of the tetrahedron instantons using
the equivariant localization theorem. In section \ref{sec:Zell},
we calculate the instanton partition function of the tetrahedron instantons
from the elliptic genus of the worldvolume theory on the D$1$-branes
probing a configuration of intersecting D$7$-branes, and match it
with the equivariant localization computation. In section \ref{sec:freefield},
we give the free field representation of the instanton partition function.
We conclude in section \ref{sec:conclusion} with some comments and
a list of interesting open questions. We have included several appendices
with relevant background material. 

\section{Tetrahedron instantons from string theory \label{sec:string}}

We begin our discussion by describing the realization of tetrahedron
instantons from string theory, as this is the most natural setting
they can be constructed.

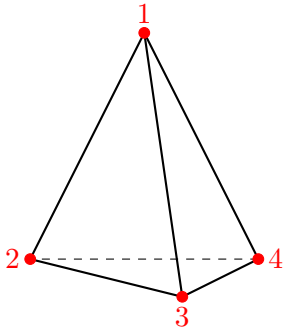
\begin{figure}
\centering
\begin{tikzpicture}
\draw[thick] (0,2) -- (1.5,-1) -- (0.5,-1.5) -- (-1.5,-1) -- cycle;
\draw[thick] (0,2) -- (0.5,-1.5);
\draw[dashed] (1.5,-1) -- (-1.5,-1);
\filldraw[red] (0,2) circle (2pt) node[anchor=south] {$1$};
\filldraw[red] (-1.5,-1) circle (2pt) node[anchor=east] {$2$};
\filldraw[red] (0.5,-1.5) circle (2pt) node[anchor=north] {$3$};
\filldraw[red] (1.5,-1) circle (2pt) node[anchor=west] {$4$};
\end{tikzpicture}

\caption{A tetrahedron with the sets $\underline{4}$ and $\underline{4}^{\vee}$
associated with vertices and faces, respectively. Each vertex is labeled
by $a\in\underline{4}$ and represents a complex plane $\mathbb{C}_{a}$.
The edge connecting two vertices labeled by $a$ and $b$ represents
a complex two-plane $\mathbb{C}_{ab}^{2}=\mathbb{C}_{a}\times\mathbb{C}_{b}$.
The face $A=\left(abc\right)\in\underline{4}^{\vee}$ has three vertices
$a$, $b$, and $c$, and represents a complex three-plane $\mathbb{C}_{A}^{3}=\prod_{a\in A}\mathbb{C}_{a}$.
The remaining vertex that is not in the face $A\in\underline{4}^{\vee}$
is denoted by $\check{A}\in\underline{4}$.}
\label{Tetrahedron}
\end{figure}

Let us identify the ten-dimensional spacetime $\mathbb{R}^{1,9}$
with $\mathbb{R}^{1,1}\times\mathbb{C}^{4}$ by choosing a complex
structure on $\mathbb{R}^{8}$. We take the coordinates on $\mathbb{R}^{1,1}$
to be $x^{0},x^{9}$. The set of coordinate labels of four complex
planes is denoted by
\begin{equation}
\underline{4}=\left\{ 1,2,3,4\right\} ,
\end{equation}
with the complex coordinate on $\mathbb{C}_{a}\subset\mathbb{C}^{4}$
being $z_{a}=x^{2a-1}+\mathrm{i}x^{2a}$. There are four complex three-planes,
$\mathbb{C}_{A}^{3}=\prod_{a\in A}\mathbb{C}_{a}\subset\mathbb{C}^{4}$
for $A\in\underline{4}^{\vee}$, where
\begin{equation}
\underline{4}^{\vee}=\begin{pmatrix}\underline{4}\\
3
\end{pmatrix}=\left\{ \left(123\right),\left(124\right),\left(134\right),\left(234\right)\right\} .
\end{equation}
For each $A\in\underline{4}^{\vee}$, we define
\begin{equation}
\check{A}=\left\{ \left.a\in\underline{4}\right|a\notin A\right\} .
\end{equation}
It is beneficial to introduce a tetrahedron (see Figure \ref{Tetrahedron})
to visualize the above data.

The tetrahedron instantons can be realized by $k$ D$1$-branes along
$\mathbb{R}^{1,1}$ probing a system of $n_{A}$ D$7_{A}$-branes
along $\mathbb{R}^{1,1}\times\mathbb{C}_{A}^{3}$ for $A\in\underline{4}^{\vee}$
in type IIB superstring theory. We summarize the configuration of
D-branes in Table \ref{D1-D7}. This can also be visualized by a tetrahedron
$\mathcal{T}_{\vec{n},k}$, where $n_{A}$ D$7_{A}$-branes sit on
the face $A$, and $k$ D$1$-branes can move on the surface of the
tetrahedron. Here we denote
\begin{equation}
\vec{n}=\left(n_{A}\right)_{A\in\underline{4}^{\vee}}.
\end{equation}
We denote the Chan-Paton spaces of the D$1$-branes and D$7_{A}$-branes
by vector spaces $\mathbf{K}$ and $\mathbf{N}_{A}$, respectively.
The presence of the D-branes breaks the ten-dimensional Lorentz group
$\mathrm{SO}\left(1,9\right)$ down to $\mathrm{SO}\left(1,1\right)_{09}\times\prod_{a\in\underline{4}}\mathrm{SO}\left(2\right)_{a}$,
where $\mathrm{SO}\left(1,1\right)_{09}$ is the two-dimensional Lorentz
group of $\mathbb{R}^{1,1}$, and $\mathrm{SO}\left(2\right)_{a}$
rotates the plane $\left(x^{2a-1},x^{2a}\right)$. 

\begin{table}
\begin{centering}
\begin{tabular}{|c|c|c|c|c|c|c|c|c|c|c|}
\hline 
\multirow{2}{*}{} & $\mathbb{R}_{t}$ & \multicolumn{2}{c|}{$\mathbb{C}_{1}$} & \multicolumn{2}{c|}{$\mathbb{C}_{2}$} & \multicolumn{2}{c|}{$\mathbb{C}_{3}$} & \multicolumn{2}{c|}{$\mathbb{C}_{4}$} & $\mathbb{R}_{\perp}$\tabularnewline
\cline{2-11} \cline{3-11} \cline{4-11} \cline{5-11} \cline{6-11} \cline{7-11} \cline{8-11} \cline{9-11} \cline{10-11} \cline{11-11} 
 & $x^{0}$ & $x^{1}$ & $x^{2}$ & $x^{3}$ & $x^{4}$ & $x^{5}$ & $x^{6}$ & $x^{7}$ & $x^{8}$ & $x^{9}$\tabularnewline
\hline 
$k$ D$1$ & $-$ & $\bullet$ & $\bullet$ & $\bullet$ & $\bullet$ & $\bullet$ & $\bullet$ & $\bullet$ & $\bullet$ & $-$\tabularnewline
\hline 
$n_{123}$ D$7_{123}$ & $-$ & $-$ & $-$ & $-$ & $-$ & $-$ & $-$ & $\bullet$ & $\bullet$ & $-$\tabularnewline
\hline 
$n_{124}$ D$7_{124}$ & $-$ & $-$ & $-$ & $-$ & $-$ & $\bullet$ & $\bullet$ & $-$ & $-$ & $-$\tabularnewline
\hline 
$n_{134}$ D$7_{134}$ & $-$ & $-$ & $-$ & $\bullet$ & $\bullet$ & $-$ & $-$ & $-$ & $-$ & $-$\tabularnewline
\hline 
$n_{234}$ D$7_{234}$ & $-$ & $\bullet$ & $\bullet$ & $-$ & $-$ & $-$ & $-$ & $-$ & $-$ & $-$\tabularnewline
\hline 
\end{tabular}
\par\end{centering}
\caption{Tetrahedron instantons constructed from D$1$-branes probing intersecting
D$7$-branes in type IIB superstring theory. Here $-$ indicates that
the D-brane extends along that direction, and $\bullet$ means that
the D-brane is located at a point in that direction. \label{D1-D7}}
\end{table}

\subsection{Condition for unbroken supersymmetry \label{subsec:SusyCondition}}

Let $Q_{L}$ and $Q_{R}$ be the supercharges which originate from
the left- and right-moving worldsheet degrees of freedom. They are
Majorana-Weyl spinors of the same chirality,
\begin{equation}
\Gamma_{c}Q_{L}=Q_{L},\quad\Gamma_{c}Q_{R}=Q_{R},\label{eq:QII}
\end{equation}
where
\begin{equation}
\Gamma_{c}=\Gamma^{0}\Gamma^{1}\cdots\Gamma^{9},\quad\Gamma_{c}^{2}=1.
\end{equation}
A set of parallel D$p$-branes along $x^{0},x^{i_{1}},\cdots,x^{i_{p}},i_{1}<\cdots<i_{p}$
preserves a linear combination $\epsilon_{L}Q_{L}+\epsilon_{R}Q_{R}$
of the supercharges with
\begin{equation}
\epsilon_{R}=\Gamma^{0}\Gamma^{i_{1}}\cdots\Gamma^{i_{p}}\epsilon_{L}.\label{eq:Dp}
\end{equation}
Hence, the presence of both the D$1$-branes and the D$7_{A}$-branes
imposes a constraint on the preserved supercharges,
\begin{equation}
\Gamma_{A}\epsilon_{L}=\epsilon_{L},\label{eq:constraint}
\end{equation}
where $\Gamma_{A}=\Gamma^{2a-1}\Gamma^{2a}\Gamma^{2b-1}\Gamma^{2b}\Gamma^{2c-1}\Gamma^{2c}$
for $A=\left(abc\right)$. The equation (\ref{eq:constraint}) has
no nonzero solutions, since $\Gamma_{A}^{2}=-1$. Hence, this configuration
is not supersymmetric.

We can also reach the conclusion that supersymmetry is completely
broken in the absence of a $B$-field by inspecting the ground state
energy. As reviewed in Appendix \ref{app:string}, the zero-point
energy in the Ramond sector is always zero due to worldsheet supersymmetry,
while that in the Neveu-Schwarz sector is given by
\begin{equation}
E^{(0)}=-\frac{1}{2}+\frac{\kappa}{8},
\end{equation}
where $\kappa$ is the number of DN and ND directions. For the D$1$-D$1$
and D$7_{A}$-D$7_{A}$ strings, $\kappa=0$ and $E^{(0)}=-\frac{1}{2}$.
This state is tachyonic and is killed by the GSO projection. The physical
ground state that survives the GSO projection has zero energy and
therefore is supersymmetric. For the D$1$-D$7_{A}$ strings, $\kappa=6$
and $E^{(0)}=\frac{1}{4}$, indicating that the ground state is stable
but not supersymmetric.

The condition for unbroken supersymmetry is modified in the presence
of a nonzero Neveu-Schwarz $B$-field \cite{Seiberg:1999vs}. We turn
on a constant $B$-field along $\mathbb{C}^{4}$ in the canonical
form, 
\begin{equation}
B=\sum_{a\in\underline{4}}b_{a}dx^{2a-1}\wedge dx^{2a},\quad b_{a}\in\mathbb{R},\label{eq:B}
\end{equation}
and define 
\begin{equation}
e^{2\pi\mathrm{i}v_{a}}=\frac{1+\mathrm{i}b_{a}}{1-\mathrm{i}b_{a}},\quad-\frac{1}{2}<v_{a}<\frac{1}{2}.
\end{equation}
Then the constraint (\ref{eq:constraint}) is modified to be
\begin{equation}
\exp\left(\sum_{a\in A}\vartheta_{a}\Gamma^{2a-1}\Gamma^{2a}\right)\epsilon_{L}=\epsilon_{L},\quad\vartheta_{a}=\pi\left(v_{a}+\frac{1}{2}\right)\in\left(0,\pi\right).\label{eq:constraintB}
\end{equation}
There are also conditions for unbroken supersymmetry due to the D$7_{acd}$-
and the D$7_{bcd}$-branes,
\begin{equation}
\exp\left(\vartheta_{a}\Gamma^{2a-1}\Gamma^{2a}\right)\epsilon_{L}=\exp\left(\vartheta_{b}\Gamma^{2b-1}\Gamma^{2b}\right)\epsilon_{L},\quad a\neq b\in\underline{4}.
\end{equation}
If we label the $32$ components of the ten-dimensional supercharges
by the eigenvalues $\left(s_{0},s_{1},s_{2},s_{3},s_{4}\right)$ of
\begin{equation}
\left(\Gamma^{0}\Gamma^{9},-\mathrm{i}\Gamma^{1}\Gamma^{2},-\mathrm{i}\Gamma^{3}\Gamma^{4},-\mathrm{i}\Gamma^{5}\Gamma^{6},-\mathrm{i}\Gamma^{7}\Gamma^{8}\right),
\end{equation}
with $s_{i}\in\left\{ \pm1\right\} $, the preserved supercharges
obey
\begin{eqnarray}
\exp\left(\mathrm{i}s_{a}\vartheta_{a}+\mathrm{i}s_{b}\vartheta_{b}+\mathrm{i}s_{c}\vartheta_{c}\right) & = & 1,\quad\forall\left(abc\right)\in\underline{4}^{\vee},\label{eq:nu1}\\
\exp\left(\mathrm{i}s_{a}\vartheta_{a}-\mathrm{i}s_{b}\vartheta_{b}\right) & = & 1,\quad\forall a\neq b\in\underline{4},\label{eq:nu2}
\end{eqnarray}
whose solutions are
\begin{eqnarray}
\vartheta_{1}=\vartheta_{2}=\vartheta_{3}=\vartheta_{4} & = & \frac{2\pi}{3},\nonumber \\
s_{1}=s_{2}=s_{3}=s_{4} & = & \pm1,\quad s_{0}=+1.
\end{eqnarray}
Hence, when the $B$-field is chosen to be
\begin{equation}
b_{1}=b_{2}=b_{3}=b_{4}=\tan\frac{\pi}{6}=\frac{1}{\sqrt{3}},\label{eq:Bsusy}
\end{equation}
or equivalently,
\begin{equation}
v_{1}=v_{2}=v_{3}=v_{4}=\frac{1}{6},\label{eq:v1234}
\end{equation}
there are two preserved supercharges, which are right-moving supercharges
$\mathcal{Q}_{+}$ and $\bar{\mathcal{Q}}_{+}$ from the viewpoint
of the common two-dimensional intersection $\mathbb{R}^{1,1}$. 

One should keep in mind that the condition (\ref{eq:Bsusy}) for unbroken
supersymmetry is valid for the original string theory vacuum. However,
a nonzero constant $B$-field away from the supersymmetric locus in
(\ref{eq:Bsusy}) can introduce instability in the form of open-string
tachyons. After tachyon condensation, the system may roll from the
original unstable and nonsupersymmetric vacuum down to a nearby supersymmetric
vacuum \cite{Seiberg:1999vs,Witten:2000mf,Nekrasov:2016gud}. Indeed,
this phenomenon happens and plays an essential role in our system.

\subsection{Low-energy spectrum}

We are interested in the low-energy spectrum of open strings ending
on D-branes. Since the system preserves $\mathcal{N}=\left(0,2\right)$
supersymmetry in two dimensions when (\ref{eq:Bsusy}) is satisfied,
it is convenient to organize fields obtained from the quantization
of open strings in terms of two-dimensional $\mathcal{N}=\left(0,2\right)$
supermultiplets. For simplicity, we assume that the $B$-field can
at most be a small deviation from the locus (\ref{eq:v1234}),
\begin{equation}
v_{a}=\frac{1}{6}+\tilde{v}_{a},\quad\left|\tilde{v}_{a}\right|\ll1.
\end{equation}
As reviewed in Appendix \ref{app:string}, we can have Neumann (N),
Dirichlet (D), and twisted (T) boundary conditions at the two ends
of the open strings.

\subsubsection{D$1$-D$1$ strings}

The D$1$-D$1$ strings satisfy NN boundary conditions along $\mathbb{R}^{1,1}$
and DD boundary conditions along $\mathbb{C}^{4}$.

In the Ramond sector, the zero-point energy vanishes. There are ten
zero modes, giving rise to $32$ degenerate ground states $\left|s_{0},s_{1},s_{2},s_{3},s_{4}\right\rangle _{\mathrm{R}}$
which form a representation of the gamma matrix algebra in $\mathbb{R}^{1,9}$.
After the GSO projection, we keep half of the ground states, which
become eight left-moving fermions and eight right-moving fermions
in two dimensions. They transform under $\mathrm{Spin}(8)\cong\mathrm{SU}(2)_{-}\times\mathrm{SU}(2)_{+}\times\mathrm{SU}(2)_{-}^{\prime}\times\mathrm{SU}(2)_{+}^{\prime}$
as $\left(\mathbf{1},\mathbf{2},\mathbf{1},\mathbf{2}\right)\oplus\left(\mathbf{1},\mathbf{2},\mathbf{2},\mathbf{1}\right)\oplus\left(\mathbf{2},\mathbf{1},\mathbf{1},\mathbf{2}\right)\oplus\left(\mathbf{2},\mathbf{1},\mathbf{2},\mathbf{1}\right)$. 

In the Neveu-Schwarz sector, the ground state $\left|0\right\rangle _{\mathrm{NS}}$
is unique and has zero-point energy $E^{(0)}=-\frac{1}{2}$. This
tachyonic mode is eliminated by the GSO projection. The excited states
$b_{-\frac{1}{2}}^{\mu}\left|0\right\rangle _{\mathrm{NS}}$ have
zero energy and survive the GSO projection. In the light-cone gauge,
they give rise to eight real scalar fields for $\mu=1,\cdots,8$.
These scalar fields describe the positions of the D$1$-branes in
$x^{1},\cdots,x^{8}$, and transform in the vector representation
of $\mathrm{Spin}(8)$. We can combine them into four complex scalars
$B_{a},a\in\underline{4}$ and their complex conjugates.

The worldvolume theory on $k$ coincident D$1$-branes is the two-dimensional
$\mathcal{N}=\left(8,8\right)$ super-Yang-Mills theory with gauge
group $\mathrm{U}\left(k\right)$, which is the dimensional reduction
of the ten-dimensional $\mathcal{N}=1$ $\mathrm{U}\left(k\right)$
super-Yang-Mills theory. 

\subsubsection{D$1$-D$7$ strings}

The boundary conditions of D$1$-D$7_{A}$ strings are NN along $\mathbb{R}^{1,1}$,
DT along $\mathbb{C}_{A}^{3}$, and DD along $\mathbb{C}_{\check{A}}$.

In the Ramond sector, the zero-point energy vanishes. In the light-cone
gauge, we have two zero modes from worldsheet fermions along $\mathbb{C}_{\check{A}}$.
Quantization of these zero modes leads to a pair of massless states
with $s_{\check{A}}=\pm1$. The GSO projection kills one of them. 

In the Neveu-Schwarz sector, the ground state has zero-point energy
\begin{equation}
E^{(0)}=\frac{1}{4}-\frac{1}{2}\sum_{a\in A}\left|v_{a}\right|=-\frac{1}{2}\sum_{a\in A}\tilde{v}_{a}.
\end{equation}
 We should consider three different cases:
\begin{enumerate}
\item When $E^{(0)}>0$, the ground state is stable, but supersymmetry is
broken.
\item When $E^{(0)}=0$, the ground state is supersymmetric. It is unique
since there are no worldsheet zero modes to be quantized. It survives
the GSO projection, and gives rise to a real scalar field which transforms
as a singlet under the $\mathrm{Spin}(8)$ group. Combining with the
similar state of D$7_{A}$-D$1$ strings and the fermions from the
Ramond sectors, we get a massless $\mathcal{N}=\left(2,2\right)$
chiral multiplet, transforming as $\left(\mathbf{k},\overline{\mathbf{n}_{A}}\right)$
under the $\mathrm{U}\left(k\right)\times\mathrm{U}\left(n_{A}\right)$
symmetry. 
\item When $E^{(0)}<0$, the ground state is tachyonic and unstable. The
lowest excited states are obtained by acting the fermionic creation
operators on it. For small $\tilde{v}_{a}$, all of these excited
states have positive energy, and it is reasonable to neglect them
when we study the low-energy theory.
\end{enumerate}

\subsubsection{D$7$-D$7$ strings}

The analysis of D$7_{A}$-D$7_{A}$ strings is similar to that of
D$1$-D$1$ strings, and we will get the dimensional reduction of
the ten-dimensional $\mathcal{N}=1$ $\mathrm{U}\left(n_{A}\right)$
supersymmetric Yang-Mills theory. The worldsheet bosons have position
and momentum zero modes along $\mathbb{R}^{1,1}\times\mathbb{C}_{A}^{3}$.
Hence the result is the eight-dimensional $\mathrm{U}\left(n_{A}\right)$
supersymmetric Yang-Mills theory with sixteen supercharges on $\mathbb{R}^{1,1}\times\mathbb{C}_{A}^{3}$.

On the other hand, the boundary conditions of D$7_{acd}$-D$7_{bcd}$
strings are NN along $\mathbb{R}^{1,1}$, TD along $\mathbb{C}_{a}$,
DT along $\mathbb{C}_{b}$, and TT along $\mathbb{C}_{c}$ and $\mathbb{C}_{d}$.
The worldsheet bosons have position and momentum zero modes along
$\mathbb{R}^{1,1}\times\mathbb{C}_{c}\times\mathbb{C}_{d}$.

In the Ramond sector, the zero-point energy vanishes. In the light-cone
gauge, we have four zero modes from worldsheet fermions along $\mathbb{C}_{c}$
and $\mathbb{C}_{d}$. Quantization of these zero modes leads to four
massless states with $\left(s_{c},s_{d}\right)=\left(\pm1,\pm1\right)$.
The GSO projection kills two states with $\left(s_{c},s_{d}\right)=\left(+1,-1\right),\left(-1,+1\right)$,
and two states with $\left(s_{c},s_{d}\right)=\left(+1,+1\right),\left(-1,-1\right)$
survive.

In the Neveu-Schwarz sector, the ground state has zero-point energy
$E^{(0)}=-\frac{1}{2}\left(\left|v_{a}\right|+\left|v_{b}\right|\right)$,
and the lowest excited states increase the energy by $\left|v_{a}\right|$
and $\left|v_{b}\right|$. Thus, the energy of the first four states
are $\frac{1}{2}\left(\pm v_{a}\pm v_{b}\right)$. The states that
survive the GSO projection have energy $\pm\frac{1}{2}\left(v_{a}-v_{b}\right)=\pm\frac{1}{2}\left(\tilde{v}_{a}-\tilde{v}_{b}\right)$.
Combining with the similar states of D$7_{bcd}$-D$7_{acd}$ strings,
we get two complex scalar fields, which are massless when $\tilde{v}_{a}=\tilde{v}_{b}.$
All the other excited states can be ignored in the low-energy theory
when $\left|\tilde{v}_{a}\right|,\left|\tilde{v}_{b}\right|\ll1$. 

Combining states in the Ramond sectors and those in the Neveu-Schwarz
sector for both D$7_{acd}$-D$7_{bcd}$ strings and D$7_{bcd}$-D$7_{acd}$
strings, we get a four-component Weyl spinor and two complex scalar
fields, which are component fields of a six-dimensional $\mathcal{N}=(1,0)$
hypermultiplet on $\mathbb{R}^{1,1}\times\mathbb{C}_{c}\times\mathbb{C}_{d}$.
These fields transform in the bifundamental representation $\left(\mathbf{n}_{acd},\overline{\mathbf{n}_{bcd}}\right)$
under $\mathrm{U}\left(n_{acd}\right)\times\mathrm{U}\left(n_{bcd}\right)$. 

\subsection{Tetrahedron instantons from the supersymmetric vacua}

We are now ready to write down the low-energy worldvolume theory on
D$1$-branes probing a configuration of intersecting D$7$-branes
in the presence of a nonzero constant $B$-field. Our goal is to find
the stable ground state of the low-energy theory. Since the D$7$-branes
are heavy from the point of view of D$1$-branes, the degrees of freedom
supported on them are frozen to their classical expectation values.
Therefore, the $\mathrm{U}\left(n_{A}\right)$ symmetry from D$7_{A}$-branes
will be treated as a global symmetry. 

\begin{table}
\begin{centering}
\begin{tabular}{|c|c|c|c|}
\hline 
Strings & $\mathcal{N}=\left(2,2\right)$ & $\mathcal{N}=\left(0,2\right)$ & $\left(\mathrm{U}\left(k\right),\mathrm{U}\left(n_{A}\right)\right)$\tabularnewline
\hline 
\multirow{4}{*}{D$1$-D$1$} & \multirow{2}{*}{Vector} & Vector $\varUpsilon$ & \multirow{4}{*}{$\left(\mathrm{\mathbf{Adj}},\mathbf{1}\right)$}\tabularnewline
\cline{3-3} 
 &  & Chiral $\varPhi_{\breve{A}}=B_{\breve{A}}+\cdots$ & \tabularnewline
\cline{2-3} \cline{3-3} 
 & \multirow{2}{*}{Chiral $(a\in A)$} & Chiral $\varPhi_{a}=B_{a}+\cdots$ & \tabularnewline
\cline{3-3} 
 &  & Fermi $\varPsi_{a,-}=\psi_{a,-}+\cdots$ & \tabularnewline
\hline 
\multirow{2}{*}{D$1$-D$7_{A}$} & \multirow{2}{*}{Chiral} & Chiral $\varPhi_{A}=I_{A}+\cdots$ & \multirow{2}{*}{$\left(\mathbf{k},\overline{\mathbf{n}_{A}}\right)$}\tabularnewline
\cline{3-3} 
 &  & Fermi $\varPsi_{A,-}=\psi_{A,-}+\cdots$ & \tabularnewline
\hline 
\end{tabular}
\par\end{centering}
\caption{Field content from D$1$-D$1$ and D$1$-D$7_{A}$ open strings in
terms of $\mathcal{N}=\left(2,2\right)$ and $\mathcal{N}=\left(0,2\right)$
supermultiplets. \label{D17}}
\end{table}

The low-energy worldvolume theory on D$1$-branes probing a single
stack of D$7_{A}$-branes with a constant $B$-field was obtained
in \cite{Witten:2000mf}. The field content is summarized in Table
\ref{D17}. In addition to the standard kinetic terms, the theory
has a superpotential
\begin{equation}
\mathcal{W}=\frac{1}{6}\epsilon^{abc}\mathrm{Tr}\varPhi_{a}\left[\varPhi_{b},\varPhi_{c}\right],\quad A=\left(abc\right),\label{eq:W}
\end{equation}
and a Fayet-Iliopoulos term with coupling 
\begin{equation}
r=\left(\sum_{a\in A}v_{a}\right)-\frac{1}{2}.
\end{equation}
We can deduce from the analysis in section \ref{subsec:SusyCondition}
that there are four preserved supercharges specified by
\begin{equation}
s_{a}=s_{b}=s_{c}=\pm1,\quad s_{\breve{A}}=\pm1,\quad s_{0}=\pm1,\quad s_{a}s_{\breve{A}}s_{0}=1.
\end{equation}
The theory at the classical level has a $\mathrm{U}(1)_{\mathcal{R}}\times\mathrm{U}(1)_{\breve{A}}$
R-symmetry, where the $\mathrm{U}(1)_{\mathcal{R}}$ ($\mathrm{U}(1)_{\breve{A}}$)
symmetry comes from rotations of $\mathbb{C}_{A}^{3}$ and $\mathbb{C}_{\breve{A}}$
in the same (opposite) directions. 

The bound states of D$1$-D$7_{A}$-branes and those of D$1$-D$7_{B}$-branes
for $A\neq B$ share the common $\mathrm{U}(1)_{\mathcal{R}}$ R-symmetry,
but the $\mathrm{U}(1)_{\breve{A}}$ symmetry and the $\mathrm{U}(1)_{\breve{B}}$
symmetry are different. Accordingly, only an $\mathcal{N}=\left(0,2\right)$
supersymmetry will be preserved if we have four stacks of D$7$-branes.
In terms of the two-dimensional $\mathcal{N}=\left(0,2\right)$ superspace,
the Lagrangian of the low-energy worldvolume theory is
\begin{eqnarray}
\mathcal{L} & = & \int d\theta^{+}d\bar{\theta}^{+}\mathrm{Tr}\left(\frac{1}{2e^{2}}\bar{\varUpsilon}\varUpsilon-\frac{\mathrm{i}}{2}\sum_{a\in\underline{4}}\bar{\varPhi}_{a}\mathcal{D}_{-}\varPhi_{a}-\frac{1}{2}\sum_{i=1}^{3}\bar{\varPsi}_{i,-}\varPsi_{i,-}\right)\nonumber \\
 &  & -\frac{1}{\sqrt{2}}\mathrm{Tr}\left(\int d\theta^{+}\left.\sum_{i=1}^{3}\varPsi_{-,i}J^{i}\right|_{\bar{\theta}^{+}=0}+c.c.\right)+\left(\frac{\mathrm{i}r}{2}\int d\theta^{+}\left.\varUpsilon\right|_{\bar{\theta}^{+}=0}+\mathrm{c.c.}\right)\nonumber \\
 &  & -\frac{1}{2}\mathrm{Tr}\sum_{A\in\underline{4}^{\vee}}\left(\mathrm{i}\bar{\varPhi}_{A}\mathcal{D}_{-}\varPhi_{A}+\bar{\varPsi}_{A,-}\varPsi_{A,-}\right),\label{eq:LD1D7}
\end{eqnarray}
where 
\begin{equation}
J^{i}=\frac{1}{2}\epsilon^{iab4}\left[\varPhi_{a},\varPhi_{b}\right],\quad E_{i}=\left[\varPhi_{4},\varPhi_{i}\right],\quad E_{A}=\varPhi_{\breve{A}}\varPhi_{A}.
\end{equation}
We also need to impose a consistency condition on the $B$-field,
\begin{equation}
v_{1}=v_{2}=v_{3}=v_{4}=\frac{1}{6}+\frac{r}{3},
\end{equation}
so that all the fields $I_{A}$ have the same mass, which may be real
or imaginary depending on the sign of the parameter $r$. This also
avoid tachyons from the quantization of the D$7$-D$7$ open strings.
Integrating out the auxiliary fields, we obtain the scalar potential
of (\ref{eq:LD1D7}),
\begin{eqnarray}
V & = & \mathrm{Tr}\left(\sum_{a\in\underline{4}}\left[B_{a},B_{a}^{\dagger}\right]+\sum_{A\in\underline{4}^{\vee}}I_{A}I_{A}^{\dagger}-r\cdot\mathds{1}_{\mathrm{U}\left(k\right)}\right)^{2}\nonumber \\
 &  & +\sum_{A\in\underline{4}^{\vee}}\mathrm{Tr}\left|B_{\breve{A}}I_{A}\right|^{2}+\sum_{a<b\in\underline{4}}\mathrm{Tr}\left|\left[B_{a},B_{b}\right]\right|^{2}.
\end{eqnarray}
Since the scalar potential $V\geq0$, the ground state is always stable.
The original string theory vacuum is given by $B_{a}=0$, $I_{A}=0$.
For $r<0$, this vacuum has positive energy, and the supersymmetry
is spontaneously broken. For $r=0$, the original string theory vacuum
preserves supersymmetry. For $r>0$, the original string theory vacuum
is not supersymmetric and does not give the global minimum of $V$.
However, the system restore supersymmetry after transitioning to a
nearby vacuum via tachyon condensation. Moreover, the theory has a
family of classical vacua, and the classical moduli space $\mathfrak{M}_{\vec{n},k}$
is given by the space of solutions to $V=0$ modulo the gauge symmetry
$\mathrm{U}\left(k\right)$,
\begin{equation}
\mathfrak{M}_{\vec{n},k}=\left.\left\{ \left.\left(\vec{B},\vec{I}\right)\right|V=0\right\} \right/\mathrm{U}(k),\label{eq:MGSI}
\end{equation}
where
\begin{equation}
\vec{B}=\left(B_{a}\right)_{a\in\underline{4}},\quad\vec{I}=\left(I_{A}\right)_{A\in\underline{4}^{\vee}}.
\end{equation}
We will call $\mathfrak{M}_{\vec{n},k}$ the moduli space of tetrahedron
instantons in the generalized gauge theory on $\bigcup_{A\in\underline{4}^{\vee}}\mathbb{C}_{A}^{3}$
with gauge groups $\left.G\right|_{A}=\mathrm{U}\left(n_{A}\right)$
and instanton number $k$.

\section{Tetrahedron instantons in noncommutative field theory \label{sec:Noncommutative}}

As shown in \cite{Seiberg:1999vs}, the dynamics of open strings connecting
D-branes in the presence of a strong constant $B$-field can be described
by a noncommutative gauge theory. The noncommutative deformation is
advantageous since the position-space uncertainty smooths out the
singularities in the conventional field theory, and it allows us to
treat uniformly the worldvolume theories of D-branes of various dimensions
\cite{Douglas:2001ba}. It also provides a natural framework for the
description of generalized gauge theories. In the section, we will
construct tetrahedron instantons as particular solutions of general
instanton equations in noncommutative field theory, to put it in perspective.

\subsection{General instanton equations}

We deform the ten-dimensional space $\mathbb{R}^{1,1}\times\mathbb{C}^{4}$
to the noncommutative space $\mathbb{R}^{1,1}\times\mathbb{C}_{\Theta}^{4}$,
where the coordinates of $\mathbb{C}_{\Theta}^{4}$ satisfy the commutation
relations
\begin{equation}
\left[z_{a},z_{b}\right]=\left[\bar{z}_{a},\bar{z}_{b}\right]=0,\quad\left[z_{a},\bar{z}_{b}\right]=-2\Theta_{a}\delta_{ab},\quad a,b\in\underline{4},\label{eq:=00005Bz=00005D}
\end{equation}
and the coordinates of $\mathbb{R}^{1,1}$ remain commutative. The
coordinates of $\mathbb{C}_{\Theta}^{4}$ are not simultaneously diagonalizable.
We introduce the creation and annihilation operators,
\begin{equation}
c_{a}^{\dagger}=\frac{z_{a}}{\sqrt{2\Theta_{a}}},\quad c_{a}=\frac{\bar{z}_{a}}{\sqrt{2\Theta_{a}}},\quad\left[c_{a},c_{b}^{\dagger}\right]=\delta_{ab},
\end{equation}
and replace the underlying spacetime manifold by the Fock module,
\begin{equation}
\mathcal{H}_{1234}=\mathbb{C}\left[c_{1}^{\dagger},c_{2}^{\dagger},c_{3}^{\dagger},c_{4}^{\dagger}\right]\left|\vec{0}\right\rangle =\bigoplus_{\vec{\mathfrak{N}}\in\mathbb{Z}_{\geq0}^{\otimes4}}\mathbb{C}\left|\vec{\mathfrak{N}}\right\rangle ,
\end{equation}
where $\left|\vec{0}\right\rangle $ is the Fock vacuum defined by
\begin{equation}
c_{a}\left|\vec{0}\right\rangle =0,a\in\underline{4},
\end{equation}
and $\vec{\mathfrak{N}}=\left(\mathfrak{N}_{1},\mathfrak{N}_{2},\mathfrak{N}_{3},\mathfrak{N}_{4}\right)$.
We define $\mathcal{H}_{S}$ for a set $S\subset\left\{ 1,2,3,4\right\} $
to be the Fock module that can be obtained from $\mathcal{H}_{1234}$
by setting $\mathfrak{N}_{a}=0$ for all $a\notin S$. We denote
\begin{equation}
\mathfrak{N}=\sum_{a\in\underline{4}}\mathfrak{N}_{a}.
\end{equation}
The creation and annihilation operators satisfy
\begin{eqnarray}
c_{a}\left|\cdots,\mathfrak{N}_{a},\cdots\right\rangle  & = & \sqrt{\mathfrak{N}_{a}}\left|\cdots,\mathfrak{N}_{a}-1,\cdots\right\rangle ,\nonumber \\
c_{a}^{\dagger}\left|\cdots,\mathfrak{N}_{a},\cdots\right\rangle  & = & \sqrt{\mathfrak{N}_{a}+1}\left|\cdots,\mathfrak{N}_{a}+1,\cdots\right\rangle .
\end{eqnarray}
The derivatives and the integrals are replaced in the noncommutative
space by
\begin{eqnarray}
\frac{\partial}{\partial z_{a}}f & \to & \frac{1}{2\Theta_{a}}\delta_{ab}\left[\bar{z}_{b},f\right],\\
\int\prod_{a\in\underline{4}}dz_{a}d\bar{z}_{a}f & \to & \prod_{a\in\underline{4}}\left(2\pi\Theta_{a}\right)\mathrm{Tr}_{\mathcal{H}_{1234}}f.
\end{eqnarray}

We now fix $\Theta_{a}=\Theta$ for all $a\in\underline{4}$. Following
\cite{Nekrasov:2002kc,Nekrasov:2016qym}, the general instanton equations
can be written as
\begin{eqnarray}
\left[\mathbf{Z}_{a},\mathbf{Z}_{b}\right]+\frac{1}{2}\epsilon_{abcd}\left[\bar{\mathbf{Z}}_{c},\bar{\mathbf{Z}}_{d}\right] & = & 0,\label{eq:ZZZZ}\\
\sum_{a\in\underline{4}}\left[\mathbf{Z}_{a},\bar{\mathbf{Z}}_{a}\right] & = & -\zeta\cdot\mathds{1}_{\mathscr{H}},\label{eq:ZZ}\\
\left[\mathbf{\Phi},\mathbf{Z}_{a}\right]=\left[\mathbf{\Phi},\bar{\mathbf{Z}}_{a}\right] & = & 0,\label{eq:PhiZ}
\end{eqnarray}
where $\mathbf{Z}_{a}$ and $\bar{\mathbf{Z}}_{a}$ are the covariant
coordinates of $\mathbb{C}_{\Theta}^{4}$,
\begin{equation}
\mathbf{Z}_{a}=c_{a}^{\dagger}+\mathrm{i}\sqrt{\frac{\Theta}{2}}\left(A_{2a-1}+\mathrm{i}A_{2a}\right),\quad\bar{\mathbf{Z}}_{a}=c_{a}-\mathrm{i}\sqrt{\frac{\Theta}{2}}\left(A_{2a-1}-\mathrm{i}A_{2a}\right),
\end{equation}
and $\mathbf{\Phi}$ is a holomorphic coordinate of $\mathbb{R}^{1,1}$.
The constant $\zeta>0$ depends on the choice of $\mathscr{H}$. 

The equations (\ref{eq:PhiZ}) are modified in the $\Omega$-background
to 
\begin{equation}
\left[\mathbf{\Phi},\mathbf{Z}_{a}\right]=\varepsilon_{a}\mathbf{Z}_{a},\quad\left[\mathbf{\Phi},\bar{\mathbf{Z}}_{a}\right]=-\varepsilon_{a}\bar{\mathbf{Z}}_{a}.\label{eq:PhiZOmega}
\end{equation}
In order to preserve the holomorphic top form $\frac{1}{4}\epsilon_{abcd}dz_{a}\wedge dz_{b}\wedge dz_{c}\wedge dz_{d}$
that is involved in (\ref{eq:ZZZZ}), we should impose the constraint
\begin{equation}
\sum_{a\in\underline{4}}\varepsilon_{a}=0.
\end{equation}

In the following, we will give various interesting solutions to the
equations (\ref{eq:ZZZZ}, \ref{eq:ZZ}, \ref{eq:PhiZOmega}) by making
different choices of $\mathscr{H}$.

\subsection{Noncommutative instantons}

The $\mathrm{U}(n)$ noncommutative instantons on $\prod_{a=1}^{p}\mathbb{C}_{a}$
correspond to the choice 
\begin{equation}
\mathscr{H}=\mathbf{N}\otimes\mathcal{H},
\end{equation}
where $\mathbf{N}\cong\mathbb{C}^{n}$, and
\begin{equation}
\mathcal{H}=\mathcal{H}_{1\cdots p}=\mathbb{C}\left[c_{1}^{\dagger},\cdots,c_{p}^{\dagger}\right]\left|0,\cdots,0\right\rangle =\bigoplus_{\vec{\mathfrak{N}}\in\mathbb{Z}_{\geq0}^{\otimes p}}\mathbb{C}\left|\mathfrak{N}_{1},\cdots,\mathfrak{N}_{p}\right\rangle .
\end{equation}
Here $p=2,3,4$ correspond to the noncommutative Yang-Mills instantons
on $\mathbb{C}_{12}^{2}$ \cite{Nekrasov:1998ss}, the noncommutative
instantons on $\mathbb{C}_{123}^{3}$ \cite{Iqbal:2003ds}, and the
instantons of the magnificent four model on $\mathbb{C}_{1234}^{4}$
\cite{Nekrasov:2017cih,Bonelli:2020gku}, respectively. They can be
obtained from the supersymmetric bound states of D$1$-branes and
$n$ D$\left(2p+1\right)$-branes with the $B$-field taken to infinity
\cite{Connes:1997cr,Seiberg:1999vs,Witten:2000mf}. The vacuum solution
is given by
\begin{eqnarray}
\mathbf{Z}_{a} & = & \begin{cases}
\mathds{1}_{\mathbf{N}}\otimes c_{a}^{\dagger}, & a=1,\cdots,p\\
0, & a=p+1,\cdots,4
\end{cases},\nonumber \\
\mathbf{\Phi} & = & \mathds{1}_{\mathbf{N}}\otimes\left(\sum_{a=1}^{p}\varepsilon_{a}c_{a}^{\dagger}c_{a}\right)-\mathrm{diag}\left(\mathtt{a}_{1},\cdots,\mathtt{a}_{n}\right)\otimes\mathds{1}_{\mathcal{H}},\label{eq:vac}
\end{eqnarray}
where $\mathtt{a}_{\alpha}$ are Coulomb parameters, and we have fixed
$\zeta=p$. In the vacuum, there is no instanton, and the gauge field
$A=0$. If we set $\varepsilon_{a}=0$ for one direction $a\in\left\{ p+1,\cdots,4\right\} $,
then $\mathbf{Z}_{a}$ is allowed to be nonzero, 
\begin{equation}
\mathbf{Z}_{a}=\mathrm{diag}\left(\mu_{1}^{(a)},\cdots,\mu_{n}^{(a)}\right)\otimes\mathds{1}_{\mathcal{H}}.
\end{equation}
For the vacuum, we have the normalized character 
\begin{equation}
\mathcal{E}_{\emptyset}=\left(\prod_{a=1}^{p}\left(1-e^{-\beta\varepsilon_{a}}\right)\right)\mathrm{Tr}_{\mathscr{H}}e^{-\beta\Phi}=\sum_{\alpha=1}^{n}e^{\beta\mathtt{a}_{\alpha}}.
\end{equation}

A large class of nontrivial solutions can be produced using the solution
generating technique \cite{Harvey:2000jb,Gross:2000ss,Kraus:2001xt}.
For simplicity, we present here only the $\mathrm{U}(1)$ case. We
make an almost gauge transformation of the vacuum solution,
\begin{eqnarray}
\mathbf{Z}_{a} & = & \begin{cases}
\mathcal{U}_{\ell}c_{a}^{\dagger}f_{\ell}\left(\sum_{a=1}^{p}c_{a}^{\dagger}c_{a}\right)\mathcal{U}_{\ell}^{\dagger}, & a=1,\cdots,p\\
0, & a=p+1,\cdots,4
\end{cases},\nonumber \\
\mathbf{\Phi} & = & \mathcal{U}_{\ell}\left(\sum_{a=1}^{p}\varepsilon_{a}c_{a}^{\dagger}c_{a}\right)\mathcal{U}_{\ell}^{\dagger}-\mathtt{a}\cdot\mathds{1}_{\mathcal{H}}.\label{eq:ZPhiU}
\end{eqnarray}
Here $\mathcal{U}_{\ell}$ is a partial isometry on $\mathcal{H}$
obeying 
\begin{equation}
\mathcal{U}_{\ell}\mathcal{U}_{\ell}^{\dagger}=\mathds{1}_{\mathcal{H}},\quad\mathcal{U}_{\ell}^{\dagger}\mathcal{U}_{\ell}=\mathds{1}_{\mathcal{H}}-\varPi_{\ell},\label{eq:UH}
\end{equation}
where $\varPi_{\ell}$ is a Hermitian projector onto a finite-dimensional
subspace of $\mathcal{H}$,
\begin{equation}
\varPi_{\ell}=\sum_{\mathfrak{N}<\ell}\left|\mathfrak{N}_{1},\cdots,\mathfrak{N}_{p}\right\rangle \left\langle \mathfrak{N}_{1},\cdots,\mathfrak{N}_{p}\right|.
\end{equation}
The real function $f_{\ell}\left(\mathfrak{N}\right)$ satisfies the
initial condition $f_{\ell}\left(\mathfrak{N}\right)=0$ for $\mathfrak{N}=0,\cdots,\ell-1$
and the finite action condition $\lim_{\mathfrak{N}\to\infty}f_{\ell}\left(\mathfrak{N}\right)=1$,
and can be found by substituting (\ref{eq:ZPhiU}) into (\ref{eq:ZZZZ},
\ref{eq:ZZ}, \ref{eq:PhiZOmega}), 
\begin{equation}
f_{\ell}\left(\mathfrak{N}\right)=\sqrt{1-\frac{\ell\left(\ell+1\right)\cdots\left(\ell+p-1\right)}{\left(\mathfrak{N}+1\right)\left(\mathfrak{N}+2\right)\cdots\left(\mathfrak{N}+p\right)}}\left(\mathds{1}_{\mathcal{H}}-\varPi_{\ell}\right).\label{eq:fln}
\end{equation}
Since $\mathcal{U}_{\ell}$ fails to be unitary only in the subspace
of $\mathcal{H}$ with $\mathfrak{N}<\ell$, (\ref{eq:ZPhiU}) is
a true gauge transformation away from a region of characteristic size
$\sqrt{\ell\Theta}$ around the origin. The solution (\ref{eq:ZPhiU})
with (\ref{eq:fln}) describes localized instantons near the origin.
These instantons would sit on top of each other if they were commutative
instantons, and the space of such configurations would have been rather
singular. The noncommutative deformation precisely resolves these
singularities, and the position-space uncertainty principle (\ref{eq:=00005Bz=00005D})
prevents the instantons from getting closer than the characteristic
size $\sqrt{\Theta}$. The topological charge is given by
\begin{equation}
k=\mathrm{ch}_{p}=\frac{\left(2\pi\Theta\right)^{p}}{p!}\mathrm{Tr}_{\mathcal{H}}\left(\frac{F}{2\pi}\right)^{p}=\frac{\ell\left(\ell+1\right)\cdots\left(\ell+p-1\right)}{p!},
\end{equation}
which is the number of states removed by the operator $\mathcal{U}_{\ell}$.
Of course, these solutions are only a subset of all the solutions,
and we can get more general solutions by relaxing the condition (\ref{eq:UH})
\cite{Kraus:2001xt}. In all these solutions, $\mathcal{U}_{\ell}$
identifies $\mathcal{H}$ with its subspace
\begin{equation}
\mathcal{H}_{\mathcal{I}}=\mathscr{I}\left(c_{1}^{\dagger},\cdots,c_{p}^{\dagger}\right)\left|0,\cdots,0\right\rangle ,\label{eq:HI}
\end{equation}
where $\mathscr{I}\left(w_{1},\cdots,w_{p}\right)\subset\mathbb{C}\left[w_{1},\cdots,w_{p}\right]$
is an ideal in the ring of polynomials, generated by monomials, and
\begin{equation}
\dim_{\mathbb{C}}\left.\mathbb{C}\left[w_{1},\cdots,w_{p}\right]\right/\mathscr{I}=k.\label{eq:dimI}
\end{equation}
Any such ideal defines a partition ($p=2$), a plane partition ($p=3$),
or a solid partition ($p=4$),
\begin{equation}
\mathscr{I}\longleftrightarrow\mathcal{Y}=\left\{ \left.\left(x_{1},\cdots,x_{p}\right)\in\mathbb{Z}_{+}^{p}\right|\prod_{a=1}^{p}w_{a}^{x_{a}-1}\notin\mathscr{I}\right\} .
\end{equation}

Let us now describe in detail the case $p=3$, which plays an important
role in this paper. The plane partition is customarily denoted by
$\pi$, and can be formed by putting $\pi_{x,y}\in\mathbb{Z}_{\ge0}$
boxes vertically at the position $(x,y)$ in a plane,
\begin{equation}
\pi=\begin{pmatrix}\pi_{1,1} & \pi_{1,2} & \pi_{1,3} & \cdots\\
\pi_{2,1} & \pi_{2,2} & \pi_{2,3} & \cdots\\
\pi_{3,1} & \pi_{3,2} & \pi_{3,3} & \cdots\\
\vdots & \vdots & \vdots & \ddots
\end{pmatrix},
\end{equation}
such that $\pi_{x,y}\geq\pi_{x+1,y},\pi_{x,y+1}$ for all $x,y\geq1$.
The volume of $\pi$ is denoted by $\left|\pi\right|$, and is given
by
\begin{equation}
\left|\pi\right|=\sum_{\left(x,y\right)}\pi_{x,y}.
\end{equation}
We can also view the plane partition $\pi$ as the set of boxes sitting
in $\mathbb{Z}_{+}^{3}$,
\begin{equation}
\pi=\left\{ \left.\left(x,y,z\right)\in\mathbb{Z}_{+}^{3}\right|1\leq z\leq\pi_{x,y}\right\} ,
\end{equation}
so that there can be at most one box at $\left(x,y,z\right)$, and
a box can occupy $\left(x,y,z\right)$ only if there are boxes in
$\left(x^{\prime},y,z\right),\left(x,y^{\prime},z\right),\left(x,y,z^{\prime}\right)$
for all $1\leq x^{\prime}<x$, $1\leq y^{\prime}<y$, $1\leq z^{\prime}<z$.
The volume of $\pi$ is then simply the total number of boxes in $\pi$.

In general, the normalized character evaluated at the solution labeled
by $\mathcal{Y}$ is given by
\begin{eqnarray}
\mathscr{E}_{\mathcal{Y}} & = & \left(\prod_{a=1}^{p}\left(1-e^{-\beta\varepsilon_{a}}\right)\right)\left.\mathrm{Tr}_{\mathscr{H}}e^{-\beta\Phi}\right|_{\mathcal{Y}}\nonumber \\
 & = & e^{\beta\mathtt{a}}-\left(\prod_{a=1}^{p}\left(1-e^{-\beta\varepsilon_{a}}\right)\right)\sum_{\left(x_{1},\cdots,x_{p}\right)\in\mathcal{Y}}e^{\beta\mathtt{a}-\beta\sum_{a=1}^{p}\varepsilon_{a}\left(x_{a}-1\right)}.
\end{eqnarray}
Once we generalize the gauge group to $\mathrm{U}(n)$, we will have
a collection of $n$ (plane, solid) partitions labeled by $\mathcal{Y}=\left\{ \mathcal{Y}^{(\alpha)},\alpha=1,\cdots,n\right\} $,
and the normalized character becomes
\begin{eqnarray}
\mathscr{E}_{\mathcal{Y}} & = & \left(\prod_{a=1}^{p}\left(1-e^{-\beta\varepsilon_{a}}\right)\right)\left.\mathrm{Tr}_{\mathscr{H}}e^{-\beta\Phi}\right|_{\mathcal{Y}}\nonumber \\
 & = & \sum_{\alpha=1}^{n}e^{\beta\mathtt{a}_{\alpha}}\nonumber \\
 &  & -\left(\prod_{a=1}^{p}\left(1-e^{-\beta\varepsilon_{a}}\right)\right)\sum_{\alpha=1}^{n}\sum_{\left(x_{1},\cdots,x_{p}\right)\in\mathcal{Y}^{(\alpha)}}e^{\beta\mathtt{a}_{\alpha}-\beta\sum_{a=1}^{p}\varepsilon_{a}\left(x_{a}-1\right)}.
\end{eqnarray}

\subsection{Spiked instantons}

We can generalize the noncommutative Yang-Mills instantons on $\mathbb{C}_{12}^{2}$
by taking
\begin{equation}
\mathscr{H}=\bigoplus_{\mathfrak{A}\in\underline{6}}\left(\mathbf{N}_{\mathfrak{A}}\otimes\mathcal{H}_{12}\right),\quad\mathbf{N}_{\mathfrak{A}}\cong\mathbb{C}^{n_{\mathfrak{A}}},\label{eq:spiked}
\end{equation}
where
\begin{equation}
\underline{6}=\begin{pmatrix}\underline{4}\\
2
\end{pmatrix}=\left\{ \left(12\right),\left(13\right),\left(14\right),\left(23\right),\left(24\right),\left(34\right)\right\} .
\end{equation}
The solutions of generalized instanton equations with (\ref{eq:spiked})
are called the spiked instantons, which can be realized in string
theory by D$1$-branes probing a stack of $n_{\mathfrak{A}}$ (anti-)D$5_{\mathfrak{A}}$-branes
in the presence of a constant $B$-field \cite{Nekrasov:2016qym,Nekrasov:2016gud}. 

The vacuum solution of spiked instantons is given by
\begin{eqnarray}
\mathbf{Z}_{1} & = & \mathds{1}_{\mathbf{N}_{12}}\otimes c_{1}^{\dagger}+\mathds{1}_{\mathbf{N}_{13}}\otimes c_{1}^{\dagger}+\mathds{1}_{\mathbf{N}_{14}}\otimes c_{1}^{\dagger},\nonumber \\
\mathbf{Z}_{2} & = & \mathds{1}_{\mathbf{N}_{12}}\otimes c_{2}^{\dagger}+\mathds{1}_{\mathbf{N}_{23}}\otimes c_{1}^{\dagger}+\mathds{1}_{\mathbf{N}_{24}}\otimes c_{1}^{\dagger},\nonumber \\
\mathbf{Z}_{3} & = & \mathds{1}_{\mathbf{N}_{13}}\otimes c_{2}^{\dagger}+\mathds{1}_{\mathbf{N}_{23}}\otimes c_{2}^{\dagger}+\mathds{1}_{\mathbf{N}_{34}}\otimes c_{1}^{\dagger},\nonumber \\
\mathbf{Z}_{4} & = & \mathds{1}_{\mathbf{N}_{14}}\otimes c_{2}^{\dagger}+\mathds{1}_{\mathbf{N}_{24}}\otimes c_{2}^{\dagger}+\mathds{1}_{\mathbf{N}_{34}}\otimes c_{2}^{\dagger},\nonumber \\
\mathbf{\Phi} & = & \bigoplus_{\mathfrak{A}\in\underline{6}}\left(\frac{1}{2}\varepsilon_{\mathfrak{A}}\cdot\mathds{1}_{\mathbf{N}_{\mathfrak{A}}}\otimes\left(\sum_{a=1}^{2}c_{a}^{\dagger}c_{a}\right)-\mathrm{diag}\left(\mathtt{a}_{\mathfrak{A},1},\cdots,\mathtt{a}_{\mathfrak{A},n_{\mathfrak{A}}}\right)\otimes\mathds{1}_{\mathcal{H}_{12}}\right).\label{eq:SI}
\end{eqnarray}
Here $\mathbf{Z}_{a}$ contains a piece in $\mathbf{N}_{\mathfrak{A}}$
if and only if $a\in\mathfrak{A}$, and $c_{a}^{\dagger}$ are assigned
to make $\left[\mathbf{Z}_{a},\mathbf{Z}_{b}\right]=0$, which are
sufficient conditions for (\ref{eq:ZZZZ}). We have no D$1$-brane
in the vacuum. The parameters $\varepsilon_{\mathfrak{A}}$ are given
in terms of the $\Omega$-deformation parameters $\varepsilon_{a}$
appearing in (\ref{eq:PhiZOmega}) by
\begin{equation}
\varepsilon_{\mathfrak{A}}=\sum_{a\in\mathfrak{A}}\varepsilon_{a}.
\end{equation}
The Coulomb parameters associated with the stack of (anti-)D$5_{\mathfrak{A}}$-branes
are $\mathtt{a}_{\mathfrak{A},\alpha}$.

We can produce nontrivial solutions of spiked instantons by substituting
in (\ref{eq:SI})
\begin{eqnarray}
\mathds{1}_{\mathbf{N}_{\mathfrak{A}}}\otimes c_{a}^{\dagger} & \to & \widetilde{\mathbf{Z}}_{\mathfrak{A},a},\\
\mathds{1}_{\mathbf{N}_{\mathfrak{A}}}\otimes\left(\sum_{a=1}^{2}c_{a}^{\dagger}c_{a}\right) & \to & \mathcal{U}_{\mathfrak{A},\ell}\left(\sum_{a=1}^{2}c_{a}^{\dagger}c_{a}\right)\mathcal{U}_{\mathfrak{A},\ell}^{\dagger}
\end{eqnarray}
where $\left(\widetilde{\mathbf{Z}}_{\mathfrak{A},1},\widetilde{\mathbf{Z}}_{\mathfrak{A},2}\right),\mathfrak{A}\in\underline{6}$
are solutions of noncommutative $\mathrm{U}\left(n_{\mathfrak{A}}\right)$
Yang-Mills instantons on $\mathbb{C}^{2}$ with $\mathscr{H}=\mathbf{N}_{\mathfrak{A}}\otimes\mathcal{H}_{12}$
and partial isometry $\mathcal{U}_{\mathfrak{A},\ell}$. Clearly,
$\left.\mathbf{Z}_{a}\right|_{\mathbf{N}_{\mathfrak{A}}}=0$ whenever
$a\notin\mathfrak{A}$. All these solutions are in one-to-one correspondence
with a collection of $\sum_{\mathfrak{A}\in\underline{6}}n_{\mathfrak{A}}$
partitions
\begin{equation}
\vec{\mathcal{Y}}=\left\{ \mathcal{Y}^{\left(\mathfrak{A},\alpha\right)},\alpha=1,\cdots,n_{\mathfrak{A}},\mathfrak{A}\in\underline{6}\right\} .
\end{equation}

\subsection{Tetrahedron instantons \label{subsec:Tetra}}

The tetrahedron instantons can be viewed as a generalization of spiked
instantons and noncommutative instantons on $\mathbb{C}^{3}$. We
take
\begin{equation}
\mathscr{H}=\bigoplus_{A\in\underline{4}^{\vee}}\left(\mathbf{N}_{A}\otimes\mathcal{H}_{123}\right),\quad\mathbf{N}_{A}\cong\mathbb{C}^{n_{A}}.
\end{equation}
The construction of the vacuum solution is similar to that for spiked
instantons,
\begin{eqnarray}
\mathbf{Z}_{1} & = & \mathds{1}_{\mathbf{N}_{123}}\otimes c_{1}^{\dagger}+\mathds{1}_{\mathbf{N}_{124}}\otimes c_{1}^{\dagger}+\mathds{1}_{\mathbf{N}_{134}}\otimes c_{1}^{\dagger},\nonumber \\
\mathbf{Z}_{2} & = & \mathds{1}_{\mathbf{N}_{123}}\otimes c_{2}^{\dagger}+\mathds{1}_{\mathbf{N}_{124}}\otimes c_{2}^{\dagger}+\mathds{1}_{\mathbf{N}_{234}}\otimes c_{1}^{\dagger},\nonumber \\
\mathbf{Z}_{3} & = & \mathds{1}_{\mathbf{N}_{123}}\otimes c_{3}^{\dagger}+\mathds{1}_{\mathbf{N}_{134}}\otimes c_{2}^{\dagger}+\mathds{1}_{\mathbf{N}_{234}}\otimes c_{2}^{\dagger},\nonumber \\
\mathbf{Z}_{4} & = & \mathds{1}_{\mathbf{N}_{124}}\otimes c_{3}^{\dagger}+\mathds{1}_{\mathbf{N}_{134}}\otimes c_{3}^{\dagger}+\mathds{1}_{\mathbf{N}_{234}}\otimes c_{3}^{\dagger},\nonumber \\
\mathbf{\Phi} & = & \bigoplus_{A\in\underline{4}^{\vee}}\left(\varepsilon_{A}\cdot\mathds{1}_{\mathbf{N}_{A}}\otimes\left(\sum_{a=1}^{3}c_{a}^{\dagger}c_{a}\right)-\mathrm{diag}\left(\mathtt{a}_{A,1},\cdots,\mathtt{a}_{A,n_{A}}\right)\otimes\mathds{1}_{\mathcal{H}_{123}}\right).\label{eq:GSI}
\end{eqnarray}
We can check that (\ref{eq:GSI}) indeed solves the equations (\ref{eq:ZZZZ},
\ref{eq:ZZ}, \ref{eq:PhiZ}). The vacuum solution describes that
there is no D$1$-brane but $n_{A}$ D$7_{A}$-branes, with the associated
Coulomb parameters given by $\mathtt{a}_{A,\alpha}$. The parameters
$\varepsilon_{A}$ are given in terms of the $\Omega$-deformation
parameters $\varepsilon_{a}$ appearing in (\ref{eq:PhiZOmega}) by
\begin{equation}
\varepsilon_{A}=\sum_{a\in A}\varepsilon_{a}.
\end{equation}

We can obtain nontrivial tetrahedron instantons by substituting in
(\ref{eq:GSI})
\begin{eqnarray}
\mathds{1}_{\mathbf{N}_{A}}\otimes c_{a}^{\dagger} & \to & \widetilde{\mathbf{Z}}_{A,a},\\
\mathds{1}_{\mathbf{N}_{A}}\otimes\left(\sum_{a=1}^{3}c_{a}^{\dagger}c_{a}\right) & \to & \mathcal{U}_{A,\ell}\left(\sum_{a=1}^{3}c_{a}^{\dagger}c_{a}\right)\mathcal{U}_{A,\ell}^{\dagger}
\end{eqnarray}
where $\left(\widetilde{\mathbf{Z}}_{A,1},\widetilde{\mathbf{Z}}_{A,2},\widetilde{\mathbf{Z}}_{A,3}\right),A\in\underline{4}^{\vee}$
are solutions of noncommutative instantons on $\mathbb{C}^{3}$ with
$\mathscr{H}=\mathbf{N}_{A}\otimes\mathcal{H}_{123}$ and partial
isometry $\mathcal{U}_{A,\ell}$. These solutions satisfy $\left.\mathbf{Z}_{\check{A}}\right|_{\mathbf{N}_{A}}=0$.
They describe bound states of D$1$-branes and $n_{A}$ D$7_{A}$-branes
in the presence of a strong $B$-field. All these solutions are in
one-to-one correspondence with a collection of $\sum_{A\in\underline{4}^{\vee}}n_{A}$
plane partitions
\begin{equation}
\vec{\pi}=\left\{ \pi^{\left(\mathcal{A}\right)},\mathcal{A}\in\underline{n}\right\} .
\end{equation}
where we combined $\left(A,\alpha\right)$ into $\mathcal{A}$, and
define
\begin{equation}
\underline{n}=\left\{ \left.\mathcal{A}=\left(A,\alpha\right)\right|\alpha=1,\cdots n_{A},A\in\underline{4}^{\vee}\right\} .
\end{equation}
We can obtain the normalized characters
\begin{eqnarray}
\mathscr{E}_{A,\vec{\pi}} & = & \left(\prod_{a\in A}\left(1-e^{-\beta\varepsilon_{a}}\right)\right)\left.\mathrm{Tr}_{\mathcal{H}_{123}\otimes\mathbf{N}_{A}}e^{-\beta\Phi}\right|_{\vec{\pi}}\nonumber \\
 & = & \sum_{\alpha=1}^{n_{A}}e^{\beta\mathtt{a}_{A,\alpha}}\nonumber \\
 &  & -\left(\prod_{a\in A}\left(1-e^{-\beta\varepsilon_{a}}\right)\right)\sum_{\alpha=1}^{n_{A}}\sum_{\left(x_{a}\right)_{a\in A}\in\pi^{\left(A,\alpha\right)}}e^{\beta\mathtt{a}_{A,\alpha}-\beta\sum_{a\in A}\varepsilon_{a}\left(x_{a}-1\right)}.\label{eq:EAPi}
\end{eqnarray}

\section{Moduli space of tetrahedron instantons \label{sec:moduli}}

In this section, we will carefully analyze the moduli space $\mathfrak{M}_{\vec{n},k}$
of tetrahedron instantons.

\subsection{Basic properties of the moduli space}

Let $\vec{B}=\left(B_{a}\right)_{a\in\underline{4}}$ and $\vec{I}=\left(I_{A}\right)_{A\in\underline{4}^{\vee}}$
be two quartets of matrices,
\begin{equation}
B_{a}\in\mathrm{End}\left(\mathbf{K}\right),\quad I_{A}\in\mathrm{Hom}\left(\mathbf{N}_{A},\mathbf{K}\right),
\end{equation}
with the vector spaces $\mathbf{K}\cong\mathbb{C}^{k}$ and $\mathbf{N}_{A}\cong\mathbb{C}^{n_{A}},A\in\underline{4}^{\vee}$.
The moduli space $\mathfrak{M}_{\vec{n},k}$ has been derived from
the string theory realization of tetrahedron instantons,
\begin{equation}
\mathfrak{M}_{\vec{n},k}=\left.\left\{ \left.\left(\vec{B},\vec{I}\right)\right|\mu^{\mathbb{R}}-r\cdot\mathds{1}_{k}=\mu^{\mathbb{C}}=\sigma=0\right\} \right/\mathrm{U}(k),
\end{equation}
where 
\begin{eqnarray}
\mu^{\mathbb{R}} & = & \sum_{a\in\underline{4}}\left[B_{a},B_{a}^{\dagger}\right]+\sum_{A\in\underline{4}^{\vee}}I_{A}I_{A}^{\dagger},\\
\mu^{\mathbb{C}} & = & \left(\mu_{ab}^{\mathbb{C}}=\left[B_{a},B_{b}\right]\right)_{a,b\in\underline{4}},\\
\sigma & = & \left(\sigma_{A}=B_{\check{A}}I_{A}\right)_{A\in\underline{4}^{\vee}},
\end{eqnarray}
and the $\mathrm{U}(k)$ symmetry acts on $B_{a}$ in the adjoint
representation and $I_{A}$ in the fundamental representation, 
\begin{equation}
\left(B_{a},I_{A}\right)\to\left(gB_{a}g^{-1},gI_{A}\right),\quad g\in\mathrm{U}(k).
\end{equation}
The metric on $\mathfrak{M}_{\vec{n},k}$ is inherited from the flat
metric on $\left(\vec{B},\vec{I}\right)$. Since the moduli space
$\mathfrak{M}_{\vec{n},k}$ is invariant under the scaling transformation
\begin{equation}
B_{a}\to\kappa B_{a},\quad I_{A}\to\kappa I_{A},\quad r\to\kappa^{2}r,\quad\kappa>0,\label{eq:rescaling}
\end{equation}
the value of $r$ is inconsequential as long as $r>0$. 

If we drop the equations $\sigma=0$, we can combine the quartet of
matrices $\vec{I}$ into a single matrix $I\in\mathrm{Hom}\left(\bigoplus_{A\in\underline{4}^{\vee}}\mathbf{N}_{A},\mathbf{K}\right)$,
and $\mathfrak{M}_{\vec{n},k}$ becomes the moduli space of instantons
in the rank $n$ magnificent four model \cite{Nekrasov:2017cih,Nekrasov:2018xsb}.

The moduli space $\mathfrak{M}_{\vec{n},k}$ admits an equivalent
description using the geometric invariant theory quotient \cite{mumford1994geometric},
\begin{equation}
\mathfrak{M}_{\vec{n},k}\cong\left.\left\{ \left.\left(\vec{B},\vec{I}\right)\right|\mu^{\mathbb{C}}=\sigma=0\right\} ^{\mathrm{stable}}\right/\mathrm{GL}(k,\mathbb{C}),\label{eq:MC}
\end{equation}
where the stability condition states that 
\begin{equation}
\sum_{A=\left(abc\right)\in\underline{4}^{\vee}}\mathbb{C}\left[B_{a},B_{b},B_{c}\right]I_{A}\left(\mathbf{N}_{A}\right)=\mathbf{K}.
\end{equation}
The virtual dimension of $\mathfrak{M}_{\vec{n},k}$ can be computed
by subtracting the number of constraints and gauge degrees of freedom
from the total number of components of the matrices,
\begin{equation}
\mathrm{vdim}_{\mathbb{C}}\mathfrak{M}_{\vec{n},k}=\left(4k^{2}+\sum_{A\in\underline{4}^{\vee}}n_{A}k\right)-\left(3k^{2}+\sum_{A\in\underline{4}^{\vee}}n_{A}k\right)-k^{2}=0.
\end{equation}
We emphasize that the vanishing virtual dimension does not mean that
the space $\mathfrak{M}_{\vec{n},k}$ is empty or a set of discrete
points. In fact, we will see that $\mathfrak{M}_{\vec{n},k}$ generally
consists of several smooth manifolds of positive actual dimensions.

We can also substitute the equations $\mu^{\mathbb{C}}=0$ with the
equations $\rho=0$ using the identity
\begin{equation}
\sum_{1\leq a<b\leq4}\mathrm{Tr}\left[B_{a},B_{b}\right]\left[B_{a},B_{b}\right]^{\dagger}=\frac{1}{2}\sum_{1\leq a<b\leq4}\mathrm{Tr}\rho_{ab}\rho_{ab}^{\dagger},
\end{equation}
where 
\begin{equation}
\rho_{ab}=\left[B_{a},B_{b}\right]+\frac{1}{2}\epsilon_{abcd}\left[B_{c}^{\dagger},B_{d}^{\dagger}\right].
\end{equation}

\subsection{Geometric interpretation}

In this subsection we discuss geometric interpretations for the moduli
space $\mathfrak{M}_{\vec{n},k}$.

Let us start with the simplest case $\vec{n}=\left(n_{123},0,0,0\right)$,
which can be realized in string theory as the bound states of $k$
D$1$-branes with $n_{123}$ D$7_{123}$-branes \cite{Iqbal:2003ds,Cirafici:2008sn,nekrasov2009instanton}.
In this case, the matrices $I_{A}$ and equations $\sigma_{A}=0$
are nontrivial only for $A=(123)$. It is useful to review two equivalent
geometric interpretations for the moduli space $\mathfrak{M}_{\left(n_{123},0,0,0\right),k}$.

Let $\mathbb{CP}^{3}=\mathbb{C}^{3}\cup\mathbb{CP}_{\infty}^{2}$
be a compactification of $\mathbb{C}^{3}$, where the homogeneous
coordinates on $\mathbb{CP}^{3}$ are $\left[z_{0}:z_{1}:z_{2}:z_{3}\right]$,
and $\mathbb{CP}_{\infty}^{2}=\left[0:z_{1}:z_{2}:z_{3}\right]$ is
the plane at infinity. We define the canonical open embedding $\iota:\mathbb{C}^{3}\hookrightarrow\mathbb{CP}^{3}$.
The moduli space $\mathfrak{M}_{\left(n_{123},0,0,0\right),k}$ coincides
with the  moduli space of $\left(\mathcal{E},\Phi\right)$, where
$\mathcal{E}$ is a torsion free sheaf on $\mathbb{CP}^{3}$ with
the Chern character
\begin{equation}
\mathrm{ch}\left(\mathcal{E}\right)=\left(n_{123},0,0,-k\right),\label{eq:chE}
\end{equation}
and the framing $\Phi$ is a trivialization of $\mathcal{E}$ on $\mathbb{CP}_{\infty}^{2}$,
\begin{equation}
\Phi:\left.\mathcal{E}\right|_{\mathbb{CP}_{\infty}^{2}}\cong\mathbf{N}_{123}\otimes\mathcal{O}_{\mathbb{CP}_{\infty}^{2}}.
\end{equation}
There is a short exact sequence
\begin{equation}
0\to\mathcal{E}\to\mathcal{F}\to\mathcal{S}_{\mathfrak{Z}}\to0,\label{eq:EFS0}
\end{equation}
where $\mathcal{F}$ is the coherent sheaf of sections of the trivial
rank $n_{123}$ holomorphic vector bundles on $\mathbb{CP}^{3}$ with
framing on $\mathbb{CP}_{\infty}^{2}$,
\begin{equation}
\mathcal{F}\cong\mathbf{N}_{123}\otimes\mathcal{O}_{\mathbb{CP}^{3}},\label{eq:F}
\end{equation}
and $\mathcal{S}_{\mathfrak{Z}}$ is a coherent sheaf supported on
the subspace $\mathfrak{Z}\subset\mathbb{C}^{3}=\mathbb{CP}^{3}\setminus\mathbb{CP}_{\infty}^{2}$,
\begin{equation}
\mathcal{S}_{\mathfrak{Z}}\cong\iota_{*}\mathcal{O}_{\mathfrak{Z}}.\label{eq:Sz}
\end{equation}
In our case, $\mathfrak{Z}$ is a union of $k$ points $\mathfrak{p}_{i}$.
The sheaf $\mathcal{F}$ is a locally free sheaf, and the torsion
free sheaf $\mathcal{E}$ fails to be locally free only along $\mathfrak{Z}$.
As a result of (\ref{eq:EFS0}), the Chern characters are related
by
\begin{equation}
\mathrm{ch}\left(\mathcal{E}\right)=\mathrm{ch}\left(\mathcal{F}\right)-\mathrm{ch}\left(\mathcal{S}_{\mathfrak{Z}}\right),\label{eq:chrelation}
\end{equation}
with 
\begin{equation}
\mathrm{ch}\left(\mathcal{F}\right)=\left(n_{123},0,0,0\right),\quad\mathrm{ch}\left(S_{\mathfrak{Z}}\right)=\left(0,0,0,\sum_{i=1}^{k}\mathrm{PD}\left[\mathfrak{p}_{i}\right]\right).
\end{equation}
Here we denote the Poincare dual of the fundamental class $\left[X\right]$
associated to $X$ by $\mathrm{PD}\left[X\right]$. From the perspective
of string theory, $\mathcal{F}$ and $\mathcal{S}_{\mathfrak{Z}}$
correspond to the D$7_{123}$-branes and the D$1$-branes, respectively.
Moreover, (\ref{eq:F}) is realized in noncommutative field theory
by the vacuum solution (\ref{eq:vac}) with $p=3$ and $\mathbf{N}=\mathbf{N}_{123}$.

As proven in \cite{Cazzaniga:2020xru}, $\mathfrak{M}_{\left(n_{123},0,0,0\right),k}$
is isomorphic to a Quot scheme
\begin{equation}
\mathfrak{M}_{\left(n_{123},0,0,0\right),k}\cong\mathrm{Quot}_{\mathbb{C}^{3}}^{k}\left(\mathcal{O}^{\oplus n_{123}}\right),
\end{equation}
which parametrizes isomorphism classes of the quotients $\mathcal{O}^{\oplus n_{123}}\twoheadrightarrow\mathcal{S}_{\mathfrak{Z}}$
such that the Hilbert-Poincare polynomial of $\mathcal{S}_{\mathfrak{Z}}$
is $k$ \cite{nitsure2005construction}. When $n_{123}=1$, this Quot
scheme is the same as the Hilbert scheme $\mathrm{Hilb}^{k}\left(\mathbb{C}^{3}\right)$
of $k$ points on $\mathbb{C}^{3}$. In noncommutative field theory,
each quotient $\mathcal{O}^{\oplus n_{123}}\twoheadrightarrow\mathcal{S}_{\mathfrak{Z}}$
corresponds to a choice of the partial isometry $\mathcal{U}_{\ell}$
with the identification (\ref{eq:HI}) satisfying (\ref{eq:dimI}).

Now we sketch a possible geometric interpretation for $\mathfrak{M}_{\vec{n},k}$
by generalizing the Quot scheme description for $\mathfrak{M}_{\left(n_{123},0,0,0\right),k}$.
We regard the worldvolume of the D$7_{123}$-branes as the physical
spacetime, and $\mathcal{F}$ is still a locally free sheaf given
by (\ref{eq:F}). The additional D$7_{A}$-branes for $A\in\underline{4}^{\vee}\setminus\left\{ (123)\right\} $
are located on the real codimension-two hyperplane $\mathfrak{h}_{A}\subset\mathbb{C}^{3}$
defined by $z_{\breve{A}}=0$, and produce real codimension-two defects
in the physical spacetime. Accordingly, $\mathfrak{Z}$ becomes a
union of hyperplanes and points,
\begin{equation}
\mathfrak{Z}=\left(\bigcup_{A\in\underline{4}^{\vee}\setminus\left\{ (123)\right\} }\mathfrak{h}_{A}\right)\cup\left(\bigcup_{i=1}^{k}\mathfrak{p}_{i}\right),
\end{equation}
and $\mathcal{S}_{\mathfrak{Z}}$ is a complex of sheaves whose entries
are $\mathbf{N}_{A}\otimes\iota_{*}\mathcal{O}_{\mathfrak{h}_{A}}$
for $A\in\underline{4}^{\vee}\setminus\left\{ (123)\right\} $, $\iota_{*}\mathcal{O}_{\mathfrak{p}_{i}}$
for $i=1,\cdots,k$, and differentials specified by strings stretching
between the D-branes. To define the Quot scheme, we need to further
specify the quotients $\mathcal{O}^{\oplus n_{123}}\twoheadrightarrow\mathcal{S}_{\mathfrak{Z}}$
by giving the Hilbert-Poincare polynomial $\mathcal{P}$, which describes
the configuration of D$1$-branes and D$7_{A}$-branes for $A\in\underline{4}^{\vee}\setminus\left\{ (123)\right\} $.
From the classical configuration of the D-branes, we can write down
their coordinate ring in a suitable basis. For example, if $\mathfrak{Z}$
arises from $n_{124}$ D$7_{124}$-branes and a single D$1$-brane,
their coordinate ring is given by
\begin{equation}
\left.\mathbb{C}\left[z_{1},z_{2},z_{3}\right]\right/\mathscr{I}_{\mathfrak{h}}\cdot\mathscr{I}_{\mathfrak{p}},
\end{equation}
where $\mathscr{I}_{\mathfrak{h}}=\left\langle Q\left(z_{3}\right)\right\rangle $
is an ideal generated by a degree $n_{124}$ polynomial $Q\left(z_{3}\right)$
which encodes the positions of D$7_{124}$-branes in $\mathbb{C}_{3}$,
and $\mathscr{I}_{\mathfrak{p}}=\left\langle z_{1}-\xi_{1},z_{2}-\xi_{2},z_{3}-\xi_{3}\right\rangle $
is an ideal which encodes the location $\left(\xi_{1},\xi_{2},\xi_{3}\right)$
of D$1$-branes in $\mathbb{C}_{123}^{3}$. From the coordinate ring,
we can calculate the Hilbert-Poincare polynomial $\mathcal{P}\left(t;n_{124}\right)$,
which is a formal power series of $t$ and depends on $n_{124}$.
In general, the Hilbert-Poincare polynomial $\mathcal{P}\left(t;n_{124},n_{134},n_{234},k\right)$
will depend on $n_{A}$ for $A\in\underline{4}^{\vee}\setminus\left\{ (123)\right\} $
and $k$. We can also read off the Chern character from $\mathcal{P}$. 

Since $\mathfrak{M}_{\vec{n},k}$ is symmetric under the permutation
of $\vec{n}$, it is natural to expect the isomorphisms such as
\begin{equation}
\mathrm{Quot}_{\mathbb{C}^{3}}^{\mathcal{P}\left(t;n_{124},n_{134},n_{234},k\right)}\left(\mathcal{O}^{\oplus n_{123}}\right)\cong\mathrm{Quot}_{\mathbb{C}^{3}}^{\mathcal{P}\left(t;n_{123},n_{134},n_{234},k\right)}\left(\mathcal{O}^{\oplus n_{124}}\right).
\end{equation}
We can interpret such isomorphisms as four possible projections of
tetrahedron instantons to the faces of the tetrahedron (see Figure
\ref{Tetrahedron}), and each shadow contains the same information.

Furthermore, the geometric interpretation for $\mathfrak{M}_{\vec{n},k}$
as the Quot scheme leads to a natural forgetful projection, 
\begin{equation}
\varrho:\mathfrak{M}_{\vec{n},k}\to\bigcup_{k^{\prime}\leq k}\mathfrak{M}_{\left(n_{123},0,0,0\right),k'},\label{eq:projection}
\end{equation}
where we drop all the information of D$7_{A}$-branes for $A\in\underline{4}^{\vee}\setminus\left\{ (123)\right\} $
in the Hilbert-Poincare polynomial.

It is rather difficult to give a geometric interpretation for $\mathfrak{M}_{\vec{n},k}$
if we want to keep the permutation symmetry of $\vec{n}$ manifest.
Here we propose a possible approach, leaving the mathematical rigor
for future work. Instead of considering four stacks of D$7_{A}$-branes
on different $\mathbb{C}_{A}^{3}$, we imagine that they would be
unified into a single D$7$-brane which wraps a complicated hyperplane
in $\mathbb{C}^{4}$. We compactify $\mathbb{C}^{4}$ into the projective
space $\mathbb{CP}^{4}=\mathbb{C}^{4}\cup\mathbb{CP}_{\infty}^{3}$
with homogeneous coordinates $\left[z_{0}:z_{1}:z_{2}:z_{3}:z_{4}\right]$,
and the hyperplane at infinity is $\mathbb{CP}_{\infty}^{3}=\left[0:z_{1}:z_{2}:z_{3}:z_{4}\right]$.
We also define $\mathbb{CP}_{A}^{3}\subset\mathbb{CP}^{4}$ and $\mathbb{CP}_{\infty,A}^{2}\subset\mathbb{CP}_{\infty}^{3}$
by $z_{\check{A}}=0$ for each $A\in\underline{4}^{\vee}$, respectively.
The hyperplane becomes an algebraic variety,
\begin{equation}
X_{\xi}=\left\{ \left[z_{0}:z_{1}:z_{2}:z_{3}:z_{4}\right]\in\mathbb{CP}^{4}\left|\left(\prod_{A\in\underline{4}^{\vee}}z_{\check{A}}^{n_{A}}\right)=\xi z_{0}^{\sum_{A\in\underline{4}^{\vee}}n_{A}}\right.\right\} ,
\end{equation}
where we introduced a small deformation parameter $\xi$ in order
to make $X_{\xi}$ a smooth manifold, and we will finally take $\xi$
to zero. We also introduce a noncompact space $\mathring{X}_{\xi}$,
which is obtained from $X_{\xi}$ by removing all points on $\mathbb{CP}_{\infty}^{3}$.
Then we can take $\mathcal{F}$ to be a rank one locally free sheaf
on $X_{\xi}$,
\begin{equation}
\mathcal{F}\cong\mathcal{O}_{X_{\xi}},
\end{equation}
and the sheaf $\mathcal{S}_{\mathfrak{Z}}$ is
\begin{equation}
\mathcal{S}_{\mathfrak{Z}}\cong\iota_{*}\mathcal{O}_{\mathfrak{Z}},
\end{equation}
where the support $\mathfrak{Z}\subset X_{\xi}\setminus\mathbb{CP}_{\infty}^{3}$
is a union of $k$ points $\mathfrak{p}_{i}$, and $\iota:\mathring{X}_{\xi}\hookrightarrow X_{\xi}$
is the natural embedding. We expect that the moduli space $\mathfrak{M}_{\vec{n},k}$
coincides with the $\xi\to0$ limit of the Hilbert scheme $\mathrm{Hilb}^{k}\left(\mathring{X}_{\xi}\right)$
of $k$ points on $\mathring{X}_{\xi}$. Equivalently, $\mathfrak{M}_{\vec{n},k}$
should also be identical to the $\xi\to0$ limit of the moduli space
of framed rank one torsion free sheaves $\mathcal{E}$ on $X_{\xi}$
with the framing 
\begin{equation}
\Phi:\left.\mathcal{E}\right|_{X_{\xi}\cap\mathbb{CP}_{\infty}^{3}}\cong\mathcal{O}_{X_{\xi}\cap\mathbb{CP}_{\infty}^{3}}.
\end{equation}
There are particularly interesting points on the moduli space $\mathfrak{M}_{\vec{n},k}$
such that the framed torsion free sheaf $\left(\mathcal{E},\Phi\right)$
admits an isomorphism,
\begin{equation}
\left(\mathcal{E},\Phi\right)\cong\bigoplus_{\mathcal{A}=(A,\alpha)\in\underline{n}}\left(\mathcal{I}_{\mathcal{A}},\Phi_{\mathcal{A}}\right),
\end{equation}
where $\mathcal{I}_{A,\alpha}$ is a rank one torsion free sheaf supported
on $\mathbb{CP}_{A}^{3}$ with the framing $\Phi_{A,\alpha}:\left.\mathcal{I}_{A,\alpha}\right|_{\mathbb{CP}_{A,\infty}^{2}}\cong\mathcal{O}_{\mathbb{CP}_{A,\infty}^{2}}$.
The tetrahedron instantons corresponding to such decompositions are
given in noncommutative field theory in section \ref{subsec:Tetra}.

\subsection{One-instanton examples}

In order to gain a better understanding of $\mathfrak{M}_{\vec{n},k}$,
we will work out explicitly the one-instanton moduli spaces step by
step. When $k=1$, the matrix $B_{a}$ is simply a complex number,
and $I_{A}=\left(I_{A,1},\cdots,I_{A,n_{A}}\right)$ is a $1\times n_{A}$
matrix if $n_{A}\geq1$. The equations $\mu^{\mathbb{C}}=0$ are satisfied
automatically, so we only need to consider
\begin{eqnarray}
\sum_{A\in\underline{4}^{\vee}}I_{A}I_{A}^{\dagger} & = & 1,\label{eq:1-instanton1}\\
B_{\check{A}}I_{A} & = & 0,\label{eq:1-instanton2}
\end{eqnarray}
where we set $r=1$ using the scaling invariance (\ref{eq:rescaling}).
Meanwhile, the group $\mathrm{U}(k)=\mathrm{U}(1)$ acts trivially
on $B_{a}$ and gives an equivalence relation $I_{A}\sim e^{\mathrm{i}\theta}I_{A}$.

\subsubsection{Instanton on $\mathbb{C}^{3}$}

We start with the rank $n$ instanton on $\mathbb{C}_{123}^{3}$ corresponding
to $\vec{n}=\left(n_{123}=n,0,0,0\right)$ \cite{nekrasov2009instanton,Cirafici:2008sn}.
There is only one $I_{A}$, namely $I_{(123)}$, and the equation
(\ref{eq:1-instanton1}) becomes
\begin{equation}
\sum_{\alpha=1}^{n}\left|I_{123,\alpha}\right|^{2}=1.\label{eq:Izeta}
\end{equation}
After modding out the $\mathrm{U}(1)$ phase, we obtain from $I_{(123)}$
a complex projective space $\mathbb{CP}^{n-1}$. Meanwhile, we get
$B_{4}=0$ from (\ref{eq:1-instanton2}), and $B_{1},B_{2},B_{3}$
are three unconstrained complex numbers. Therefore, the one-instanton
moduli space of the rank $n$ instanton on $\mathbb{C}_{123}^{3}$
is given by
\begin{equation}
\mathfrak{M}_{\left(n,0,0,0\right),1}\cong\mathbb{C}^{3}\times\mathbb{CP}^{n-1}.
\end{equation}
Here the factor $\mathbb{C}^{3}$ stands for the center of the instanton,
and the factor $\mathbb{CP}^{n-1}$ stands for the size and the gauge
orientation of the instanton.

\subsubsection{Generalized folded instanton}

We go one step further by allowing $\vec{n}$ to have two nonzero
elements, 
\begin{equation}
\vec{n}=\left(n_{123}=n,n_{124}=n^{\prime},0,0\right),
\end{equation}
which can be viewed as a generalization of the folded instantons \cite{Nekrasov:2016qym}.
In this case, the nonzero $I_{A}$ are $I_{123}$ and $I_{124}$.
$B_{1},B_{2}$ are unconstrained complex numbers. When $B_{3}$ and
$B_{4}$ are both nonzero, we know from (\ref{eq:1-instanton2}) that
$I_{123}=I_{124}=0$, which contradicts (\ref{eq:1-instanton1}).
When $B_{4}=0$ and $B_{3}\neq0$, $I_{124}=0$ and $I_{123}$ satisfies
(\ref{eq:Izeta}). Modding out the $\mathrm{U}(1)$ phase, we get
a $\mathbb{CP}^{n-1}$ from $I_{123}$. Similarly, by exchanging $3\leftrightarrow4$,
we get a $\mathbb{CP}^{n^{\prime}-1}$ from $I_{124}$. When $B_{3}=B_{4}=0$,
we have 
\begin{equation}
\sum_{\alpha=1}^{n}\left|I_{123,\alpha}\right|^{2}+\sum_{\alpha=1}^{n^{\prime}}\left|I_{124,\alpha}\right|^{2}=1,
\end{equation}
which gives a $\mathbb{CP}^{n+n^{\prime}-1}$ after modding out the
$\mathrm{U}(1)$ phase. Therefore, the moduli space $\mathfrak{M}_{\left(n,n^{\prime},0,0\right),1}$
consists of three smooth manifolds with different actual dimensions
for generic $n$ and $n^{\prime}$,
\begin{equation}
\mathfrak{M}_{\left(n,n^{\prime},0,0\right),1}\cong\mathbb{C}^{2}\times\mathbb{C}^{*}\times\mathbb{CP}^{n-1}\bigcup\mathbb{C}^{2}\times\mathbb{C}^{*}\times\mathbb{CP}^{n^{\prime}-1}\bigcup\mathbb{C}^{2}\times\mathbb{CP}^{n+n^{\prime}-1}.
\end{equation}
The first and the second components of $\mathfrak{M}_{\left(n,n^{\prime},0,0\right),1}$
correspond to the instanton being only on $\mathbb{C}_{123}^{3}$
and $\mathbb{C}_{124}^{3}$, respectively. The factor $\mathbb{C}^{2}\times\mathbb{C}^{*}$
parametrizes the center of the instanton, while $\mathbb{CP}^{n-1}$
or $\mathbb{CP}^{n^{\prime}-1}$ parametrizes the size and the gauge
orientation of the instanton. The last component of $\mathfrak{M}_{\left(n,n^{\prime},0,0\right),1}$
corresponds to the instanton being on the intersection $\mathbb{C}_{12}^{2}=\mathbb{C}_{123}^{3}\cap\mathbb{C}_{124}^{3}$,
and the center of the instanton gives the factor $\mathbb{C}^{2}$. 

Recall that the moduli space of vortices with charge $k$ in the $\mathrm{U}\left(n+n'\right)$
gauge theory is given by the symplectic quotient \cite{Hanany:2003hp}
\begin{equation}
\mathcal{V}_{n+n^{\prime},k}\cong\left.\left\{ \left(\mathtt{B},\mathtt{I}\right)\left|\left[\mathtt{B},\mathtt{B}^{\dagger}\right]+\mathtt{I}\mathtt{I}^{\dagger}=r_{v}\cdot\mathds{1}_{k}\right.\right\} \right/\mathrm{U}(k),\quad r_{v}>0,
\end{equation}
where $\mathtt{B}\in\mathrm{End}\left(\mathbb{C}^{k}\right)$, $\mathtt{I}\in\mathrm{Hom}\left(\mathbb{C}^{n+n^{\prime}},\mathbb{C}^{k}\right)$,
and the $\mathrm{U}(k)$ action is
\begin{equation}
\left(\mathtt{B},\mathtt{I}\right)\to\left(g\mathtt{B}g^{-1},g\mathtt{I}\right),\quad g\in\mathrm{U}(k).
\end{equation}
We introduce the following actions on $\mathcal{V}_{n+n^{\prime},k}$,
\begin{eqnarray}
\mathbb{T}_{1} & : & \left(\mathtt{B},\mathtt{I}\right)\to\left(q\mathtt{B},\mathtt{I}\right),\quad q\in\mathbb{C}^{\ast}\\
\mathbb{T}_{2} & : & \left(\mathtt{B},\mathtt{I}\right)\to\left(\mathtt{B},\mathtt{I}h^{-1}\right),\quad h=\mathrm{diag}\left(\stackrel{n}{\overbrace{1,\cdots,1}},\stackrel{n^{\prime}}{\overbrace{-1,\cdots,-1}}\right).
\end{eqnarray}
Now we focus on the simple case $k=1$. The fixed points of $\mathcal{V}_{n+n^{\prime},1}$
under the $\mathbb{T}_{1}$ action satisfy
\begin{equation}
\mathtt{B}=0,\quad\mathtt{I}\mathtt{I}^{\dagger}=r_{v}\cdot\mathds{1}_{k},
\end{equation}
and therefore the $\mathbb{T}_{1}$-fixed points of $\mathcal{V}_{n+n^{\prime},1}$
form a manifold
\begin{equation}
\mathcal{V}_{n+n^{\prime},1}^{\mathbb{T}_{1}}\cong\mathbb{CP}^{n+n'-1}.
\end{equation}
On the other hand, if we write
\begin{equation}
\mathtt{I}=\begin{pmatrix}\mathtt{I}_{n} & 0\\
0 & \mathtt{I}_{n^{\prime}}
\end{pmatrix},
\end{equation}
then the fixed points of $\mathcal{V}_{n+n^{\prime},k}$ under the
$\mathbb{T}_{2}$ action satisfy
\begin{equation}
\left\{ \mathtt{I}_{n}\mathtt{I}_{n}^{\dagger}=r_{v}\cdot\mathds{1}_{k},\mathtt{I}_{n^{\prime}}=0\right\} \ \mathrm{or}\ \left\{ \mathtt{I}_{n^{\prime}}\mathtt{I}_{n^{\prime}}^{\dagger}=r_{v}\cdot\mathds{1}_{k},\mathtt{I}_{n}=0\right\} ,
\end{equation}
and consequently the $\mathbb{T}_{2}$-fixed points of $\mathcal{V}_{n+n^{\prime},1}$
are given by 
\begin{equation}
\mathcal{V}_{n+n^{\prime},1}^{\mathbb{T}_{2}}\cong\mathbb{C}\times\left(\mathbb{CP}^{n-1}\cup\mathbb{CP}^{n^{\prime}-1}\right).
\end{equation}
It is interesting that the moduli space $\mathfrak{M}_{\left(n,n^{\prime},0,0\right),1}$
can be rewritten as
\begin{equation}
\mathfrak{M}_{\left(n,n^{\prime},0,0\right),1}\cong\mathbb{C}^{2}\times\left(\mathcal{V}_{n+n^{\prime},1}^{\mathbb{T}_{1}}\cup\mathcal{V}_{n+n^{\prime},1}^{\mathbb{T}_{2}}\right),
\end{equation}
which is manifestly symmetric between $n$ and $n^{\prime}$. Here
we used the fact that 
\begin{equation}
\mathcal{V}_{n+n^{\prime},1}^{\mathbb{T}_{1}}\cap\mathcal{V}_{n+n^{\prime},1}^{\mathbb{T}_{2}}\cong\left\{ 0\right\} \times\left(\mathbb{CP}^{n-1}\cup\mathbb{CP}^{n^{\prime}-1}\right).
\end{equation}
It is straightforward to generalize this relation between $\mathfrak{M}_{\left(n,n^{\prime},0,0\right),k}$
and $\mathcal{V}_{n+n^{\prime},k}$ to any positive integer $k$.

\subsubsection{Generic tetrahedron instanton}

Now it is clear how to obtain the one-instanton moduli space $\mathfrak{M}_{\vec{n},1}$
for generic $\vec{n}$. The equations (\ref{eq:1-instanton1}) and
(\ref{eq:1-instanton2}) have no solutions when all $B_{a}$ are nonzero.
When there are $r$ nonzero $B_{\check{A}}$ with $r=3,2,1,0$, the
equations (\ref{eq:1-instanton2}) require the corresponding $r$
of $I_{A}$ to be zero, and the remaining $(4-r)$ of $I_{A}$ are
constrained by (\ref{eq:1-instanton1}), producing a complex projective
space after modding out the $\mathrm{U}(1)$ phase. Combining all
the possibilities, we get
\begin{eqnarray}
\mathfrak{M}_{\vec{n},1} & \cong & \left[\bigcup_{A\in\underline{4}^{\vee}}\left(\mathbb{C}^{*}\right)^{3}\times\mathbb{CP}^{n_{A}-1}\right]\cup\left[\bigcup_{A\neq B\in\underline{4}^{\vee}}\left(\mathbb{C}^{*}\right)^{2}\times\mathbb{CP}^{n_{A,B}-1}\right]\cup\nonumber \\
 &  & \cup\left[\bigcup_{A\neq B\neq C\in\underline{4}^{\vee}}\mathbb{C}^{*}\times\mathbb{CP}^{n_{A,B,C}-1}\right]\cup\left[\mathbb{CP}^{n_{\underline{4}^{\vee}}-1}\right],
\end{eqnarray}
where
\begin{equation}
n_{S}=\sum_{A\in S}n_{A},\quad S\subset\underline{4}^{\vee}.
\end{equation}
We see that $\mathfrak{M}_{\vec{n},1}$ for generic $\vec{n}$ consists
of $2^{4}-1=15$ smooth manifolds of different actual dimensions.
The interpretation of each component of $\mathfrak{M}_{\vec{n},1}$
is a straightforward generalization of that of $\mathfrak{M}_{\left(n,n^{\prime},0,0\right),1}$.

\subsection{Symmetries of the moduli space}

In the definition of the moduli space $\mathfrak{M}_{\vec{n},k}$,
we have the freedom to pick the basis for the vector space $\mathbf{N}_{A}$.
This induces a $\mathrm{U}\left(n_{A}\right)$ symmetry, which acts
on $I_{A}$ in the anti-fundamental representation and acts trivially
on other operators,
\begin{equation}
B_{a}\to B_{a},\quad I_{B}\to\delta_{A,B}I_{B}h^{-1},\quad h\in\mathrm{U}\left(n_{A}\right).
\end{equation}
We parametrize the Cartan subalgebra of the Lie algebra of $\mathrm{U}\left(n_{A}\right)$
by
\begin{equation}
\mathtt{a}_{A}=\mathrm{diag}\left(\mathtt{a}_{A,1},\cdots,\mathtt{a}_{A,n_{A}}\right).
\end{equation}
Since the common center $\mathrm{U}(1)_{c}$ of $\prod_{A\in\underline{4}^{\vee}}\mathrm{U}\left(n_{A}\right)$
is contained in $\mathrm{U}(k)$, it is the group
\begin{equation}
\mathrm{PU}\left(\vec{n}\right)=\frac{\prod_{A\in\underline{4}^{\vee}}\mathrm{U}\left(n_{A}\right)}{\mathrm{U}(1)_{c}}
\end{equation}
that acts nontrivially on $\mathfrak{M}_{\vec{n},k}$. Accordingly,
the parameters $\mathtt{a}_{A,\alpha}$ are defined up to the simultaneous
shift $\mathtt{a}_{A,\alpha}\to\mathtt{a}_{A,\alpha}+\xi$, where
$\xi$ is a constant number. Sometimes it is useful to separate the
$\mathrm{U}\left(n_{A}\right)$ into the $\mathrm{U}(1)$ part and
the $\mathrm{SU}\left(n_{A}\right)$ part, and their respective Cartan
subalgebras are parametrized by 
\begin{equation}
\bar{\mathtt{a}}_{A}=\frac{1}{n_{A}}\sum_{\alpha=1}^{n_{A}}\mathtt{a}_{A,\alpha},\quad\tilde{\mathtt{a}}_{A,\alpha}=\mathtt{a}_{A,\alpha}-\bar{\mathtt{a}}_{A}.
\end{equation}

In addition, $\mathfrak{M}_{\vec{n},k}$ has an $\mathrm{SU}(4)$
symmetry which acts on $\left(\vec{B},\vec{I}\right)$ as
\begin{equation}
B_{a}\to U_{ab}B_{b},\quad I_{A}\to I_{A},\quad U\in\mathrm{SU}(4).
\end{equation}
This $\mathrm{SU}(4)$ symmetry is induced from the rotation symmetry
of $\mathbb{C}^{4}$ that leaves the holomorphic top form invariant.
We parametrize the Cartan subalgebra of the Lie algebra of $\mathrm{SU}(4)$
by
\begin{equation}
\varepsilon=\mathrm{diag}\left(\varepsilon_{1},\varepsilon_{2},\varepsilon_{3},\varepsilon_{4}\right),\quad\sum_{a\in\underline{4}}\varepsilon_{a}=0.
\end{equation}
For any $S\subset\underline{4}$, we define
\begin{equation}
\varepsilon_{S}=\sum_{a\in S}\varepsilon_{a}.
\end{equation}

In total, the symmetry group of $\mathfrak{M}_{\vec{n},k}$ is $\mathrm{PU}\left(\vec{n}\right)\times\mathrm{SU}(4)$.
If we adopt the holomorphic description (\ref{eq:MC}), the symmetry
group gets complexified, and its maximal torus is
\begin{equation}
\mathbf{T}=\mathbf{T}_{\vec{a}}\times\mathbf{T}_{\vec{\varepsilon}}=\mathrm{GL}(1,\mathbb{C})^{n_{\underline{4}^{\vee}}-1}\times\mathrm{GL}(1,\mathbb{C})^{3}.
\end{equation}
We denote
\begin{equation}
\vec{\mathtt{a}}=\left\{ \mathtt{a}_{\mathcal{A}},\mathcal{A}\in\underline{n}\right\} ,\quad\vec{\varepsilon}=\left\{ \varepsilon_{a},a\in\underline{4}\right\} ,
\end{equation}
and 
\begin{equation}
\vec{t}=\left\{ t_{\mathcal{A}}=e^{\beta\mathtt{a}_{\mathcal{A}}},\mathcal{A}\in\underline{n}\right\} ,\quad\vec{q}=\left\{ q_{a}=e^{\beta\varepsilon_{a}},a\in\underline{4}\right\} .
\end{equation}

\section{Instanton partition function from equivariant localization \label{sec:Zloc}}

In this section, we will compute the instanton partition function
using equivariant localization theorem. 

\subsection{Fixed points}

Generalizing the arguments of \cite{nakajima1999lectures,Nekrasov:2002qd},
we can find the set $\mathfrak{M}_{\vec{n},k}^{\mathbf{T}}$ of all
$\mathbf{T}$-fixed points of $\mathfrak{M}_{\vec{n},k}$. It is convenient
to work with the holomorphic description (\ref{eq:MC}). We also assume
that all the parameters $\vec{a},\vec{\varepsilon}$ take generic
values. The nongeneric case is more complicated but can still be handled
following \cite{Nekrasov:2016qym}.

We choose suitable bases for $\mathbf{N}_{A},A\in\underline{4}^{\vee}$
so that they decompose into one-dimensional vector spaces,
\begin{equation}
\mathbf{N}_{A}=\bigoplus_{\alpha=1}^{n_{A}}\mathbf{N}_{A,\alpha},
\end{equation}
with $\mathbf{N}_{A,\alpha}$ being the eigenspace of $\mathbf{T}_{\vec{a}}$
action with eigenvalue $t_{A,\alpha}$. If $\left(\vec{B},\vec{I}\right)$
is a $\mathbf{T}$-fixed point, it must be invariant under the combination
of an arbitrary $\mathbf{T}$-transformation and a related $\mathrm{GL}(k,\mathbb{C})$
gauge transformation, 
\begin{eqnarray}
B_{a} & = & q_{a}gB_{a}g^{-1},\quad a\in\underline{4},\nonumber \\
I_{A,\alpha} & = & gI_{A,\alpha}t_{A,\alpha}^{-1},\quad A\in\underline{4}^{\vee}.\label{eq:fixed}
\end{eqnarray}
Hence $g\left(\vec{t},\vec{q}\right)=e^{\beta\phi}\in\mathrm{GL}(k,\mathbb{C})$
defines a representation $\mathbf{T}\to\mathrm{GL}(k,\mathbb{C})$.
Since every irreducible complex representation of an abelian group
is one-dimensional, we can decompose $\mathbf{K}$ into the orthogonal
direct sum
\begin{equation}
\mathbf{K}=\bigoplus_{A\in\underline{4}^{\vee}}\mathbf{K}_{A}=\bigoplus_{\mathcal{A}\in\underline{n}}\mathbf{K}_{A,\alpha},
\end{equation}
where $\mathbf{K}_{A,\alpha}$ is the eigenspace of $\mathbf{T}_{\vec{a}}$
action with eigenvalue $t_{A,\alpha}$, and can be further decomposed
into a direct sum of eigenspaces of $\mathbf{T}_{\vec{\varepsilon}}$.
From (\ref{eq:fixed}), we have 
\begin{eqnarray}
 &  & gB_{a}^{x-1}B_{b}^{y-1}B_{c}^{z-1}I_{A}\left(\mathbf{N}_{A,\alpha}\right)\nonumber \\
 & = & q_{a}^{1-x}q_{b}^{1-y}q_{c}^{1-z}t_{A,\alpha}B_{a}^{x-1}B_{b}^{y-1}B_{c}^{z-1}I_{A}\left(\mathbf{N}_{A,\alpha}\right),\quad x,y,z\geq1.
\end{eqnarray}
Thus, $B_{a}^{x-1}B_{b}^{y-1}B_{c}^{z-1}I_{A}\left(\mathbf{N}_{A,\alpha}\right)$
is an eigenspace of $\mathbf{T}$ with eigenvalue $q_{a}^{1-x}q_{b}^{1-y}q_{c}^{1-z}t_{A,\alpha}$.
Due to the stability condition, we must have
\begin{equation}
\mathbf{K}_{A=(abc),\alpha}=\bigoplus_{\left(x,y,z\right)\in\pi^{(A,\alpha)}}B_{a}^{x-1}B_{b}^{y-1}B_{c}^{z-1}I_{A}\left(\mathbf{N}_{A,\alpha}\right),
\end{equation}
where the set $\pi^{(A,\alpha)}\subset\mathbb{Z}_{+}^{3}$ contains
$k_{A,\alpha}=\dim\mathbf{K}_{A,\alpha}$ elements. It has been shown
explicitly in \cite{Cirafici:2008sn} that all possible $\pi^{(A,\alpha)}$
are in one-to-one correspondence with plane partitions. Hence, each
$\mathbf{T}$-fixed points of $\mathfrak{M}_{\vec{n},k}$ is labeled
by a collection of plane partitions
\begin{equation}
\vec{\pi}=\left\{ \pi^{(\mathcal{A})},\mathcal{A}\in\underline{n}\right\} ,\label{eq:pp}
\end{equation}
such that the total volume of $\vec{\pi}$ is $k$,
\begin{equation}
k=\left|\vec{\pi}\right|=\sum_{\mathcal{A}\in\underline{n}}\left|\pi^{(\mathcal{A})}\right|.
\end{equation}
From the point of view of noncommutative field theory, each $\mathbf{T}$-fixed
point is given by a tetrahedron instanton sitting near the origin
of the spacetime whose solution is labeled with $\vec{\pi}$. On the
other hand, in the geometric language, each $\mathbf{T}$-fixed point
corresponds to a decomposition $\bigoplus_{\mathcal{A}=(A,\alpha)\in\underline{n}}\left(\mathcal{I}_{\mathcal{A}},\Phi_{\mathcal{A}}\right)$,
where $\mathcal{I}_{\mathcal{A}}$ is an $\mathbf{T}_{\vec{\varepsilon}}$-invariant
ideal sheaf supported on the $\mathbf{T}_{\vec{\varepsilon}}$-fixed
zero-dimensional subscheme contained in $\mathbb{C}_{A}^{3}=\mathbb{CP}_{A}^{3}\setminus\mathbb{CP}_{A,\infty}^{2}$,
and the framing $\Phi_{A,\alpha}:\left.\mathcal{I}_{A,\alpha}\right|_{\mathbb{CP}_{A,\infty}^{2}}\cong\mathcal{O}_{\mathbb{CP}_{A,\infty}^{2}}$.

\subsection{Tangent space}

Now let us look at the holomorphic tangent space $T_{\vec{\pi}}\mathfrak{M}_{\vec{n},k}$,
where $\vec{\pi}$ labels a fixed point $\left(\vec{B},\vec{I}\right)\in\mathfrak{M}_{\vec{n},k}^{\mathbf{T}}$.
If $\left(\vec{B}+\vec{b},\vec{I}+\vec{i}\right)\in\mathfrak{M}_{\vec{n},k}$
is a nearby point, then $\left(\vec{b},\vec{i}\right)$ should obey
the linearization of the equations $\mu^{\mathbb{C}}=\sigma=0$, 
\begin{equation}
d_{2}\left(\vec{b},\vec{i}\right)\equiv\left(\left[b_{a},B_{b}\right]+\left[B_{a},b_{b}\right],b_{\check{A}}I_{A}+B_{\check{A}}i_{A}\right)=0,
\end{equation}
up to an infinitesimal $\mathrm{GL}(k,\mathbb{C})$-transformation,
\begin{equation}
\left(b_{a},i_{A}\right)\sim\left(b_{a},i_{A}\right)+d_{1}\left(\phi\right),\quad d_{1}\left(\phi\right)\equiv\left(\left[\phi,B_{a}\right],\phi I_{A}\right),\quad\phi\in\mathfrak{gl}(k,\mathbb{C}).
\end{equation}
We have the following deformation complex, 
\begin{eqnarray}
0 & \to & \mathrm{End}\left(\mathbf{K}_{\vec{\pi}}\right)\xrightarrow{d_{1}}\left(\mathrm{End}\left(\mathbf{K}_{\vec{\pi}}\right)\otimes\mathbb{C}^{4}\right)\oplus\left(\bigoplus_{A\in\underline{4}^{\vee}}\mathrm{Hom}\left(\mathbf{N}_{A},\mathbf{K}_{\vec{\pi}}\right)\right)\nonumber \\
 & \xrightarrow{d_{2}} & \left(\mathrm{End}\left(\mathbf{K}_{\vec{\pi}}\right)\otimes\wedge^{2,+}\mathbb{C}^{4}\right)\oplus\left(\bigoplus_{A\in\underline{4}^{\vee}}\mathrm{Hom}\left(\mathbf{N}_{A},\mathbf{K}_{\vec{\pi}}\right)\otimes\wedge^{3}\mathbb{C}_{A}^{3}\right)\to0,
\end{eqnarray}
whose middle cohomology group is isomorphic to the tangent space $T_{\vec{\pi}}\mathfrak{M}_{\vec{n},k}$.
We can compute the $\mathbf{T}$-equivariant Chern character of $T_{\vec{\pi}}\mathfrak{M}_{\vec{n},k}$,
\begin{eqnarray}
\chi_{\vec{\pi}} & = & \mathrm{Ch}_{\mathbf{T}}\left(T_{\vec{\pi}}\mathfrak{M}_{\vec{n},k}\right)\nonumber \\
 & = & -K_{\vec{\pi}}^{*}K_{\vec{\pi}}+K_{\vec{\pi}}^{*}K_{\vec{\pi}}\mathrm{Ch}_{\mathbf{T}}\left(\mathbb{C}^{4}\right)+N_{A}^{*}K_{\vec{\pi}}-\nonumber \\
 &  & -K_{\vec{\pi}}^{*}K_{\vec{\pi}}\mathrm{Ch}_{\mathbf{T}}\left(\wedge^{2,+}\mathbb{C}^{4}\right)-\sum_{A\in\underline{4}^{\vee}}N_{A}^{*}K_{\vec{\pi}}\mathrm{Ch}_{\mathbf{T}}\left(\wedge^{3}\mathbb{C}_{A}^{3}\right)\nonumber \\
 & = & -K_{\vec{\pi}}^{*}K_{\vec{\pi}}L+\sum_{A\in\underline{4}^{\vee}}N_{A}^{*}K_{\vec{\pi}}\left(1-q_{A}^{-1}\right),\label{eq:character}
\end{eqnarray}
where $\left(e^{\beta w}\right)^{*}=e^{-\beta w}$, and
\begin{eqnarray}
N_{A} & = & \mathrm{Ch}_{\mathbf{T}}\left(\mathbf{N}_{A}\right)=\sum_{\alpha=1}^{n_{A}}t_{A,\alpha},\\
K_{\vec{\pi}} & = & \mathrm{Ch}_{\mathbf{T}}\left(\mathbf{K}_{\vec{\pi}}\right)=\left.\sum_{i=1}^{k}e^{\beta\phi}\right|_{\vec{\pi}}\nonumber \\
 & = & \sum_{A=(a,b,c)\in\underline{4}^{\vee}}\sum_{\alpha=1}^{n_{A}}t_{A,\alpha}\sum_{(x,y,z)\in\pi^{(A,\alpha)}}q_{a}^{1-x}q_{b}^{1-y}q_{c}^{1-z},\\
L & = & 1-\mathrm{Ch}_{\mathbf{T}}\left(\mathbb{C}^{4}\right)+\mathrm{Ch}_{\mathbf{T}}\left(\wedge^{2,+}\mathbb{C}^{4}\right)\nonumber \\
 & = & 1-\sum_{a\in\underline{4}}q_{a}^{-1}+q_{1}^{-1}q_{2}^{-1}+q_{1}^{-1}q_{3}^{-1}+q_{2}^{-1}q_{3}^{-1}.
\end{eqnarray}
Notice that the normalized character (\ref{eq:EAPi}) computed in
noncommutative field theory can be related to $N_{A}$ and $K_{\vec{\pi}}$
by
\begin{equation}
\mathscr{E}_{A,\vec{\pi}}=N_{A}-\left(\prod_{a\in A}\left(1-q_{a}^{-1}\right)\right)\left.K_{\vec{\pi}}\right|_{A}.
\end{equation}

\subsection{Equivariant integrals}

The $\mathbf{T}$-equivariant symplectic volume of $\mathfrak{M}_{\vec{n},k}$
is defined as the integral of the $\mathbf{T}$-equivariant cohomology
class $1\in H_{\mathbf{T}}^{*}\left(\mathfrak{M}_{\vec{n},k}\right)$
over the virtual fundamental cycle \cite{behrend1997intrinsic,li1998virtual}
of $\mathfrak{M}_{\vec{n},k}$,
\begin{equation}
\mathcal{Z}_{k}\left(\vec{\mathtt{a}};\vec{\varepsilon}\right)=\int_{\left[\mathfrak{M}_{\vec{n},k}\right]^{\mathrm{vir}}}1,\label{eq:volume}
\end{equation}
where $\left(\vec{\mathtt{a}},\vec{\varepsilon}\right)$ are generators
of $H_{\mathbf{T}}^{*}\left(\mathrm{pt}\right)$. Since $\mathfrak{M}_{\vec{n},k}$
is noncompact and is a union of manifolds of different actual dimensions,
we should apply the Atiyah-Bott equivariant localization theorem \cite{Atiyah:1984px}
in the virtual approach \cite{graber1999localization} to evaluate
the $\mathbf{T}$-equivariant integral,
\begin{equation}
\mathcal{Z}_{k}\left(\vec{\mathtt{a}};\vec{\varepsilon}\right)=\sum_{\vec{\pi},\left|\vec{\pi}\right|=k}\frac{1}{\mathrm{e_{\mathbf{T}}}\left(T_{\vec{\pi}}\mathfrak{M}_{\vec{n},k}\right)}=\sum_{\vec{\pi},\left|\vec{\pi}\right|=k}\mathbb{E}\left\{ -\chi_{\vec{\pi}}\right\} ,
\end{equation}
where $\mathrm{e_{\mathbf{T}}}\left(T_{\vec{\pi}}\mathfrak{M}_{\vec{n},k}\right)$
is the $\mathbf{T}$-equivariant Euler class of the tangent space
of $\mathfrak{M}_{\vec{n},k}$ at $\vec{\pi}$, and the operator $\mathbb{E}$
converts additive Chern characters to multiplicative classes,
\begin{equation}
\mathbb{E}\left\{ \sum_{i}m_{i}e^{\beta w_{i}}\right\} =\sideset{}{_{i}^{\prime}}\prod w_{i}^{m_{i}},\label{eq:E}
\end{equation}
where the $w_{i}=0$ term should be excluded in the product. The instanton
partition function is the generating function of $\mathcal{Z}_{k}\left(\vec{\mathtt{a}},\vec{\varepsilon}\right)$,
\begin{equation}
\mathcal{Z}\left(\vec{\mathtt{a}};\vec{\varepsilon};\mathtt{q}\right)=\sum_{k=0}^{\infty}\mathtt{q}^{k}\mathcal{Z}_{k}\left(\vec{\mathtt{a}};\vec{\varepsilon}\right)=\sum_{\vec{\pi}}\mathtt{q}^{\left|\vec{\pi}\right|}\mathbb{E}\left\{ -\chi_{\vec{\pi}}\right\} ,\label{eq:generating}
\end{equation}
where $\mathtt{q}$ is the instanton counting parameter. Notice that
$\chi_{\vec{\pi}}$ is not invariant under the permutations of $q_{a}$.
However, we have
\begin{equation}
L+L^{*}=\prod_{a\in\underline{4}}\left(1-q_{a}^{-1}\right).
\end{equation}
Therefore, $\mathbb{E}\left\{ -\chi_{\vec{\pi}}\right\} $ is invariant
under the permutations of $q_{a}$, up to an overall $\pm$ sign that
depends on the ordering in $a\in\underline{4}$. The orientation problem
also appeared in the study of magnificent four model \cite{Nekrasov:2017cih,Nekrasov:2018xsb}.

We can obtain the K-theoretic and elliptic versions of the instanton
partition function by replacing the integrand in (\ref{eq:volume})
from $1$ to the arithmetic genus $\hat{A}_{\beta}\left(\mathfrak{M}_{\vec{n},k}\right)$
and the elliptic genus $\varphi_{\mathrm{ell}}\left(\mathfrak{M}_{\vec{n},k}\right)$,
respectively \cite{Losev:1997tp,Baulieu:1997nj}. Correspondingly,
the definition of the operator $\mathbb{E}$ becomes
\begin{equation}
\mathbb{E}\left\{ \sum_{i}m_{i}e^{\beta w_{i}}\right\} =\begin{cases}
\sideset{}{^{\prime}}\prod_{i}\left(1-e^{\beta w_{i}}\right)^{m_{i}}, & \mathrm{K-theoretical}\\
\sideset{}{^{\prime}}\prod_{i}\theta_{1}\left(\left.w_{i}\right|\tau\right)^{m_{i}}, & \mathrm{elliptic}
\end{cases}.\label{eq:E-K-ell}
\end{equation}

In fact, the result (\ref{eq:generating}) suggests a more refined
version of the instanton partition function with four independent
instanton counting parameters $\mathtt{q}_{A}$ for $A\in\underline{4}^{\vee}$,
\begin{equation}
\mathcal{Z}^{\mathrm{ref}}\left(\vec{\mathtt{a}};\vec{\varepsilon};\vec{\mathtt{q}}\right)=\sum_{\vec{\pi}}\prod_{A\in\underline{4}^{\vee}}\mathtt{q}_{A}^{\left|\pi^{(A)}\right|}\mathbb{E}\left\{ -\chi_{\vec{\pi}}\right\} ,
\end{equation}
where $\vec{\mathtt{q}}=\left\{ \mathtt{q}_{A},A\in\underline{4}^{\vee}\right\} $
and $\left|\pi^{(A)}\right|=\sum_{\alpha=1}^{n_{A}}\left|\pi^{(A,\alpha)}\right|$.

\section{Instanton partition function from elliptic genus \label{sec:Zell}}

In this section, we will compute the instanton partition function
from the elliptic genus of the low-energy worldvolume theory on D$1$-branes,
where all the heavy stringy modes are decoupled.

\subsection{Definition via elliptic genus}

We have shown that the low-energy worldvolume theory on D$1$-branes
probing a system of intersecting D$7$-branes is a two-dimensional
$\mathcal{N}=\left(0,2\right)$ supersymmetric gauge theory, with
two supercharges $Q_{+}$ and $\bar{Q}_{+}$. This theory has a $\mathrm{U}(1)^{4}$
global symmetry induced from $\prod_{a\in\underline{4}}\mathrm{SO}\left(2\right)_{a}$
rotating $\mathbb{C}^{4}$. The corresponding bosonic generators $\mathcal{J}_{a}$
commute with each other, but do not commute with $Q_{+}$ and $\bar{Q}_{+}$,
\begin{equation}
\left[\mathcal{J}_{a},Q_{+}\right]=-Q_{+},\quad\left[\mathcal{J}_{a},\bar{Q}_{+}\right]=\bar{Q}_{+}.\label{eq:JQ}
\end{equation}
We can choose three linearly independent combinations of $\mathcal{J}_{a}$,
for instance
\begin{equation}
\left(\mathcal{J}_{1}-\mathcal{J}_{4},\mathcal{J}_{2}-\mathcal{J}_{4},\mathcal{J}_{3}-\mathcal{J}_{4}\right),
\end{equation}
which commute with $Q_{+}$ and $\bar{Q}_{+}$. They generate a group
$\mathrm{U}(1)^{3}\subset\mathrm{U}(1)^{4}$, and can be identified
with $\mathbf{T}_{\vec{\varepsilon}}$. 

The elliptic genus of the two-dimensional worldvolume theory on $k$
D$1$-branes probing intersecting D$7$-branes is defined to be
\begin{eqnarray}
 &  & Z_{k}\left(\tau;\vec{\mathtt{a}};\varepsilon_{1},\varepsilon_{2},\varepsilon_{3}\right)\nonumber \\
 & = & \mathrm{Tr}_{\mathcal{H}_{k}}\left[\left(-1\right)^{F}q^{H_{L}}\bar{q}^{H_{R}}\prod_{a=1}^{3}e^{2\pi\mathrm{i}\varepsilon_{a}\left(\mathcal{J}_{a}-\mathcal{J}_{4}\right)}\prod_{\mathcal{A}\in\underline{n}}e^{2\pi\mathrm{i}\mathtt{a}_{\mathcal{A}}T_{\mathcal{A}}}\right],\label{eq:elliptic}
\end{eqnarray}
where the trace is taken in the RR sector of the Hilbert space $\mathcal{H}_{k}$
of the worldvolume theory, $H_{L}$ and $H_{R}$ are the left- and
right-moving Hamiltonians respectively, and $T_{\mathcal{A}=\left(A,\alpha\right)},\alpha=1,\cdots,n_{A}$
are the Cartan generators of the symmetry group $\mathrm{U}\left(n_{A}\right)$.
The parameter 
\begin{equation}
q=e^{2\pi\mathrm{i}\tau}
\end{equation}
specifies the complex structure $\tau$ of a torus. The fugacities
$\left(\varepsilon_{1},\varepsilon_{2},\varepsilon_{3}\right)$ and
$\mathtt{a}_{\mathcal{A}=\left(A,\alpha\right)}$ are associated with
$\mathrm{U}(1)^{3}$ and $\mathrm{U}\left(n_{A}\right)$, respectively.
We can introduce $\varepsilon_{4}=-\varepsilon_{1}-\varepsilon_{2}-\varepsilon_{3}$
to make the expression more symmetric,
\begin{equation}
Z_{k}\left(\tau;\vec{\mathtt{a}};\vec{\varepsilon}\right)=\mathrm{Tr}_{\mathcal{H}_{k}}\left[\left(-1\right)^{F}q^{H_{L}}\bar{q}^{H_{R}}\prod_{a\in\underline{4}}e^{2\pi\mathrm{i}\varepsilon_{a}\mathcal{J}_{a}}\prod_{\mathcal{A}\in\underline{n}}e^{2\pi\mathrm{i}\mathtt{a}_{\mathcal{A}}T_{\mathcal{A}}}\right]_{\sum_{a\in\underline{4}}\varepsilon_{a}=0}.
\end{equation}
It is clear that $\varepsilon_{a},a\in\underline{4}$ can be identified
with the standard $\Omega$-deformation parameters \cite{Nekrasov:2002qd}.
The instanton partition function is then the grand canonical partition
function of the elliptic genus,
\begin{equation}
Z^{\mathrm{inst}}\left(\tau;\vec{\mathtt{a}};\vec{\varepsilon};\mathtt{q}\right)=1+\sum_{k=1}^{\infty}\mathtt{q}^{k}Z_{k}\left(\tau;\vec{\mathtt{a}};\vec{\varepsilon}\right).\label{eq:Zinstelliptic}
\end{equation}

The elliptic genus can be calculated using the supersymmetric localization
techniques, and is given by contour integrals \cite{Benini:2013xpa},
\begin{equation}
Z_{k}=\frac{1}{k!}\int\prod_{i=1}^{k}d\phi_{i}\left(Z_{k}^{1-1}\prod_{A\in\underline{4}^{\vee}}Z_{k}^{1-7_{A}}\right),\label{eq:Zkell}
\end{equation}
where $k!$ is the order of the Weyl group of $\mathrm{U}(k)$. The
contributions from the D$1$-D$1$ strings and D$1$-D$7_{A}$ strings
are \cite{Benini:2018hjy}
\begin{eqnarray}
Z_{k}^{1-1} & = & \left[\frac{2\pi\eta(\tau)^{3}\prod_{1\leq a<b\leq3}\theta\left(\varepsilon_{ab}\right)}{\prod_{a\in\underline{4}}\theta\left(\varepsilon_{a}\right)}\right]^{k}\times\nonumber \\
 &  & \times\prod_{\substack{i,j=1\\
i\neq j
}
}^{k}\frac{\theta\left(\phi_{ij}\right)\prod_{1\leq a<b\leq3}\theta\left(\phi_{ij}+\varepsilon_{ab}\right)}{\prod_{a\in\underline{4}}\theta\left(\phi_{ij}+\varepsilon_{a}\right)},\\
Z_{k}^{1-7_{A}} & = & \prod_{i=1}^{k}\prod_{\alpha=1}^{n_{A}}\frac{\theta\left(\phi_{i}-\mathtt{a}_{A,\alpha}-\varepsilon_{A}\right)}{\theta\left(\phi_{i}-\mathtt{a}_{A,\alpha}\right)},
\end{eqnarray}
where $\phi_{ij}=\phi_{i}-\phi_{j}$, and we use the abbreviation
\footnote{Hopefully, this does not create any confusion. Especially, this is
different from the notation $\theta_{i}\left(\tau\right)\equiv\theta_{i}\left(\left.0\right|\tau\right)$
that often appears in the literature.}
\begin{equation}
\theta\left(z\right)\equiv\theta_{1}\left(\left.z\right|\tau\right).\label{eq:theta}
\end{equation}
We emphasize that the detailed description of $\mathfrak{M}_{\vec{n},k}$
is not used in the computation. We only need to know the supermultiplets
that appear in the worldvolume theory, as well as their charges under
the symmetry group $\mathbf{T}$.

We see that $Z^{\mathrm{inst}}$ is invariant under an overall shift,
\begin{equation}
\phi_{i}\to\phi_{i}-\xi,\quad\mathtt{a}_{A,\alpha}\to\mathtt{a}_{A,\alpha}+\xi.\label{eq:shift}
\end{equation}
This confirms the claim that the center $\mathrm{U}(1)_{c}$ of $\prod_{A\in\underline{4}^{\vee}}\mathrm{U}\left(n_{A}\right)$
acts trivially, and the final result of the partition function should
not dependent on the overall shift of $\mathtt{a}_{A,\alpha}$.

The integral in (\ref{eq:Zkell}) make sense only when the integrand
is invariant under the large gauge transformations $\phi_{i}\to\phi_{i}+r+s\tau$
for $r,s\in\mathbb{Z}$ \cite{Benini:2018hjy,Benini:2013xpa}. From
the transformation property of the Jacobi theta function $\theta_{1}\left(\left.z\right|\tau\right)$
under shifts of $z$,
\begin{equation}
\theta_{1}\left(\left.z+r+s\tau\right|\tau\right)=\left(-1\right)^{r+s}\exp\left(-\pi\mathrm{i}s^{2}\tau-2\pi\mathrm{i}sz\right)\theta_{1}\left(\left.z\right|\tau\right),\quad r,s\in\mathbb{Z},
\end{equation}
we obtain that 
\begin{equation}
Z_{k}\to\left(\prod_{A\in\underline{4}^{\vee}}\prod_{i=1}^{k}\prod_{\alpha=1}^{n_{A}}e^{2\pi\mathrm{i}s\varepsilon_{A}}\right)Z_{k}.
\end{equation}
To get rid of the extra phase factor for all $k\in\mathbb{Z}^{+}$,
we should impose the consistency condition 
\begin{equation}
\sum_{A\in\underline{4}^{\vee}}n_{A}\varepsilon_{A}\in\mathbb{Z},\label{eq:nepsilon}
\end{equation}
which generalizes the similar condition obtained in \cite{Benini:2018hjy}.

\subsection{$k=1$}

When $k=1$, the elliptic genus is given by
\begin{equation}
Z_{1}=\left[\frac{2\pi\eta(\tau)^{3}\prod_{1\leq a<b\leq3}\theta\left(\varepsilon_{ab}\right)}{\prod_{a\in\underline{4}}\theta\left(\varepsilon_{a}\right)}\right]\int d\phi\prod_{\mathcal{A}\in\underline{n}}\frac{\theta\left(\phi-\mathtt{a}_{\mathcal{A}}-\varepsilon_{\mathcal{A}}\right)}{\theta\left(\phi-\mathtt{a}_{\mathcal{A}}\right)}.
\end{equation}
It is straightforward to evaluate this integral explicitly. The set
of poles in the integrand are 
\begin{equation}
\mathscr{M}_{\ast}^{\mathrm{sing}}=\left\{ \left.\phi\right|\phi-\mathtt{a}_{\mathcal{A}}=0\mod\mathbb{Z}+\tau\mathbb{Z}\right\} .
\end{equation}
We should take all of them and the result is given by
\begin{eqnarray}
Z_{1} & = & \left[\frac{2\pi\eta(\tau)^{3}\prod_{1\leq a<b\leq3}\theta\left(\varepsilon_{ab}\right)}{\prod_{a\in\underline{4}}\theta\left(\varepsilon_{a}\right)}\right]\sum_{\phi_{\ast}\in\mathscr{M}^{\mathrm{sing}}}\oint_{\phi_{\ast}}d\phi\prod_{\mathcal{A}\in\underline{n}}\frac{\theta\left(\phi-\mathtt{a}_{\mathcal{A}}-\varepsilon_{\mathcal{A}}\right)}{\theta\left(\phi-\mathtt{a}_{\mathcal{A}}\right)}\nonumber \\
 & = & \sum_{\mathcal{A}=\left(A,\alpha\right)\in\underline{n}}\left[\frac{\prod_{a<b\in A}\theta\left(\varepsilon_{ab}\right)}{\prod_{a\in A}\theta\left(\varepsilon_{a}\right)}\prod_{\mathcal{B}\in\underline{n}\setminus\left\{ \mathcal{A}\right\} }\frac{\theta\left(\mathtt{a}_{\mathcal{A}}-\mathtt{a}_{\mathcal{B}}-\varepsilon_{\mathcal{B}}\right)}{\theta\left(\mathtt{a}_{\mathcal{A}}-\mathtt{a}_{\mathcal{B}}\right)}\right],
\end{eqnarray}
where we have used $\sum_{a\in\underline{4}}\varepsilon_{a}=0$. Due
to the product over $a<b\in A$, the result depends on the ordering
of $a\in\underline{4}$.

\subsection{General $k$}

Now we proceed with general $k$. As shown in \cite{Benini:2013xpa},
we should apply the JK residue formula \cite{jeffrey1995localization}
to evaluate the contour integrals in (\ref{eq:Zkell}).

\subsubsection{Classification of potential poles in terms of trees }

We first classify all the potential poles in the integrand that can
have nonzero JK-residues, temporarily ignoring the numerator. 

The denominator of the integral (\ref{eq:Zkell}) becomes zero along
the hyperplanes
\begin{eqnarray}
H_{A,ij,a} & = & \left\{ \phi_{i}-\phi_{j}=-\varepsilon_{a}\right\} ,\label{eq:HA}\\
H_{F,i,\mathcal{A}} & = & \left\{ \phi_{i}=\mathtt{a}_{\mathcal{A}}\right\} ,\label{eq:HF}
\end{eqnarray}
where the identifications up to $\mathbb{Z}+\tau\mathbb{Z}$ are understood.
We introduce the standard basis $\left\{ \mathbf{e}_{i}\right\} _{i=1,\cdots,k}$
of $\mathbb{R}^{k}$,
\begin{equation}
\mathbf{e}_{i}=\left(0,\cdots,0,\overset{i}{1},0,\cdots,\overset{k}{0}\right).
\end{equation}
The charge vectors associated with (\ref{eq:HA}) and (\ref{eq:HF})
are $\mathbf{h}_{A,ij}=\mathbf{e}_{i}-\mathbf{e}_{j}$ and $\mathbf{h}_{F,i}=\mathbf{e}_{i}$,
respectively.

A singularity is called nondegenerate if exactly $k$ linearly independent
hyperplanes intersect at the point, and is called degenerate if the
total number of hyperplanes through the point is greater than $k$.
A practical way to deal with the degenerate singularities is to blowup
them into nondegenerate ones by introducing small generic nonphysical
fugacities to deform the hyperplane arrangement. In the end of the
computation, we remove the deformation by sending the nonphysical
fugacities to zero in a continuous way. In the following, we will
only consider the situation where all singularities are nondegenerate. 

We denote the charge vectors of the $k$ hyperplanes by
\begin{equation}
\mathbf{Q}=\begin{pmatrix}\mathbf{Q}_{1}\\
\vdots\\
\mathbf{Q}_{k}
\end{pmatrix},\quad\mathbf{Q}_{I}\in\left\{ \mathbf{h}_{A},\mathbf{h}_{F}\right\} .
\end{equation}
The JK-residue can be nonzero only if $\boldsymbol{\eta}\in\mathrm{Cone}\left(\mathbf{Q}\right)$,
i.e.,
\begin{equation}
\sum_{I=1}^{k}\lambda_{I}\mathbf{Q}_{I}=\boldsymbol{\eta},\quad\lambda_{I}>0.\label{eq:cone}
\end{equation}
In our problem, the result will depend on $\boldsymbol{\eta}$, and
we should take the standard choice \footnote{One could take other choices of $\boldsymbol{\eta}$ if one includes
a $P$ field as in \cite{Benini:2018hjy}.}
\begin{equation}
\boldsymbol{\eta}=\sum_{i=1}^{k}\mathbf{e}_{i}=\left(1,1,\cdots,1\right).
\end{equation}
Since charge vectors of type $\mathbf{h}_{A}$ only generate at most
a $(k-1)$-dimensional subspace of $\mathbb{R}^{k}$, $\mathbf{Q}$
must contain $M\geq1$ charge vectors of type $\mathbf{h}_{F}$, which
are taken to be $\mathbf{e}_{1},\cdots,\mathbf{e}_{M}$ using Weyl
permutations. We will show that it is possible to divide $\mathbf{Q}$
into $M$ subsets in such a way that each subset contains exactly
one charge vector of type $\mathbf{h}_{F}$.\textbf{ }Let us start
with $\mathbf{e}_{1}$. If $\mathbf{Q}_{j_{1}}=\mathbf{e}_{1}-\mathbf{e}_{j_{1}}$
is also in $\mathbf{Q}$, the condition (\ref{eq:cone}) gives
\begin{equation}
\left(\lambda_{1}+\lambda_{j_{1}}\right)\mathbf{e}_{1}+\sum_{I\neq1,j_{1}}\lambda_{I}\mathbf{Q}_{I}=\lambda_{j_{1}}\mathbf{e}_{j_{1}}+\sum_{i=1}^{k}\mathbf{e}_{i}.
\end{equation}
Since the coefficient of $\mathbf{e}_{j_{1}}$ on the right-hand side
is positive, $\mathbf{Q}$ must contain $\mathbf{e}_{j_{1}}-\mathbf{e}_{j_{2}}$
for at least one $j_{2}$. Notice that $\mathbf{e}_{j_{1}}$ cannot
be in $\mathbf{Q}$, since it is not linearly independent with $\mathbf{e}_{1}$
and $\mathbf{e}_{1}-\mathbf{e}_{j_{1}}$ that are already in $\mathbf{Q}$.
Then the same argument for $\mathbf{e}_{j_{2}}$ leads to the requirement
that $\mathbf{Q}$ must contain $\mathbf{e}_{j_{2}}-\mathbf{e}_{j_{3}}$
for at least one $j_{3}\neq1,j_{1}$. Since there are only a finite
number of elements in $\mathbf{Q}$, this procedure cannot be carried
on forever, and finally it is impossible to match the coefficient
of one $\mathbf{e}_{i}$. Therefore, $\mathbf{e}_{1}-\mathbf{e}_{j}$
is not allowed to be in $\mathbf{Q}$.

On the contrary, $\mathbf{Q}$ can contain one or more charge vectors
$\mathbf{e}_{j_{1}^{(\mu)}}-\mathbf{e}_{1}$, which are labeled by
$\mu=1,\cdots$ , and we require $j_{1}^{(\mu)}>M$ in order to avoid
linearly dependent combinations of charge vectors. We can draw an
oriented rooted tree. The root vertex is labeled by $\mathbf{e}_{1}$.
For each $\mathbf{e}_{j_{1}^{(\mu)}}-\mathbf{e}_{1}\in\mathbf{Q}$,
we put an arrow from $\mathbf{e}_{1}$ to the vertex $\mathbf{e}_{j_{1}^{(\mu)}}$.
We can go on and add $\mathbf{e}_{j_{2}^{(\nu)}}-\mathbf{e}_{j_{1}^{(\mu)}}$
in $\mathbf{Q}$, with $j_{2}^{(\nu)}$ being different from $1,\cdots,M$
and $j_{1}^{(\mu)}$ so that there are no linear relations among selected
charge vectors. The tree grows by adding the vertices $\mathbf{e}_{j_{2}^{(\nu)}}$
and arrows from $\mathbf{e}_{j_{1}^{(\mu)}}$ to $\mathbf{e}_{j_{2}^{(\nu)}}$.
We can repeat this construction until no charge vectors can be further
added in this way, ending up with an oriented rooted tree with root
$\mathbf{e}_{1}$ and arrows corresponding to charge vectors of type
$\mathbf{h}_{A,ij},i>j$. The linearly independent condition ensures
that there can be no cycles. Subsequently, we can proceed with $\mathbf{e}_{2}$,
and produce a similar oriented rooted tree. The trees with root $\mathbf{e}_{1}$
and $\mathbf{e}_{2}$ must be disconnected, otherwise there will be
linear relations among charge vectors. After performing this construction
for all $\mathbf{e}_{1},\cdots,\mathbf{e}_{M}$, we divide all the
charge vectors in $\mathbf{Q}$ into a disjoint union of $M$ oriented
rooted trees, with $k$ vertices in total.

It is convenient to perform a Weyl permutation of $\phi_{i}$ so that
$\mathbf{Q}$ form a block diagonal matrix, 
\begin{equation}
\mathbf{Q}=\mathrm{diag}\left(\mathbf{Q}^{(1)},\cdots,\mathbf{Q}^{(M)}\right),
\end{equation}
where the block $\mathbf{Q}^{(m)}$ is a square matrix of order $k_{m}$,
\begin{equation}
\mathbf{Q}^{(m)}=\begin{pmatrix}1 & 0 & 0 & 0 & \cdots & 0\\
-1 & 1 & 0 & 0 & \cdots & 0\\
\ast & \ast & 1 & 0 & \cdots & 0\\
\ast & \ast & \ast & 1 & \cdots & 0\\
\vdots & \vdots & \vdots & \vdots & \ddots & \vdots\\
\ast & \ast & \ast & \ast & \cdots & 1
\end{pmatrix},\quad m=1,\dots,M,\quad\sum_{m=1}^{M}k_{m}=k.
\end{equation}
The first row in $\mathbf{Q}^{(m)}$ corresponds to the root of the
$m$-th tree, and the remaining rows correspond to the other vertices
of the $m$-th tree. Each $\ast$ can be either $0$ or $-1$, and
there is exactly one $-1$ in each row containing $\ast$. We relabel
the poles $\phi_{i}$ by
\begin{equation}
\phi_{m,l}\equiv\phi_{l+\sum_{j=1}^{m-1}k_{j}},\quad l=1,\dots,k_{m}.
\end{equation}
The positions of poles are solutions to the equations
\begin{equation}
\mathbf{Q}^{(m)}\begin{pmatrix}\phi_{m,1}\\
\vdots\\
\phi_{m,k_{m}}
\end{pmatrix}=\begin{pmatrix}\gamma_{m,1}\\
\vdots\\
\gamma_{m,k_{m}}
\end{pmatrix},
\end{equation}
where
\begin{equation}
\gamma_{m,l}\in\begin{cases}
\left\{ \mathtt{a}_{\mathcal{A}},\mathcal{A}\in\underline{n}\right\} , & l=1\\
\left\{ -\varepsilon_{a},a\in\underline{4}\right\} , & l>1
\end{cases}.
\end{equation}
In particular, we can have an injective map
\begin{equation}
\varrho:\left\{ 1,\cdots,M\right\} \to\underline{n},
\end{equation}
and the pole corresponding to the root of the $m$-th tree is 
\begin{equation}
\phi_{m,1}=\mathtt{a}_{\varrho(m)}.
\end{equation}
We can decorate the trees associated with $\mathbf{Q}$ into trees
describing potential poles that can have nonvanishing JK-residues
by assigning $\varrho(m)$ to the root of the $m$-th tree, and painting
each arrow by the $a$-th color if the pole associated with the target
vertex differs from the pole associated with the source vertex by
$-\varepsilon_{a}$. 

\subsubsection{Classification of genuine poles in terms of colored plane partitions}

There is an important flaw in the above classification of poles that
can give nonvanishing JK-residues, because the denominator can have
extra zeros from linearly dependent hyperplanes, and the zeros in
the numerator will cancel some zeros in the denominator. We define
the genuine poles to be the poles that indeed give nonvanishing JK-residues.
These genuine poles must be contained in the set of potential poles
found above. 

We claim that the genuine poles $\phi_{m,l}$ are completely classified
by a collection of colored plane partitions,
\begin{equation}
\vec{\pi}=\left\{ \pi^{(\mathcal{A})},\mathcal{A}\in\underline{n}\right\} ,
\end{equation}
where each $\pi^{(\mathcal{A})}$ is restricted to be a plane partition,
and we allow some of $\pi^{(\mathcal{A})}$ to be empty. If there
are $M$ non-empty plane partitions in $\vec{\pi}$, then we can introduce
a bijective map
\begin{equation}
\varrho:\left\{ 1,\cdots,M\right\} \to\left\{ \left.\mathcal{A}\in\underline{n}\right|\pi^{(\mathcal{A})}\neq\emptyset\right\} ,
\end{equation}
and the poles labeled by $\vec{\pi}$ are at
\begin{equation}
\phi_{m,s}=\mathtt{a}_{\mathcal{A}}+(1-x)\varepsilon_{a}+(1-y)\varepsilon_{b}+(1-z)\varepsilon_{c},
\end{equation}
where $\varrho(m)=\mathcal{A}=\left(abc,\alpha\right)$ and $s=\left(x,y,z\right)\in\pi^{(\mathcal{A})}$.
This claim can be proved by induction on $k$ as follows.

For $k=1$, all the allowed poles are at $\left\{ \mathtt{a}_{\mathcal{A}},\mathcal{A}\in\underline{n}\right\} $.
For each given pole, there is only one nonempty plane partition in
$\vec{\pi}$, and is given by $\left\{ (1,1,1)\right\} \in\mathbb{Z}_{+}^{3}$.
We have shown that they give nonvanishing contributions to $Z_{1}$.
Hence, the claim holds for the base case.

We assume that the claim is true for $k-1$ and examine it for $k$.
If all the blocks of $\mathbf{Q}$ are one dimensional, then the poles
are at 
\begin{equation}
\phi_{i}=\mathtt{a}_{\varrho(i)},
\end{equation}
with the map $\varrho:\left\{ 1,\cdots,k\right\} \to\left\{ \left.\mathcal{A}\in\underline{n}\right|\pi^{(\mathcal{A})}\neq\emptyset\right\} $.
There are $k$ nonempty plane partitions in $\vec{\pi}$, with each
one being $\left\{ (1,1,1)\right\} \in\mathbb{Z}_{+}^{3}$. All of
them will give nonzero contributions to the JK-residue, and the claim
holds. 

We then consider the case when $\mathbf{Q}$ is that it contains at
least one charge vector of type $\mathbf{h}_{A,ij},i>j$. Up to Weyl
permutations, we can always arrange the $k$ hyperplanes so that the
charge vectors of the first $(k-1)$ hyperplanes only contain $\mathbf{e}_{1},\cdots,\mathbf{e}_{k-1}$,
and the charge vector of the last hyperplane $H_{k}$ is $\mathbf{Q}_{k}=\mathbf{e}_{k}-\mathbf{e}_{J}$
with a fixed $J$. From the picture of trees, $H_{k}$ is associated
with the arrow from $\mathbf{e}_{J}$ to $\mathbf{e}_{k}$ and $\mathbf{e}_{k}$
is not the source of any other arrow. In other words, $\mathbf{e}_{k}$
corresponds to an end of a tree with multiple vertices. The integrand
which contains $\phi_{1},\cdots,\phi_{k-1}$ but not $\phi_{k}$ is
precisely the integrand for the instanton number $k-1$. The poles
$\phi_{1},\cdots,\phi_{k}$ can contribute to the JK-residue if $\sum_{I=1}^{k}\lambda_{I}\mathbf{Q}_{I}=\boldsymbol{\eta}$
with $\lambda_{I}>0$, which leads to
\begin{equation}
\sum_{I=1}^{k-1}\lambda_{I}\mathbf{Q}_{I}=\left(\sum_{i=1}^{k}\mathbf{e}_{i}\right)-\lambda_{k}\left(\mathbf{e}_{k}-\mathbf{e}_{J}\right).
\end{equation}
Because the left-hand side does not contain $\mathbf{e}_{k}$, we
need $\lambda_{k}=1$ and
\begin{equation}
\sum_{I=1}^{k-1}\lambda_{I}\mathbf{Q}_{I}=\left(\sum_{i=1}^{k-1}\mathbf{e}_{i}\right)+\mathbf{e}_{J}.
\end{equation}
Since the right-hand side is in the same chamber as $\left(\sum_{i=1}^{k-1}\mathbf{e}_{i}\right)$,
we know that the poles $\phi_{1},\cdots,\phi_{k-1}$ must also contribute
to the JK-residue. Therefore, the genuine poles for $\phi_{1},\cdots,\phi_{k}$
can be obtained by first giving the genuine poles for $\phi_{1},\cdots,\phi_{k-1}$,
and then determining the proper position of the pole $\phi_{k}$ by
choosing $H_{k}$. By the induction hypothesis, the genuine poles
for $\phi_{1},\cdots,\phi_{k-1}$ are labeled by a collection $\vec{\pi}$
of colored plane partitions with $\left|\vec{\pi}\right|=k-1$. We
need to show that there is a bijection between the possible choices
of $H_{k}$ giving nonzero $k$-dimensional JK-residue and the ways
of making a collection $\vec{\pi}^{\prime}$ of colored plane partitions
with $\left|\vec{\pi}^{\prime}\right|=k$ from $\vec{\pi}$ by adding
a box. Without loss of generality, we assume that $\phi_{k}$ is in
a tree whose root vertex corresponds to the pole at $\mathtt{a}_{123,1}=\mathtt{a}_{\ast}$.
Based on our assumption of $H_{k}$, the potential pole for $\phi_{k}$
is $\phi_{k}=\phi_{J}-\varepsilon_{a}$ for $a\in\underline{4}$.
Accordingly, adding $H_{k}$ can only deform $\pi^{(123,1)}=\pi_{\ast}$,
leaving the other colored plane partitions invariant. We can factorize
the integrand of $Z_{k}$ into two parts,
\begin{equation}
Z_{k}^{1-1}\prod_{A\in\underline{4}^{\vee}}Z_{k}^{1-7_{A}}=\left(Z_{k}^{1-1}\prod_{A\in\underline{4}^{\vee}}Z_{k}^{1-7_{A}}\right)^{\mathrm{reg}}\times I_{k}.
\end{equation}
Here the regular part contains neither zeros nor poles in the neighborhood
of $\phi_{k}\to\phi_{J}-\varepsilon_{a}$, and $I_{k}$ is given by
\begin{equation}
I_{k}=\frac{f\left(0\right)f\left(\varepsilon_{12}\right)f\left(\varepsilon_{13}\right)f\left(\varepsilon_{23}\right)}{\prod_{a\in\underline{4}}f\left(\varepsilon_{a}\right)}\times\theta\left(\phi_{k}-\mathtt{a}_{\ast}-\varepsilon_{123}\right),
\end{equation}
where 
\begin{equation}
f(x)=\prod_{s\in\pi_{\ast}}\left(\theta\left(\phi_{k}-c_{s}+x\right)\theta\left(c_{s}-\phi_{k}+x\right)\right),
\end{equation}
and
\begin{equation}
c_{s=(x,y,z)}=\mathtt{a}_{\ast}+(1-x)\varepsilon_{1}+(1-y)\varepsilon_{2}+(1-z)\varepsilon_{3}.
\end{equation}
If $\phi_{J}=\mathtt{a}_{\ast}$ corresponding to the box $(1,1,1)\in\pi_{\ast}$,
then the factor $\theta\left(\phi_{k}-\mathtt{a}_{\ast}-\varepsilon_{123}\right)$
in the numerator cancels the factor $\theta\left(\phi_{k}-\mathtt{a}_{\ast}+\varepsilon_{4}\right)$
in the denominator using the constraint $\sum_{a\in\underline{4}}\varepsilon_{a}=0$,
and the genuine poles are $\phi_{k}=\mathtt{a}_{\ast}-\varepsilon_{a}$
for $a\in\left\{ 1,2,3\right\} $. In the following, we assume that
$\phi_{J}$ corresponds to the box $\left(x,y,z\right)\in\pi_{\ast}\setminus\left\{ (1,1,1)\right\} $,
then the potential poles are 
\begin{equation}
\phi_{k}=\mathtt{a}_{\ast}+\left(1-x^{\prime}\right)\varepsilon_{1}+\left(1-y^{\prime}\right)\varepsilon_{2}+\left(1-z^{\prime}\right)\varepsilon_{3},
\end{equation}
with four possibilities
\begin{equation}
\left(x^{\prime},y^{\prime},z^{\prime}\right)\in\left\{ \left(x+1,y,z\right),\left(x,y+1,z\right),\left(x,y,z+1\right),\left(x-1,y-1,z-1\right)\right\} .
\end{equation}
When the box $\left(x^{\prime},y^{\prime},z^{\prime}\right)$ is already
contained in $\pi_{\ast}$, the numerator of $I_{k}$ contains a double
zero from 
\begin{equation}
\theta\left(\phi_{k}-c_{\left(x^{\prime},y^{\prime},z^{\prime}\right)}\right)\theta\left(c_{\left(x^{\prime},y^{\prime},z^{\prime}\right)}-\phi_{k}\right),
\end{equation}
and the residue vanishes. Therefore, there can be at most one box
at each $\left(x,y,z\right)\in\pi_{\ast}^{\prime}$. We denote the
combination of the plane partition $\pi_{\ast}$ and the box $\left(x^{\prime},y^{\prime},z^{\prime}\right)$
by $\pi_{\ast}^{\prime}$. We need to show that if $\pi_{\ast}^{\prime}$
is not a plane partition, then the residue is zero.

If $\left(x^{\prime},y^{\prime},z^{\prime}\right)$ is one of the
boxes $\left(x+1,y,z\right)$, $\left(x,y+1,z\right)$, and $\left(x,y,z+1\right)$,
the box $\left(x,y,z\right)\in\pi_{\ast}\setminus\left\{ (1,1,1)\right\} $
must sit on the boundary of $\pi_{\ast}$. We can focus on the case
$\left(x^{\prime},y^{\prime},z^{\prime}\right)=\left(x+1,y,z\right)$,
and the other cases can be obtained by simple permutations. We want
to count the order $\Delta$ of singularity for a potential pole $\phi_{k}$,
which is the number of poles from the denominator minus the number
of zeros from the numerator. The residue is nonzero when $\Delta=1$.
We need to further make the following distinction:
\begin{itemize}
\item When $y=z=1$, $\pi_{\ast}^{\prime}$ is a plane partition. $I_{k}$
only contains a pole from $\theta\left(\phi_{k}-c_{\left(x,1,1\right)}+\varepsilon_{1}\right)$,
and therefore the residue is nonzero. 
\item When $y>1$ and $z=1$ (by exchanging $y$ and $z$ we can get results
for $z>1$ and $x=y=1$), $\pi_{\ast}^{\prime}$ is a plane partition
if and only if 
\begin{equation}
\left(x+1,y-1,1\right)\in\pi_{\ast}.
\end{equation}
The poles and the zero of $I_{k}$ are
\begin{eqnarray}
\mathrm{poles} & : & \begin{cases}
\theta\left(\phi_{k}-c_{\left(x,y,1\right)}+\varepsilon_{1}\right),\\
\theta\left(\phi_{k}-c_{\left(x+1,y-1,1\right)}+\varepsilon_{2}\right), & \mathrm{if}\ \left(x+1,y-1,1\right)\in\pi_{\ast}^{\prime}
\end{cases}\nonumber \\
\mathrm{zero} & : & \theta\left(\phi_{k}-c_{\left(x,y-1,1\right)}+\varepsilon_{12}\right).
\end{eqnarray}
If $\pi_{\ast}^{\prime}$ is a plane partition, $\Delta=1$, and the
residue is nonzero. On the other hand, if $\left(x+1,y-1,1\right)\notin\pi_{\ast}$
so that $\pi_{\ast}^{\prime}$ is not a plane partition, $\Delta=0$,
and the residue vanishes.
\item When $y,z>1$, $\pi_{\ast}^{\prime}$ is a plane partition if and
only if 
\begin{equation}
\left(x+1,y-1,z\right),\left(x+1,y,z-1\right)\in\pi_{\ast}.
\end{equation}
The poles and zeros of $I_{k}$ are
\begin{eqnarray}
\mathrm{poles} & : & \begin{cases}
\theta\left(\phi_{k}-c_{\left(x,y,z\right)}+\varepsilon_{1}\right),\\
\theta\left(c_{\left(x,y-1,z-1\right)}-\phi_{k}+\varepsilon_{4}\right),\\
\theta\left(\phi_{k}-c_{\left(x+1,y-1,z\right)}+\varepsilon_{2}\right), & \mathrm{if}\ \left(x+1,y-1,z\right)\in\pi_{\ast}\\
\theta\left(\phi_{k}-c_{\left(x+1,y,z-1\right)}+\varepsilon_{3}\right), & \mathrm{if}\ \left(x+1,y,z-1\right)\in\pi_{\ast}
\end{cases}\nonumber \\
\mathrm{zeros} & : & \begin{cases}
\theta\left(\phi_{k}-c_{\left(x,y-1,z\right)}+\varepsilon_{12}\right),\\
\theta\left(\phi_{k}-c_{\left(x,y,z-1\right)}+\varepsilon_{13}\right),\\
\theta\left(\phi_{k}-c_{\left(x+1,y-1,z-1\right)}+\varepsilon_{23}\right), & \mathrm{if}\ \left(x+1,y-1,z-1\right)\in\pi_{\ast}
\end{cases}.
\end{eqnarray}
Since $\left(x+1,y-1,z-1\right)\in\pi_{\ast}$ is automatically satisfied
when $\left(x+1,y-1,z\right)\in\pi_{\ast}$ or $\left(x+1,y,z-1\right)\in\pi_{\ast}$,
we can have $\Delta=1$ so that the residue is nonzero only if $\pi_{\ast}^{\prime}$
is a plane partition.
\end{itemize}
If $\left(x^{\prime},y^{\prime},z^{\prime}\right)=\left(x-1,y-1,z-1\right)$,
the residue can be nonzero only if $\left(x-1,y,z\right)\notin\pi_{\ast}$,
since the numerator would contain a zero from $\theta\left(c_{\left(x-1,y,z\right)}-\phi_{k}+\varepsilon_{23}\right)$
otherwise. Similarly, $\pi_{\ast}$ cannot contain $\left(x,y-1,z\right)$
and $\left(x,y,z-1\right)$. However, this is in contradiction to
the assumption that $\left(x,y,z\right)\in\pi_{\ast}\setminus\left\{ (1,1,1)\right\} $
and $\pi_{\ast}$ is a plane partition. Therefore, taking $\left(x^{\prime},y^{\prime},z^{\prime}\right)=\left(x-1,y-1,z-1\right)$
will always lead to a vanishing residue. 

In summary, we have shown that all the genuine poles of $\phi_{k}$
are in one-to-one correspondence with the possibilities of adding
a box to $\vec{\pi}$ to make a collection of colored plane partitions. 

\subsubsection{Expression}

Eventually, we obtain the elliptic genus $Z_{k}$,
\begin{equation}
Z_{k}=\sum_{\vec{\pi},\left|\vec{\pi}\right|=k}Z_{\vec{\pi}}.
\end{equation}
We define
\begin{equation}
\mathcal{C}_{\mathcal{A},s}=\mathtt{a}_{\mathcal{A}}+(1-x)\varepsilon_{a}+(1-y)\varepsilon_{b}+(1-z)\varepsilon_{c},
\end{equation}
for $\mathcal{A}=\left(abc,\alpha\right)\in\underline{n}$ and $s=\left(x,y,z\right)\in\pi^{(\mathcal{A})}$,
and
\begin{equation}
\mathcal{D}_{\mathcal{B},t}^{\mathcal{A},s}=\mathcal{C}_{\mathcal{A},s}-\mathcal{C}_{\mathcal{B},t}.
\end{equation}
We also introduce the notation
\begin{equation}
\mathcal{R}\left\{ \theta\left(\left.x\right|\tau\right)\right\} =\frac{\theta\left(x\right)\prod_{1\leq a<b\leq3}\theta\left(x+\varepsilon_{ab}\right)}{\prod_{a\in\underline{4}}\theta\left(x+\varepsilon_{a}\right)}.
\end{equation}
Then $Z_{\vec{\pi}}$ can be expressed as 
\begin{equation}
Z_{\vec{\pi}}=\left(\prod_{\mathcal{A}\in\underline{n}}Z_{\vec{\pi}}^{(\mathcal{A})}\right)\left(\prod_{\mathcal{A}\neq\mathcal{B}\in\underline{n}}Z_{\vec{\pi}}^{(\mathcal{A},\mathcal{B})}\right),
\end{equation}
where
\begin{eqnarray}
Z_{\vec{\pi}}^{\left(\mathcal{A}=\left(A,\alpha\right)\right)} & = & \theta\left(\varepsilon_{\check{A}}\right)\left(\frac{\prod_{a<b\in A}\theta\left(\varepsilon_{ab}\right)}{\prod_{a\in\underline{4}}\theta\left(\varepsilon_{a}\right)}\right)^{\left|\pi^{(\mathcal{A})}\right|}\left(\sideset{}{^{\prime}}\prod_{s\neq t\in\pi^{(\mathcal{A})}}\mathcal{R}\left\{ \theta\left(\mathcal{D}_{\mathcal{A},t}^{\mathcal{A},s}\right)\right\} \right)\times\nonumber \\
 &  & \times\left(\prod_{s\in\pi^{(\mathcal{A})}\setminus(1,1,1)}\frac{\theta\left(\mathcal{C}_{\mathcal{A},s}-\mathtt{a}_{\mathcal{A}}-\varepsilon_{\mathcal{A}}\right)}{\theta\left(\mathcal{C}_{\mathcal{A},s}-\mathtt{a}_{\mathcal{A}}\right)}\right),
\end{eqnarray}
and
\begin{equation}
Z_{\vec{\pi}}^{(\mathcal{A},\mathcal{B})}=\left(\prod_{s\in\pi^{(\mathcal{A})}}\prod_{t\in\pi^{(\mathcal{B})}}\mathcal{R}\left\{ \theta\left(\mathcal{D}_{\mathcal{B},t}^{\mathcal{A},s}\right)\right\} \right)\left(\prod_{s\in\pi^{(\mathcal{A})}}\frac{\theta\left(\mathcal{C}_{\mathcal{A},s}-\mathtt{a}_{\mathcal{B}}-\varepsilon_{\mathcal{B}}\right)}{\theta\left(\mathcal{C}_{\mathcal{A},s}-\mathtt{a}_{\mathcal{B}}\right)}\right).
\end{equation}
The instanton partition function is
\begin{equation}
Z=\sum_{\vec{\pi}}\mathtt{q}^{\left|\vec{\pi}\right|}Z_{\vec{\pi}},
\end{equation}
which is identical to $\mathcal{Z}^{\mathrm{inst}}$ in (\ref{eq:generating})
if we use the elliptic version (\ref{eq:E-K-ell}) of the operator
$\mathbb{E}$.

\subsubsection{Example: $k=2$ and $\vec{n}=\left(1,1,0,0\right)$}

Let us present here explicitly the result for the simplest nontrivial
example, $k=2$ and $\vec{n}=\left(n_{123}=1,n_{124}=1,0,0\right)$.
We are dealing with the integral
\begin{eqnarray}
Z_{2} & = & \frac{1}{2}\left[\frac{2\pi\eta(\tau)^{3}\prod_{1\leq a<b\leq3}\theta\left(\varepsilon_{ab}\right)}{\prod_{a\in\underline{4}}\theta\left(\varepsilon_{a}\right)}\right]^{2}\nonumber \\
 &  & \int d\phi_{1}d\phi_{2}\frac{\theta^{2}\left(\phi_{1}-\phi_{2}\right)\prod_{1\leq a<b\leq3}\theta\left(\phi_{1}-\phi_{2}\pm\varepsilon_{ab}\right)}{\prod_{a\in\underline{4}}\theta\left(\phi_{1}-\phi_{2}\pm\varepsilon_{a}\right)}\times\nonumber \\
 &  & \times\prod_{i=1}^{2}\frac{\theta\left(\phi_{i}-\mathtt{a}_{123}-\varepsilon_{123}\right)}{\theta\left(\phi_{i}-\mathtt{a}_{123}\right)}\frac{\theta\left(\phi_{i}-\mathtt{a}_{124}-\varepsilon_{124}\right)}{\theta\left(\phi_{i}-\mathtt{a}_{124}\right)}.\label{eq:Z2}
\end{eqnarray}
Due to the invariance under the overall shift (\ref{eq:shift}), the
result can only depend on the difference
\begin{equation}
\delta=\mathtt{a}_{123}-\mathtt{a}_{124}.
\end{equation}

The genuine poles are completely classified by a collection of two
(possibly empty) colored plane partitions,
\begin{equation}
\vec{\pi}=\left\{ \pi^{(123)},\pi^{(124)}\right\} ,
\end{equation}
and the total number of boxes 
\begin{equation}
\left|\pi^{(123)}\right|+\left|\pi^{(124)}\right|=k=2.
\end{equation}
There are three possibilities:
\begin{enumerate}
\item If $\left|\pi^{(123)}\right|=2,\left|\pi^{(124)}\right|=0$, the genuine
poles are at $\left(\mathtt{a}_{123},\mathtt{a}_{123}-\varepsilon_{a}\right),a=1,2,3$,
and the corresponding residues are 
\begin{eqnarray}
z_{\left(\mathtt{a}_{123},\mathtt{a}_{123}-\varepsilon_{1}\right)} & = & \frac{\theta\left(\varepsilon_{12}\right)\theta\left(\varepsilon_{13}\right)\theta\left(\varepsilon_{23}\right)}{\theta\left(\varepsilon_{1}\right)\theta\left(\varepsilon_{2}\right)\theta\left(\varepsilon_{3}\right)}\frac{\theta\left(\delta+\varepsilon_{3}\right)\theta\left(\delta-\varepsilon_{1}+\varepsilon_{3}\right)}{\theta\left(\delta\right)\theta\left(\delta-\varepsilon_{1}\right)}\times\nonumber \\
 &  & \times\frac{\theta\left(2\varepsilon_{1}+\varepsilon_{2}\right)\theta\left(2\varepsilon_{1}+\varepsilon_{3}\right)\theta\left(\varepsilon_{23}-\varepsilon_{1}\right)}{\theta\left(2\varepsilon_{1}\right)\theta\left(\varepsilon_{1}-\varepsilon_{2}\right)\theta\left(\varepsilon_{1}-\varepsilon_{3}\right)},\\
z_{\left(\mathtt{a}_{123},\mathtt{a}_{123}-\varepsilon_{2}\right)} & = & \frac{\theta\left(\varepsilon_{12}\right)\theta\left(\varepsilon_{13}\right)\theta\left(\varepsilon_{23}\right)}{\theta\left(\varepsilon_{1}\right)\theta\left(\varepsilon_{2}\right)\theta\left(\varepsilon_{3}\right)}\frac{\theta\left(\delta+\varepsilon_{3}\right)\theta\left(\delta-\varepsilon_{2}+\varepsilon_{3}\right)}{\theta\left(\delta\right)\theta\left(\delta-\varepsilon_{2}\right)}\times\nonumber \\
 &  & \times\frac{\theta\left(2\varepsilon_{2}+\varepsilon_{1}\right)\theta\left(2\varepsilon_{2}+\varepsilon_{3}\right)\theta\left(\varepsilon_{13}-\varepsilon_{2}\right)}{\theta\left(2\varepsilon_{2}\right)\theta\left(\varepsilon_{2}-\varepsilon_{1}\right)\theta\left(\varepsilon_{2}-\varepsilon_{3}\right)},\\
z_{\left(\mathtt{a}_{123},\mathtt{a}_{123}-\varepsilon_{3}\right)} & = & \frac{\theta\left(\varepsilon_{12}\right)\theta\left(\varepsilon_{13}\right)\theta\left(\varepsilon_{23}\right)}{\theta\left(\varepsilon_{1}\right)\theta\left(\varepsilon_{2}\right)\theta\left(\varepsilon_{3}\right)}\frac{\theta\left(\delta+\varepsilon_{3}\right)}{\theta\left(\delta-\varepsilon_{3}\right)}\times\nonumber \\
 &  & \times\frac{\theta\left(2\varepsilon_{3}+\varepsilon_{1}\right)\theta\left(2\varepsilon_{3}+\varepsilon_{2}\right)\theta\left(\varepsilon_{12}-\varepsilon_{3}\right)}{\theta\left(2\varepsilon_{3}\right)\theta\left(\varepsilon_{3}-\varepsilon_{1}\right)\theta\left(\varepsilon_{3}-\varepsilon_{2}\right)},
\end{eqnarray}
\item If $\left|\pi^{(123)}\right|=0,\left|\pi^{(124)}\right|=2$, the genuine
poles are at $\left(\mathtt{a}_{124},\mathtt{a}_{124}-\varepsilon_{a}\right),a=1,2,4$,
and the corresponding residues are
\begin{eqnarray}
z_{\left(\mathtt{a}_{124},\mathtt{a}_{124}-\varepsilon_{1}\right)} & = & \frac{\theta\left(\varepsilon_{12}\right)\theta\left(\varepsilon_{14}\right)\theta\left(\varepsilon_{24}\right)}{\theta\left(\varepsilon_{1}\right)\theta\left(\varepsilon_{2}\right)\theta\left(\varepsilon_{4}\right)}\frac{\theta\left(\delta-\varepsilon_{4}\right)\theta\left(\delta+\varepsilon_{1}-\varepsilon_{4}\right)}{\theta\left(\delta\right)\theta\left(\delta+\varepsilon_{1}\right)}\times\nonumber \\
 &  & \times\frac{\theta\left(2\varepsilon_{1}+\varepsilon_{2}\right)\theta\left(2\varepsilon_{1}+\varepsilon_{4}\right)\theta\left(\varepsilon_{24}-\varepsilon_{1}\right)}{\theta\left(2\varepsilon_{1}\right)\theta\left(\varepsilon_{1}-\varepsilon_{2}\right)\theta\left(\varepsilon_{1}-\varepsilon_{4}\right)},\\
z_{\left(\mathtt{a}_{124},\mathtt{a}_{124}-\varepsilon_{2}\right)} & = & \frac{\theta\left(\varepsilon_{12}\right)\theta\left(\varepsilon_{14}\right)\theta\left(\varepsilon_{24}\right)}{\theta\left(\varepsilon_{1}\right)\theta\left(\varepsilon_{2}\right)\theta\left(\varepsilon_{4}\right)}\frac{\theta\left(\delta-\varepsilon_{4}\right)\theta\left(\delta+\varepsilon_{2}-\varepsilon_{4}\right)}{\theta\left(\delta\right)\theta\left(\delta+\varepsilon_{2}\right)}\times\nonumber \\
 &  & \times\frac{\theta\left(2\varepsilon_{2}+\varepsilon_{1}\right)\theta\left(2\varepsilon_{2}+\varepsilon_{4}\right)\theta\left(\varepsilon_{14}-\varepsilon_{2}\right)}{\theta\left(2\varepsilon_{2}\right)\theta\left(\varepsilon_{2}-\varepsilon_{1}\right)\theta\left(\varepsilon_{2}-\varepsilon_{4}\right)},\\
z_{\left(\mathtt{a}_{124},\mathtt{a}_{124}-\varepsilon_{4}\right)} & = & \frac{\theta\left(\varepsilon_{12}\right)\theta\left(\varepsilon_{14}\right)\theta\left(\varepsilon_{24}\right)}{\theta\left(\varepsilon_{1}\right)\theta\left(\varepsilon_{2}\right)\theta\left(\varepsilon_{4}\right)}\frac{\theta\left(\delta-\varepsilon_{4}\right)}{\theta\left(\delta+\varepsilon_{4}\right)}\times\nonumber \\
 &  & \times\frac{\theta\left(2\varepsilon_{4}+\varepsilon_{1}\right)\theta\left(2\varepsilon_{4}+\varepsilon_{2}\right)\theta\left(\varepsilon_{12}-\varepsilon_{4}\right)}{\theta\left(2\varepsilon_{2}\right)\theta\left(\varepsilon_{2}-\varepsilon_{1}\right)\theta\left(\varepsilon_{2}-\varepsilon_{4}\right)},
\end{eqnarray}
\item If $\left|\pi^{(123)}\right|=\left|\pi^{(124)}\right|=1$, the genuine
pole can only be at $\left(\mathtt{a}_{123},\mathtt{a}_{124}\right)$,
and the corresponding residue is
\begin{eqnarray}
z_{\left(\mathtt{a}_{123},\mathtt{a}_{124}\right)} & = & \frac{\theta^{2}\left(\varepsilon_{12}\right)\theta^{2}\left(\varepsilon_{13}\right)\theta^{2}\left(\varepsilon_{23}\right)}{\theta^{2}\left(\varepsilon_{1}\right)\theta^{2}\left(\varepsilon_{2}\right)\theta\left(\varepsilon_{3}\right)\theta\left(\varepsilon_{4}\right)}\times\nonumber \\
 &  & \times\frac{\theta\left(\pm\delta+\varepsilon_{12}\right)\theta\left(\pm\delta+\varepsilon_{13}\right)\theta\left(\pm\delta+\varepsilon_{23}\right)}{\theta\left(\pm\delta+\varepsilon_{1}\right)\theta\left(\pm\delta+\varepsilon_{2}\right)\theta\left(-\delta+\varepsilon_{3}\right)\theta\left(\delta+\varepsilon_{4}\right)}.
\end{eqnarray}
\end{enumerate}
In the above calculations, we have taken a particular ordering of
$\phi_{1}$ and $\phi_{2}$ to cancel the factor of $2$ in the denominator
of (\ref{eq:Z2}). Summing up these contributions, we get
\begin{equation}
Z_{2}=\sum_{a\in\left(123\right)}z_{\left(\mathtt{a}_{123},\mathtt{a}_{123}-\varepsilon_{a}\right)}+\sum_{a\in\left(124\right)}z_{\left(\mathtt{a}_{124},\mathtt{a}_{124}-\varepsilon_{a}\right)}+z_{\left(\mathtt{a}_{123},\mathtt{a}_{124}\right)}.\label{eq:Z2sum}
\end{equation}
One can check that (\ref{eq:Z2sum}) matches the general expression
given in the previous subsection. 

\subsection{Expectation value of real codimension-two defects \label{subsec:Projection}}

Up to now, we treat all D$7_{A}$-branes on equal footing, but the
string theory construction of tetrahedron instantons and the geometric
interpretation of the moduli space suggest a different point of view
of the instanton partition function. We choose the physical spacetime
to be $\mathbb{R}^{1,1}\times\mathbb{C}_{123}^{3}$, so that the bound
states of D$1$- and D$7_{123}$-branes give rise to instantons on
$\mathbb{C}_{123}^{3}$. The remaining D$7_{A}$-branes for $A\in\underline{4}^{\vee}\setminus\left\{ (123)\right\} $
will produce real codimension-two defects from the viewpoint of the
physical spacetime. This provides the physical realization of the
projection of the moduli space $\mathfrak{M}_{\vec{n},k}$ of tetrahedron
instantons to the moduli spaces $\mathfrak{M}_{\left(n_{123},0,0,0\right),k^{\prime}}$
of instantons on $\mathbb{C}_{123}^{3}$ discussed in section \ref{sec:moduli}.
Thus we identify the instanton partition function as the expectation
value of real codimension-two defects $\mathcal{O}_{A}$ in the instanton
partition function of the Donaldson-Thomas theory,
\begin{eqnarray}
Z & = & \sum_{k=0}^{\infty}\frac{\mathtt{q}^{k}}{k!}\int\prod_{i=1}^{k}d\phi_{i}\left[\left(Z_{k}^{1-1}Z_{k}^{1-7_{123}}\right)\left(\prod_{A\in\underline{4}^{\vee}\setminus\left\{ \left(123\right)\right\} }Z_{k}^{1-7_{A}}\right)\right]\nonumber \\
 & = & \left\langle \prod_{A\in\underline{4}^{\vee}\setminus\left\{ \left(123\right)\right\} }\mathcal{O}_{A}\right\rangle _{\mathrm{DT}},\label{eq:Zproj}
\end{eqnarray}
where the bracket denotes the unnormalized vacuum expectation value
in the Donaldson-Thomas theory on $\mathbb{C}_{123}^{3}$ whose instanton
partition function is given by
\begin{equation}
Z_{\mathrm{DT}}=\sum_{k=0}^{\infty}\frac{\mathtt{q}^{k}}{k!}\int\prod_{i=1}^{k}d\phi_{i}Z_{k}^{1-1}Z_{k}^{1-7_{123}}.
\end{equation}

\subsection{Dimensional reductions}

We now briefly discuss dimensional reductions of the system.

Performing a T-duality along $x^{9}$ of the configuration in Table
\ref{D1-D7}, we get D$0$-branes probing a configuration of intersecting
D$6$-branes in type IIA superstring theory. The generating function
of the generalized Witten indices of the supersymmetric gauged quantum
mechanical models on D$0$-branes is the K-theoretical version of
the instanton partition function of tetrahedron instantons. Since
there are no anomalies of large gauge transformations, we no longer
impose the constraint (\ref{eq:nepsilon}). Taking the limit $q\to0$
of $Z$, we get the dimensionally reduced instanton partition function
$Z^{\downarrow}$, 
\begin{eqnarray}
Z^{\downarrow} & = & \sum_{\vec{\pi}}\mathtt{q}^{\left|\vec{\pi}\right|}Z_{\vec{\pi}}^{\downarrow}\nonumber \\
 & = & \sum_{\vec{\pi}}\mathtt{q}^{\left|\vec{\pi}\right|}\left(\prod_{\mathcal{A}\in\underline{n}}Z_{\vec{\pi}}^{\downarrow(\mathcal{A})}\right)\left(\prod_{\mathcal{A}\neq\mathcal{B}\in\underline{n}}Z_{\vec{\pi}}^{\downarrow(\mathcal{A},\mathcal{B})}\right),\label{eq:Zreduce}
\end{eqnarray}
where $Z_{\vec{\pi}}^{\downarrow(\mathcal{A})}$ and $Z_{\vec{\pi}}^{\downarrow(\mathcal{A},\mathcal{B})}$
are obtained from $Z_{\vec{\pi}}^{(\mathcal{A})}$ and $Z_{\vec{\pi}}^{(\mathcal{A},\mathcal{B})}$
by substituting
\begin{equation}
\theta\left(z\right)\to2\sinh\left(\frac{\beta z}{2}\right),
\end{equation}
and $\beta$ is the circumference of the circle of the supersymmetric
quantum mechanics. The instanton partition function $Z^{\downarrow}$
matches $\mathcal{Z}$ in (\ref{eq:generating}) with the K-theoretical
version (\ref{eq:E-K-ell}) of the operator $\mathbb{E}$. 

We can further perform a T-duality along $x^{0}$ direction to get
D-instantons probing a configuration of intersecting D$5$-branes
in type IIB superstring theory. The instanton partition function $Z^{\Downarrow}$
is obtained by 
\begin{equation}
Z^{\Downarrow}=\sum_{\vec{\pi}}\mathtt{q}^{\left|\vec{\pi}\right|}Z_{\vec{\pi}}^{\Downarrow}=\sum_{\vec{\pi}}\mathtt{q}^{\left|\vec{\pi}\right|}\left(\prod_{\mathcal{A}\in\underline{n}}Z_{\vec{\pi}}^{\Downarrow(\mathcal{A})}\right)\left(\prod_{\mathcal{A}\neq\mathcal{B}\in\underline{n}}Z_{\vec{\pi}}^{\Downarrow(\mathcal{A},\mathcal{B})}\right).
\end{equation}
Here $Z_{\vec{\pi}}^{\Downarrow(\mathcal{A})}$ and $Z_{\vec{\pi}}^{\Downarrow(\mathcal{A},\mathcal{B})}$
are obtained from $Z_{\vec{\pi}}^{(\mathcal{A})}$ and $Z_{\vec{\pi}}^{(\mathcal{A},\mathcal{B})}$
by substituting $\theta\left(z\right)\to z$,
\begin{eqnarray}
Z_{\vec{\pi}}^{\Downarrow\left(\mathcal{A}=\left(A,\alpha\right)\right)} & = & \varepsilon_{\check{A}}\left(\frac{\prod_{a<b\in A}\varepsilon_{ab}}{\prod_{a\in\underline{4}}\varepsilon_{a}}\right)^{\left|\pi^{(\mathcal{A})}\right|}\left(\prod_{s\neq t\in\pi^{(\mathcal{A})}}\mathcal{R}\left\{ \mathcal{D}_{\mathcal{A},t}^{\mathcal{A},s}\right\} \right)\times\nonumber \\
 &  & \times\left(\prod_{s\in\pi^{(\mathcal{A})}\setminus(1,1,1)}\frac{\mathcal{C}_{\mathcal{A},s}-a_{\mathcal{A}}-\varepsilon_{\mathcal{A}}}{\mathcal{C}_{\mathcal{A},s}-a_{\mathcal{A}}}\right),\\
Z_{\vec{\pi}}^{\Downarrow(\mathcal{A},\mathcal{B})} & = & \left(\prod_{s\in\pi^{(\mathcal{A})}}\prod_{t\in\pi^{(\mathcal{B})}}\mathcal{R}\left\{ \mathcal{D}_{\mathcal{B},t}^{\mathcal{A},s}\right\} \right)\left(\prod_{s\in\pi^{(\mathcal{A})}}\frac{\mathcal{C}_{\mathcal{A},s}-a_{\mathcal{B}}-\varepsilon_{\mathcal{B}}}{\mathcal{C}_{\mathcal{A},s}-a_{\mathcal{B}}}\right),
\end{eqnarray}
where
\begin{equation}
\mathcal{R}\left\{ x\right\} =\frac{x\left(x+\varepsilon_{12}\right)\left(x+\varepsilon_{13}\right)\left(x+\varepsilon_{23}\right)}{\left(x+\varepsilon_{1}\right)\left(x+\varepsilon_{2}\right)\left(x+\varepsilon_{3}\right)\left(x+\varepsilon_{4}\right)}.
\end{equation}
The partition function $Z^{\Downarrow}$ matches $\mathcal{Z}$ in
(\ref{eq:generating}) exactly.

\section{Free field representation \label{sec:freefield}}

Following \cite{Kazakov:1998ji,Nekrasov:2002qd,Losev:2003py,Nekrasov:2003rj,Benini:2018hjy},
we give a free field representation of the instanton partition function.
This is in the general spirit of the BPS/CFT correspondence \cite{Nekrasov:2015wsu}. 

Recall that the torus propagator for a free massless $r$-component
scalar field $\varphi=\left(\varphi_{1},\cdots,\varphi_{r}\right)$
is given by \cite{DiFrancesco:1997nk}
\begin{eqnarray}
G_{i,j}\left(z,\bar{z}\right) & = & \left\langle \varphi_{i}\left(z,\bar{z}\right)\varphi_{j}\left(0,0\right)\right\rangle _{\mathbb{T}^{2}}\nonumber \\
 & = & -\log\left|\frac{\theta_{1}\left(\left.z\right|\tau\right)}{2\pi\eta(\tau)^{3}}\exp\left(-\frac{\pi\left(\mathrm{Im}z\right)^{2}}{\mathrm{Im}\tau}\right)\right|^{2}\delta_{i,j},\quad i,j=1,\cdots,r,
\end{eqnarray}
where the torus $\mathbb{T}^{2}$ is described by a complex $z$-plane
with the identification $z\cong z+1\cong z+\tau$, and $G_{i,j}\left(z,\bar{z}\right)$
is the normalized doubly periodic solution of the Laplacian on $\mathbb{T}^{2}$,
\begin{equation}
-\Delta G_{i,j}\left(z,\bar{z}\right)=\left(2\pi\delta^{2}(z)-\frac{4\pi}{\mathrm{Im}\tau}\right)\delta_{i,j}.
\end{equation}
The basic vertex operators of the theory are the exponential fields
parameterized by a $r$-component vector parameter $\alpha=\left(\alpha_{1},\cdots,\alpha_{r}\right)$,
\begin{equation}
\mathbb{V}_{\alpha}\left(z,\bar{z}\right)=:e^{\mathrm{i}\sum_{i=1}^{r}\alpha_{i}\varphi_{i}\left(z,\bar{z}\right)}:.
\end{equation}
We require that the complex structure $\tau$ of $\mathbb{T}^{2}$
is the same as that in the definition of the elliptic genus (\ref{eq:elliptic}).
Then we will use the abbreviation (\ref{eq:theta}) in the following. 

We shall take $r=7$ and introduce a slightly deformed vertex operator
\begin{equation}
\mathscr{V}_{\alpha,\rho}\left(z,\bar{z}\right)=:e^{\mathrm{i}\sum_{i=1}^{7}\alpha_{i}\varphi_{i}\left(z+\rho_{i},\bar{z}+\rho_{i}\right)}::e^{-\mathrm{i}\sum_{i=1}^{7}\alpha_{i}\varphi_{i}\left(z-\rho_{i},\bar{z}-\rho_{i}\right)}:,
\end{equation}
where
\begin{equation}
\alpha=\left(\mathrm{i},\mathrm{i},\mathrm{i},\mathrm{i},1,1,1\right),\quad\rho=\frac{1}{2}\left(\varepsilon_{1},\varepsilon_{2},\varepsilon_{3},\varepsilon_{4},\varepsilon_{12},\varepsilon_{13},\varepsilon_{23}\right).
\end{equation}
It is an important fact that when $\sum_{a\in\underline{4}}\varepsilon_{a}=0$
we have
\begin{equation}
\sum_{i=1}^{7}\alpha_{i}^{2}\left(\mathrm{Im}\left(\rho_{i}\right)\right)^{2}=0.
\end{equation}
Performing the Wick contraction, we can get 
\begin{equation}
\mathscr{V}_{\alpha,\rho}\left(z,\bar{z}\right)=\left|\frac{2\pi\eta(\tau)^{3}\prod_{1\leq a<b\leq3}\theta\left(\varepsilon_{ab}\right)}{\prod_{a\in\underline{4}}\theta\left(\varepsilon_{a}\right)}\right|^{2}:\mathscr{V}_{\alpha,\rho}\left(z,\bar{z}\right):,
\end{equation}
and 
\begin{eqnarray}
 &  & \left\langle :\mathscr{V}_{\alpha,\rho}\left(z,\bar{z}\right)::\mathscr{V}_{\alpha,\rho}\left(w,\bar{w}\right):\right\rangle _{\mathbb{T}^{2}}\nonumber \\
 & = & \left|\frac{\theta^{2}\left(z-w\right)\prod_{1\leq a<b\leq3}\theta\left(z-w\pm\varepsilon_{ab}\right)}{\prod_{a\in\underline{4}}\theta\left(z-w\pm\varepsilon_{a}\right)}\right|^{2},\label{eq:VV}
\end{eqnarray}
where $\theta\left(z\pm\tilde{\varepsilon}\right)=\theta\left(z+\tilde{\varepsilon}\right)\theta\left(z-\tilde{\varepsilon}\right)$.
Since (\ref{eq:VV}) takes the form of an absolute square, we can
define the holomorphic part as
\begin{eqnarray}
 &  & \left\langle :\mathscr{V}_{\alpha,\rho}\left(z,\bar{z}\right)::\mathscr{V}_{\alpha,\rho}\left(w,\bar{w}\right):\right\rangle _{\mathbb{T}^{2}}^{\mathrm{hol}}\nonumber \\
 & = & \frac{\theta^{2}\left(z-w\right)\prod_{1\leq a<b\leq3}\theta\left(z-w\pm\varepsilon_{ab}\right)}{\prod_{a\in\underline{4}}\theta\left(z-w\pm\varepsilon_{a}\right)}.
\end{eqnarray}

We further introduce a linear source operator,
\begin{equation}
\varUpsilon=\frac{1}{2\pi\mathrm{i}}\oint_{\Gamma}dz\sum_{A\in\underline{4}^{\vee}}\varpi_{A}(z)\partial_{z}\varphi_{\check{A}}(z),
\end{equation}
where the contour $\Gamma$ is chosen to be a loop around $z=0$ encircling
all $\pm\rho_{i}$ for $i=1,\cdots,7$, and $\varpi_{A}(z)$ is a
locally analytic function inside $\Gamma$,
\begin{equation}
\varpi_{A}(z)=\sum_{\alpha=1}^{n_{A}}\log\theta\left(z-\mathtt{a}_{A,\alpha}-\frac{1}{2}\varepsilon_{A}\right).
\end{equation}
Then
\begin{equation}
\left\langle e^{\varUpsilon}:\mathscr{V}_{\alpha,\rho}(z):\right\rangle _{\mathbb{T}^{2}}=\prod_{\mathcal{A}=\left(A,\alpha\right)\in\underline{n}}\frac{\theta\left(z-\mathtt{a}_{\mathcal{A}}-\varepsilon_{A}\right)}{\theta\left(z-\mathtt{a}_{\mathcal{A}}\right)},
\end{equation}
which is already holomorphic. 

Therefore, we have the expansion
\begin{eqnarray}
 &  & \left\langle e^{\varUpsilon}e^{\mathtt{q}\oint_{\mathcal{C}}\mathscr{V}_{\alpha,\rho}(z)dz}\right\rangle _{\mathbb{T}^{2}}^{\mathrm{hol}}\nonumber \\
 & = & \sum_{k=0}^{\infty}\frac{\mathtt{q}^{k}}{k!}\left[\frac{2\pi\eta(\tau)^{3}\prod_{1\leq a<b\leq3}\theta\left(\varepsilon_{ab}\right)}{\prod_{a\in\underline{4}}\theta\left(\varepsilon_{a}\right)}\right]^{k}\times\nonumber \\
 &  & \times\oint_{\mathcal{C}}dz_{1}\cdots\oint_{\mathcal{C}}dz_{k}\prod_{i=1}^{k}\prod_{\mathcal{A}=\left(A,\alpha\right)\in\underline{n}}\frac{\theta\left(z-\mathtt{a}_{\mathcal{A}}-\varepsilon_{A}\right)}{\theta\left(z-\mathtt{a}_{\mathcal{A}}\right)}\times\nonumber \\
 &  & \times\prod_{\substack{i,j=1\\
i\neq j
}
}^{k}\frac{\theta\left(z_{ij}\right)\prod_{1\leq a<b\leq3}\theta\left(z_{ij}+\varepsilon_{ab}\right)}{\prod_{a\in\underline{4}}\theta\left(z_{ij}+\varepsilon_{a}\right)},
\end{eqnarray}
which coincides with the instanton partition function (\ref{eq:Zinstelliptic})
if the contour $\mathcal{C}$ is chosen to give the Jeffrey-Kirwan
residues. We see that the contributions from the D$1$-D$1$ and D$1$-D$7$
strings are reproduced by the Wick contractions within the exponentiated
integrated vertex, and the Wick contractions between the exponentiated
integrated vertex and the linear source, respectively. 

\section{Conclusions and future directions \label{sec:conclusion}}

In this paper, we introduced tetrahedron instantons and explained
how to construct them from string theory and from noncommutative field
theory. We analyzed the moduli space of tetrahedron instantons and
discussed its geometric interpretations. We computed the instanton
partition function in two different approaches: the infrared approach
which computes the partition function via equivariant localization
on the moduli space of tetrahedron instantons, and the ultraviolet
approach which computes the partition function as the elliptic genus
of the worldvolume theory on the D$1$-branes probing a configuration
of intersecting D$7$-branes. Both approaches lead to the same result.
Our instanton partition function can also be viewed as the expectation
value of the most general real codimension-two defects in the instanton
partition function of the Donaldson-Thomas theory. Finally, we find
a free field representation of the instanton partition function, indicating
the existence of a novel kind of symmetry acting on the cohomology
of the moduli spaces of tetrahedron instantons. 

There are still many interesting aspects of tetrahedron instantons
that remain to be better understood. Some of the future directions
in which this work could be continued are listed in the following.
\begin{enumerate}
\item According to \cite{Witten:2000mf}, a supersymmetric bound state can
be formed by the D$1$-D$9$ system if we turn on a constant $B$-field
satisfying
\begin{equation}
\sum_{a\in\underline{4}}v_{a}\geq1.
\end{equation}
Therefore, we can generalize the tetrahedron instantons by adding
a stack of D$9$-branes without further breaking the supersymmetry
if the $B$-field satisfies the condition
\begin{equation}
v_{1}=v_{2}=v_{3}=v_{4}\geq\frac{1}{4}.
\end{equation}
This generalization can also be viewed as instantons in the magnificent
four model with all possible real codimension-two defects. Furthermore,
it is fascinating to also incorporate in the system the spiked instantons,
which can be realized by D$1$-branes probing six stacks of (anti-)D$5$-branes
in type IIB superstring theory. As analyzed in \cite{Nekrasov:2016gud},
supersymmetry is completely broken when we have six stacks of D$5$-branes,
and two supercharges can be preserved when we have four stacks of
D$5$-branes and two stacks of anti-D$5$-branes in the presence of
a $B$-field obeying 
\begin{equation}
v_{1}=-v_{2}=v_{3}=-v_{4}.
\end{equation}
In both cases, no supersymmetry will be preserved when we put together
the magnificent four model, the tetrahedron instantons and the spiked
instantons. On the other hand, the configuration of D$1$-branes with
six stacks of anti-D$5$-branes preserve two supercharges when the
$B$-field obeys \footnote{We thank the reviewer for pointing to us this configuration.}
\begin{equation}
v_{1}=v_{2}=v_{3}=v_{4}.
\end{equation}
In this case, we can study the supersymmetric combination of the magnificent
four model, the tetrahedron instantons and the spiked instantons.
This combined system can be understood as instantons in the magnificent
four model with all possible real codimension-two and real codimension-four
defects.
\item It was proposed that the partition function of the magnificent four
model is the mother of all instanton partition functions \cite{Nekrasov:2017cih,Nekrasov:2018xsb}.
In particular, it was shown in \cite{Nekrasov:2018xsb} that the partition
function of the magnificent four model at a degenerate limit reduces
to the instanton partition function of the Donaldson-Thomas theory
on $\mathbb{C}^{3}$. The magnificent four model can be realized in
string theory using D$0$-branes probing a collection of D$8$- and
anti-D$8$-branes wrapping a Calabi-Yau fourfold, with an appropriate
$B$-field. Here the D$0$-D$8$ system gives an ADHM-type construction
for instantons in the eight-dimensional gauge theory, while the presence
of the anti-D$8$-branes introduces certain fundamental matter fields.
The degenerate limit corresponds to a fine-tuned position of the anti-D$8$-branes,
and it was conjectured that anti-D$8$-branes will annihilate the
D$8$-branes, leaving a configuration of D$6$-branes after the tachyon
condensation. It is natural to imagine that by taking more general
degenerate limits, the instanton partition function of our model can
always be obtained from that of the magnificent four model. The matching
of the instanton partition function will then be a highly nontrivial
test of the tachyon condensation in nontrivial string backgrounds. 
\item It is well known that the partition function of the Donaldson-Thomas
theory on a toric Calabi-Yau threefold and the partition function
of the magnificent four model play important roles in the study of
the compactification of M-theory on Calabi-Yau fivefolds \cite{Haupt:2008nu,Nekrasov:2014nea,Zotto:2021xah}.
An equivalence between the Donaldson-Thomas invariants and Gromov-Witten
invariants was conjectured \cite{maulik2006gromov1,maulik2006gromov2,maulik2011gromov}.
Together with the Gopakumar-Vafa invariants \cite{Gopakumar:1998ii,Gopakumar:1998ki,Gopakumar:1998jq},
they arise from different expansions of the same topological string
amplitude. A fascinating direction is to explore our model from this
viewpoint. We consider the bound state of $k$ D$0$-branes and $n_{A}$
D$6_{A}$-branes on $\mathbb{S}^{1}\times\mathbb{C}_{A}^{3}$ for
all $A\in\underline{4}^{\vee}$, which can be lifted to M-theory as
a bound state of $k$ graviton Kaluza-Klein modes on $\mathbb{S}^{1}\times\mathscr{X}$,
where $\mathscr{X}$ is a noncompact Calabi-Yau fivefolds. When only
one of the $n_{A}$ is nonzero, $\mathscr{X}$ becomes $\mathbb{C}_{A}^{3}\times\mathrm{TN}_{n_{A}}$,
where $\mathrm{TN}_{n_{A}}$ is the $n_{A}$-centered Taub-NUT space.
After introducing the $\Omega$-deformation, the eleven-dimensional
spacetime $\mathbb{S}^{1}\times\mathscr{X}$ is replaced by a fiber
bundle over $\mathbb{S}^{1}$ with fiber $\mathscr{X}$, such that
the fiber is rotated by an element $g\in\mathrm{SU}(5)$ as we go
around $\mathbb{S}^{1}$. The eleven-dimensional supergravity partition
function on this background is defined as the twisted Witten index,
\begin{eqnarray}
Z^{\mathrm{sugra}}\left[\mathbb{S}^{1}\rtimes_{g}\mathscr{X}\right]\left(\mathfrak{q}_{1},\cdots,\mathfrak{q}_{5}\right) & = & \mathrm{Tr}_{\mathcal{H}(\mathscr{X})}(-1)^{F}e^{-\beta\left\{ \mathfrak{Q},\mathfrak{Q}^{*}\right\} }g\nonumber \\
 & = & \exp\left[\sum_{\ell=1}^{\infty}\frac{1}{\ell}\mathcal{F}^{\mathrm{sugra}}\left(\mathfrak{q}_{1}^{\ell},\cdots,\mathfrak{q}_{5}^{\ell}\right)\right],
\end{eqnarray}
where $\mathcal{H}(\mathscr{X})$ is the Hilbert space of the supergravity
theory on $\mathscr{X}$, $F$ is the fermion number operator, $\beta$
is the circumference of $\mathbb{S}^{1}$, $\mathfrak{Q}$ is a preserved
supercharge that commutes with $g$, and $\mathfrak{q}_{1},\cdots,\mathfrak{q}_{5}$
satisfying $\prod_{i=1}^{5}\mathfrak{q}_{i}=1$ are the fugacities
associated with the $\mathrm{SU}(5)$ action. We can decompose the
single-particle index $\mathcal{F}$ into two parts,
\begin{equation}
\mathcal{F}^{\mathrm{sugra}}=\mathcal{F}^{\mathrm{sugra,pert}}+\mathcal{F}^{\mathrm{sugra,inst}},
\end{equation}
where $\mathcal{F}^{\mathrm{sugra,pert}}$ is the perturbative contribution
from D$6$-branes in the absence of D$0$-branes, and $\mathcal{F}^{\mathrm{sugra,inst}}$
should coincide with the single-particle index (\ref{eq:character})
of the instanton partition function. An extraordinary feature of this
correspondence is that the instanton counting parameter $\mathtt{q}$
in the instanton partition function will be expressed in terms of
the fugacities $\mathfrak{q}_{1},\cdots,\mathfrak{q}_{5}$ in $\mathcal{F}^{\mathrm{sugra,inst}}$.
\item In this paper we only considered the simplest spacetime geometry $\mathbb{R}^{1,1}\times\mathbb{C}^{4}$.
It is definitely interesting to generalize our analysis to $\mathbb{R}^{1,1}\times\mathscr{Y}$,
where $\mathscr{Y}$ is an arbitrary toric Calabi-Yau fourfold. For
example, one can consider the orbifold $\mathscr{Y}=\mathbb{C}^{4}/\Gamma$,
where $\Gamma$ is a finite subgroup of $\mathrm{SU}(4)$. The moduli
space will be a generalization of Nakajima quiver varieties \cite{Nakajima:1994nid,Nakajima1998,Nekrasov:2016qym,Bonelli:2020gku}
and the chain-saw quiver \cite{feigin2011yangians,finkelberg2014quantization,Kanno:2011fw}.
The instanton partition function on the orbifold can be obtained by
projecting onto the $\Gamma$-invariant part. Another nature choice
is to blowup the origin of $\mathbb{C}^{4}$ in the spirit of \cite{Nakajima:2003pg,Nakajima:2003uh,Nakajima:2005fg},
and it may be useful for the study of BPS/CFT correspondence \cite{Nekrasov:2020qcq,Jeong:2020uxz}.
These instanton partition functions should lie between the Donaldson-Thomas
invariants of toric Calabi-Yau threefolds \cite{maulik2006gromov1,maulik2006gromov2}
and fourfolds \cite{Cao:2017swr,Cao:2018rbp,Cao:2020otr}. We can
even generalize our computations by including extra D-branes wrapping
compact divisors.
\item The instanton partition function of the Donaldson-Thomas theory was
identified with the classical statistical mechanics of melting crystal
\cite{Okounkov:2003sp}, and can be interpreted as a quantum gravitational
path integral involving fluctuations of geometry and topology \cite{Iqbal:2003ds}.
It would be wonderful if one can provide a similar interpretation
for the instanton partition function of tetrahedron instantons, in
particular from the expression (\ref{eq:Zproj}).
\item It would be interesting if we can have a better understanding of the
free field representation of the instanton partition function, generalizing
the discussion in \cite{Dijkgraaf:2007sw}.
\item We may consider the tetrahedron instantons with supergroups by adding
negative branes \cite{Vafa:2001qf,Okuda:2006fb,Dijkgraaf:2016lym}
in our construction. We can then calculate the instanton partition
function as in \cite{Kimura:2019msw}. 
\end{enumerate}


\acknowledgments
We are grateful to Giulio Bonelli, Nikita Nekrasov, Mauricio Romo, Peng Shan, and Dingxin Zhang for comments and discussions. We thank Jie Zhou in particular for valuable comments and suggestions on the manuscript. WY would also like to thank the education from the course \emph{Geometric Representation Theory} at Tsinghua University. EP and XZ are partially supported by the GIF Research Grant I-1515-303./2019. WY is supported by the Young overseas high-level talents introduction plan, national key research and development program of China (NO. 2020YFA0713000), and NNSF of China with Grant NO: 11847301 and 12047502. 


\appendix

\section{Open strings in the presence of a constant $B$-field \label{app:string}}

The closed string background on which the open strings propagate is
the flat spacetime $\mathbb{R}^{1,1}\times\mathbb{C}^{4}$ with metric
$G_{\mu\nu}=\eta_{\mu\nu}$ and a constant $B$-field whose nonzero
components are given by (\ref{eq:B}). The worldsheet action of the
open string on such background in the conformal gauge is
\begin{eqnarray}
S & = & \frac{1}{4\pi\ell_{s}^{2}}\int d\tau\int_{0}^{\pi}d\sigma\,G_{\mu\nu}\left(\partial_{\tau}X^{\mu}\partial_{\tau}X^{\nu}-\partial_{\sigma}X^{\mu}\partial_{\sigma}X^{\nu}+2\mathrm{i}\psi_{-}^{\mu}\partial_{+}\psi_{-}^{\nu}+2\mathrm{i}\psi_{+}^{\mu}\partial_{-}\psi_{+}^{\nu}\right)\nonumber \\
 &  & -\frac{1}{2\pi\ell_{s}^{2}}\int d\tau\,\left[B_{\mu\nu}\left(\left(\partial_{\tau}X^{\mu}\right)X^{\nu}+\mathrm{i}\psi_{-}^{\mu}\psi_{-}^{\nu}+\mathrm{i}\psi_{+}^{\mu}\psi_{+}^{\nu}\right)\right]_{\sigma=0}^{\sigma=\pi},\label{eq:action}
\end{eqnarray}
where $\ell_{s}$ is the string length, and $\sigma^{\pm}=\tau\pm\sigma$
are the light-cone coordinates with $\partial_{\pm}=\frac{1}{2}\left(\partial_{\tau}\pm\partial_{\sigma}\right)$.
From the variations of the action (\ref{eq:action}), we can obtain
the equations of motion for $X^{\mu}$ and $\psi_{\pm}$,
\begin{equation}
\partial_{+}\partial_{-}X^{\mu}=0,\quad\partial_{+}\psi_{-}^{\mu}=\partial_{-}\psi_{+}^{\mu}=0,\label{eq:eom}
\end{equation}
as well as the boundary conditions
\begin{eqnarray}
\left[\left(G_{\mu\nu}\partial_{\sigma}X^{\mu}+B_{\mu\nu}\partial_{\tau}X^{\mu}\right)\delta X^{\nu}\right]_{\sigma=0}^{\sigma=\pi} & = & 0,\label{eq:bc1}\\
\left[\delta\psi_{-}^{\mu}\left(G_{\mu\nu}-B_{\mu\nu}\right)\psi_{-}^{\nu}-\delta\psi_{+}^{\mu}\left(G_{\mu\nu}+B_{\mu\nu}\right)\psi_{+}^{\nu}\right]_{\sigma=0}^{\sigma=\pi} & = & 0.\label{eq:bc2}
\end{eqnarray}
Hence, there are two possible boundary conditions for the worldsheet
bosons $X^{\mu}$ at $\sigma=0$ or $\sigma=\pi$: the Dirichlet (D)
boundary condition 
\begin{equation}
\left.\delta X^{\mu}\right|_{\sigma=0,\pi}=0\Leftrightarrow\left.\partial_{\tau}X^{\mu}\right|_{\sigma=0,\pi}=0,\label{eq:Dbc}
\end{equation}
and the twisted (T) boundary condition
\begin{equation}
\left.\left(G_{\mu\nu}\partial_{\sigma}X^{\mu}+B_{\mu\nu}\partial_{\tau}X^{\mu}\right)\right|_{\sigma=0,\pi}=0.\label{eq:Tbc}
\end{equation}
The boundary condition (\ref{eq:Tbc}) becomes the Neumann (N) boundary
condition for $B=0$. The worldsheet supersymmetry transformations
in the bulk are
\begin{equation}
\delta X^{\mu}=\mathrm{i}\epsilon_{+}\psi_{-}^{\mu}-\mathrm{i}\epsilon_{-}\psi_{+}^{\mu},\quad\delta\psi_{\pm}^{\mu}=\pm2\epsilon_{\mp}\partial_{\pm}X^{\mu}.\label{eq:wssusy}
\end{equation}

Since we introduce D$1$-branes along $\mathbb{R}^{1,1}$ and D$7_{A}$-branes
along $\mathbb{R}^{1,1}\times\mathbb{C}_{A}^{3}$ with $A\in\underline{4}^{\vee}$,
open strings always satisfy NN boundary conditions along $\mathbb{R}^{1,1}$,
\begin{equation}
\left.\partial_{\sigma}X^{0,9}\right|_{\sigma=0}=\left.\partial_{\sigma}X^{0,9}\right|_{\sigma=\pi}=0.
\end{equation}
For the remaining $8$ directions, let us introduce the complex bosons
\begin{equation}
Z^{a}=X^{2a-1}+\mathrm{i}X^{2a},\quad\bar{Z}^{a}=X^{2a-1}-\mathrm{i}X^{2a},\quad a\in\underline{4}.
\end{equation}
The general solution of the equation of motion of $Z^{a}$ is given
by $Z^{a}=Z_{L}^{a}\left(\sigma^{+}\right)+Z_{R}^{a}\left(\sigma^{-}\right)$,
where
\begin{eqnarray}
Z_{L}^{a}\left(\sigma^{+}\right) & = & \frac{z_{L}^{a}}{2}+\frac{\ell_{s}^{2}}{2}p_{L}^{a}\sigma^{+}+\frac{\mathrm{i}\ell_{s}}{\sqrt{2}}\sum_{n\neq0}\frac{\alpha_{n}^{a}}{n}e^{-\mathrm{i}n\sigma^{+}},\nonumber \\
Z_{R}^{a}\left(\sigma^{-}\right) & = & \frac{z_{R}^{a}}{2}+\frac{\ell_{s}^{2}}{2}p_{R}^{a}\sigma^{-}+\frac{\mathrm{i}\ell_{s}}{\sqrt{2}}\sum_{n\neq0}\frac{\tilde{\alpha}_{n}^{a}}{n}e^{-\mathrm{i}n\sigma^{-}},
\end{eqnarray}
and the boundary condition can be written uniformly as
\begin{equation}
\left.\left(\partial_{+}-e^{-2\pi\mathrm{i}\nu_{a}}\partial_{-}\right)Z^{a}\right|_{\sigma=0}=\left.\left(\partial_{+}-e^{-2\pi\mathrm{i}\nu_{a}^{\prime}}\partial_{-}\right)Z^{a}\right|_{\sigma=\pi}=0.\label{eq:bcZ}
\end{equation}
Here $\nu_{a}=v_{a}$ ($\nu_{a}^{\prime}=v_{a}$) if the $\sigma=0$
($\sigma=\pi$) end of the open string is on D$7_{A}$-brane with
$a\in A$, and $\nu_{a}=\frac{1}{2}$ ($\nu_{a}^{\prime}=\frac{1}{2}$)
otherwise. The mode expansions of $Z^{a}$ when $\nu_{a}=\nu_{a}^{\prime}$
is
\begin{equation}
Z^{a}=z_{a}+\ell_{s}^{2}p^{a}\left(\sigma^{+}+e^{2\pi\mathrm{i}\nu_{a}}\sigma^{-}\right)+\frac{\mathrm{i}\ell_{s}}{\sqrt{2}}\sum_{n\in\mathbb{Z}\setminus\left\{ 0\right\} }\frac{\alpha_{n}^{a}}{n}\left(e^{-\mathrm{i}n\sigma^{+}}+e^{2\pi\mathrm{i}\nu_{a}}e^{-\mathrm{i}n\sigma^{-}}\right),
\end{equation}
and when $\nu_{a}^{\prime}-\nu_{a}=\delta\neq0$ is
\begin{equation}
Z^{a}=z_{a}+\frac{\mathrm{i}\ell_{s}}{\sqrt{2}}\sum_{r\in\mathbb{Z}+\delta}\frac{\alpha_{r}^{a}}{r}\left(e^{-\mathrm{i}r\sigma^{+}}+e^{2\pi\mathrm{i}\nu_{a}}e^{-\mathrm{i}r\sigma^{-}}\right).
\end{equation}

Meanwhile, we introduce the complex combinations of fermions
\begin{equation}
\Psi_{\pm}^{a}=\psi_{\pm}^{2a-1}+\mathrm{i}\psi_{\pm}^{2a},\quad\bar{\Psi}_{\pm}^{a}=\psi_{\pm}^{2a-1}-\mathrm{i}\psi_{\pm}^{2a}.
\end{equation}
The boundary conditions compatible with (\ref{eq:bcZ}) can be chosen
as
\begin{equation}
\left.\left(\Psi_{+}^{a}-(-1)^{\xi}e^{-2\pi\mathrm{i}\nu_{a}}\Psi_{-}^{a}\right)\right|_{\sigma=0}=\left.\left(\Psi_{+}^{a}-e^{-2\pi\mathrm{i}\nu_{a}^{\prime}}\Psi_{-}^{a}\right)\right|_{\sigma=\pi}=0,\label{eq:bcPsi}
\end{equation}
with $\xi=0$ for the Ramond sector and $\xi=1$ for the Neveu-Schwarz
sector. The Ramond sector preserves half of the worldsheet supersymmetry
(\ref{eq:wssusy}) with $\epsilon=\epsilon_{-}=-\epsilon_{+}$, while
the Neveu-Schwarz sector breaks all the worldsheet supersymmetry.
We combine $\Psi_{+}^{a}$ and $\Psi_{-}^{a}$ into a single field
$\Psi^{a}$ with the extended range $0\leq\sigma\leq2\pi$, 
\begin{equation}
\Psi^{a}\left(\tau,\sigma\right)=\begin{cases}
\Psi_{+}^{a}\left(\tau,\sigma\right) & 0\leq\sigma\leq\pi\\
e^{-2\pi\mathrm{i}\nu_{a}^{\prime}}\Psi_{-}^{a}\left(\tau,2\pi-\sigma\right) & \pi\leq\sigma\leq2\pi
\end{cases},
\end{equation}
whose field equation is $\partial_{-}\Psi^{a}=0$. The boundary condition
(\ref{eq:bcPsi}) at $\sigma=\pi$ ensures that $\Psi^{a}\left(\tau,\sigma\right)$
is continuous, while the boundary condition (\ref{eq:bcPsi}) at $\sigma=0$
leads to
\begin{equation}
\Psi^{a}\left(\tau,2\pi\right)=\exp\left(-2\pi\mathrm{i}\left(\delta-\frac{1}{2}\xi\right)\right)\Psi^{a}\left(\tau,0\right).
\end{equation}
Therefore, the mode expansion of $\Psi^{a}$ in the Ramond sector
is
\begin{equation}
\Psi^{a}\left(\tau,\sigma\right)=\ell_{s}\sum_{r\in\mathbb{Z}+\delta}d_{r}^{a}e^{-\mathrm{i}r\sigma^{+}},
\end{equation}
and that in the Neveu-Schwarz sector is
\begin{equation}
\Psi^{a}\left(\tau,\sigma\right)=\ell_{s}\sum_{r\in\mathbb{Z}+\delta-\frac{1}{2}}b_{r}^{a}e^{-\mathrm{i}r\sigma^{+}}.
\end{equation}

The zero-point energy of $Z^{a}$ is
\begin{equation}
\mathscr{V}_{Z}\left(\delta\right)=\sum_{n=0}^{\infty}\left(n+\left|\delta\right|\right)\overset{\mathrm{reg}}{=}\zeta_{H}\left(-1,\left|\delta\right|\right)=\frac{1}{24}-\frac{1}{2}\left(\left|\delta\right|-\frac{1}{2}\right)^{2},
\end{equation}
and that of $\Psi^{a}$ is
\begin{equation}
\mathscr{V}_{\Psi}\left(\delta\right)=\begin{cases}
-\sum_{n=0}^{\infty}\left(n+\left|\delta\right|\right)\overset{\mathrm{reg}}{=}-\zeta_{H}\left(-1,\left|\delta\right|\right)=-\frac{1}{24}+\frac{1}{2}\left(\left|\delta\right|-\frac{1}{2}\right)^{2} & \mathrm{R}\\
-\sum_{n=0}^{\infty}\left(n+\left|\left|\delta\right|-\frac{1}{2}\right|\right)\overset{\mathrm{reg}}{=}-\zeta_{H}\left(-1,\left|\left|\delta\right|-\frac{1}{2}\right|\right)=-\frac{1}{24}+\frac{1}{2}\left(\left|\left|\delta\right|-\frac{1}{2}\right|-\frac{1}{2}\right)^{2} & \mathrm{NS}
\end{cases},
\end{equation}
where $\zeta_{H}\left(s,a\right)=\sum_{n=0}^{\infty}(n+a)^{-s}$ is
the Hurwitz zeta function. The sum of the zero-point energy $\mathscr{V}=\mathscr{V}_{Z}+\mathscr{V}_{\Psi}$
is 
\begin{equation}
\mathscr{V}\left(\delta\right)=\begin{cases}
0, & \mathrm{R}\\
\frac{1}{8}-\frac{1}{2}\left|\left|\delta\right|-\frac{1}{2}\right|, & \mathrm{NS}
\end{cases}.
\end{equation}
The vanishing of the zero-point energy in the Ramond sector is guaranteed
by the unbroken worldsheet supersymmetry.

In the absence of a constant $B$-field, we have $\delta=0$ for DD
or NN directions, and $\left|\delta\right|=\frac{1}{2}$ for DN and
ND directions. The total zero-point energy of the D$p$-D$p^{\prime}$
strings in the Neveu-Schwarz sector is given by
\begin{equation}
E^{(0)}=\frac{\kappa}{2}\mathscr{V}\left(\frac{1}{2}\right)+\frac{8-\kappa}{2}\mathscr{V}\left(0\right)=-\frac{1}{2}+\frac{\kappa}{8},
\end{equation}
where $\kappa$ is the number of DN and ND directions.

For the $B$-field given by (\ref{eq:B}), the physical ground states
of D$1$-D$1$ and D$7_{A}$-D$7_{A}$ strings still have zero energy.
The total zero-point energy of the D$1$-D$7_{A}$ strings in the
Neveu-Schwarz sector becomes
\begin{equation}
E^{(0)}=\sum_{a\in A}\mathscr{V}\left(\frac{1}{2}-v_{a}\right)+\mathscr{V}\left(0\right)=\frac{1}{4}-\frac{1}{2}\sum_{a\in A}\left|v_{a}\right|,
\end{equation}
and that of the D$7_{(acd)}$-D$7_{(bcd)}$ string becomes
\begin{equation}
E^{(0)}=\mathscr{V}\left(\frac{1}{2}-v_{a}\right)+\mathscr{V}\left(v_{b}-\frac{1}{2}\right)+2\mathscr{V}\left(0\right)=-\frac{1}{2}\left(\left|v_{a}\right|+\left|v_{b}\right|\right).
\end{equation}

\section{Two-dimensional supersymmetric gauge theory}

In this appendix, we review two-dimensional $\mathcal{N}=\left(2,2\right)$
and $\mathcal{N}=\left(0,2\right)$ supersymmetric gauge theories
\cite{Witten:1993yc,Hori:2003ic}.

\subsection{$\mathcal{N}=\left(2,2\right)$ supersymmetry}

The $\mathcal{N}=\left(2,2\right)$ supersymmetry algebra in two-dimensional
Minkowski spacetime $\mathbb{R}^{1,1}$ with coordinates $x^{\mu},\mu=0,1$
is generated by four supercharges $Q_{\pm}$ and $\bar{Q}_{\pm}=Q_{\pm}^{\dagger}$,
spacetime translations $H$, $P$, the Lorentz boost $M=M_{01}$,
and $\mathrm{U}(1)_{V}$ and $\mathrm{U}(1)_{A}$ R-symmetries $F_{V}$
and $F_{A}$. They satisfy the (anti-)commutation relations,
\begin{eqnarray}
Q_{\pm}^{2}=\bar{Q}_{\pm}^{2}=0, & \quad & \left\{ Q_{\pm},\bar{Q}_{\pm}\right\} =2\left(H\mp P\right),\nonumber \\
\left\{ \bar{Q}_{+},\bar{Q}_{-}\right\} =2Z, & \quad & \left\{ Q_{+},Q_{-}\right\} =2Z^{*},\nonumber \\
\left\{ \bar{Q}_{+},Q_{-}\right\} =2\tilde{Z}, & \quad & \left\{ Q_{+},\bar{Q}_{-}\right\} =2\tilde{Z}^{*},\nonumber \\
\left[M,Q_{\pm}\right]=\mp Q_{\pm}, & \quad & \left[M,\bar{Q}_{\pm}\right]=\mp\bar{Q}_{\pm},\nonumber \\
\left[F_{V},Q_{\pm}\right]=-Q_{\pm}, & \quad & \left[F_{V},\bar{Q}_{\pm}\right]=+\bar{Q}_{\pm},\nonumber \\
\left[F_{A},Q_{\pm}\right]=\mp Q_{\pm}, & \quad & \left[F_{A},\bar{Q}_{\pm}\right]=\pm\bar{Q}_{\pm},
\end{eqnarray}
where $Z$ and $\tilde{Z}$ commute with all operators in the theory
and are called central charges. A central charge can be nonzero if
there is a soliton that interpolates different vacua or if the theory
has a continuous abelian symmetry. In superconformal field theory,
both central charges must vanish.

In terms of the $\mathcal{N}=\left(2,2\right)$ superspace with coordinates
$\left(x^{\mu},\theta^{\pm},\bar{\theta}^{\pm}\right)$, the supercharges
are given by
\begin{eqnarray}
Q_{\pm} & = & \frac{\partial}{\partial\theta^{\pm}}+2\mathrm{i}\bar{\theta}^{\pm}\partial_{\pm},\nonumber \\
\bar{Q}_{\pm} & = & -\frac{\partial}{\partial\bar{\theta}^{\pm}}-2\mathrm{i}\theta^{\pm}\partial_{\pm},
\end{eqnarray}
where $\partial_{\pm}=\frac{1}{2}\left(\partial_{0}\pm\partial_{1}\right)$.
They anti-commute with the super-derivatives
\begin{eqnarray}
\mathrm{D}_{\pm} & = & \frac{\partial}{\partial\theta^{\pm}}-2\mathrm{i}\bar{\theta}^{\pm}\partial_{\pm},\nonumber \\
\bar{\mathrm{D}}_{\pm} & = & -\frac{\partial}{\partial\bar{\theta}^{\pm}}+2\mathrm{i}\theta^{\pm}\partial_{\pm},
\end{eqnarray}
which also obey anti-commutation relations 
\begin{equation}
\mathrm{D}_{\pm}^{2}=\bar{\mathrm{D}}_{\pm}^{2}=0,\quad\left\{ \mathrm{D}_{\pm},\bar{\mathrm{D}}_{\pm}\right\} =4\mathrm{i}\partial_{\pm}.
\end{equation}
R-symmetries act on a superfield $\mathcal{F}\left(x^{\mu},\theta^{\pm},\bar{\theta}^{\pm}\right)$
with vector R-charge $q_{V}$ and axial R-charge $q_{A}$ as
\begin{eqnarray}
e^{\mathrm{i}\alpha F_{V}}\mathcal{F}\left(x^{\mu},\theta^{\pm},\bar{\theta}^{\pm}\right) & = & e^{\mathrm{i}\alpha q_{V}}\mathcal{F}\left(x^{\mu},e^{-\mathrm{i}\alpha}\theta^{\pm},e^{\mathrm{i}\alpha}\bar{\theta}^{\pm}\right),\\
e^{\mathrm{i}\alpha F_{A}}\mathcal{F}\left(x^{\mu},\theta^{\pm},\bar{\theta}^{\pm}\right) & = & e^{\mathrm{i}\alpha q_{A}}\mathcal{F}\left(x^{\mu},e^{\mp\mathrm{i}\alpha}\theta^{\pm},e^{\pm\mathrm{i}\alpha}\bar{\theta}^{\pm}\right).
\end{eqnarray}

There are three basic types of $\mathcal{N}=\left(2,2\right)$ superfields.
A chiral superfield $\Phi$ satisfies
\begin{equation}
\bar{\mathrm{D}}_{\pm}\Phi=0,
\end{equation}
which can be expanded as
\begin{equation}
\Phi\left(x^{\mu},\theta^{\pm},\bar{\theta}^{\pm}\right)=\phi\left(y^{\pm}\right)+\sqrt{2}\theta^{\alpha}\psi_{\alpha}\left(y^{\pm}\right)+2\theta^{+}\theta^{-}F\left(y^{\pm}\right),
\end{equation}
where $y^{\pm}=x^{\pm}-2\mathrm{i}\theta^{\pm}\bar{\theta}^{\pm}$,
and $F$ is a complex auxiliary field. The complex conjugate of $\Phi$
is an anti-chiral superfield, $\mathrm{D}_{\pm}\bar{\Phi}=0$. 

A twisted chiral superfield $\Lambda$ satisfies
\begin{equation}
\bar{\mathrm{D}}_{+}\Lambda=\mathrm{D}_{-}\Lambda=0,
\end{equation}
which can be expanded as
\begin{equation}
\Lambda=\varphi\left(\tilde{y}^{\pm}\right)+\sqrt{2}\theta^{+}\bar{\chi}_{+}\left(\tilde{y}^{\pm}\right)+\sqrt{2}\bar{\theta}^{-}\chi_{-}\left(\tilde{y}^{\pm}\right)+2\theta^{+}\bar{\theta}^{-}\widetilde{F}\left(\tilde{y}^{\pm}\right),
\end{equation}
where $\tilde{y}^{\pm}=x^{\pm}\mp2\mathrm{i}\theta^{\pm}\bar{\theta}^{\pm}$,
and $\widetilde{F}$ is a complex auxiliary field. The complex conjugate
of $\Lambda$ is a twisted anti-chiral superfield, $\bar{\mathrm{D}}_{-}\Lambda=\mathrm{D}_{+}\Lambda=0$.

We can also introduce a vector multiplet, which consists of a vector
field $A_{\pm}$, Dirac fermions $\lambda_{\pm}$, $\bar{\lambda}_{\pm}$
which are conjugate to each other, and a complex scalar $\sigma$
in the adjoint representation of the gauge group. The vector superfield
$V$ is a real superfield and can be expanded in the Wess-Zumino gauge
as

\begin{eqnarray}
V & = & \theta^{-}\bar{\theta}^{-}\left(A_{0}-A_{1}\right)+\theta^{+}\bar{\theta}^{+}\left(A_{0}+A_{1}\right)-\theta^{-}\bar{\theta}^{+}\sigma-\theta^{+}\bar{\theta}^{-}\bar{\sigma}+\nonumber \\
 &  & +\sqrt{2}\mathrm{i}\left(\theta^{-}\theta^{+}\bar{\theta}^{-}\bar{\lambda}_{-}+\theta^{-}\theta^{+}\bar{\theta}^{+}\bar{\lambda}_{+}+\bar{\theta}^{+}\bar{\theta}^{-}\theta^{-}\lambda_{-}+\bar{\theta}^{+}\bar{\theta}^{-}\theta^{+}\lambda_{+}\right)+\nonumber \\
 &  & +2\theta^{-}\theta^{+}\bar{\theta}^{+}\bar{\theta}^{-}D,
\end{eqnarray}
where $D$ is a real auxiliary field. To couple a matter superfield
to the gauge field, we simply replace the super-derivatives $\mathrm{D}_{\pm}^{2},\bar{\mathrm{D}}_{\pm}^{2}$
by the gauge-covariant super-derivatives 
\begin{equation}
\mathbb{D}_{\pm}=e^{-V}D_{\pm}e^{V},\quad\bar{\mathbb{D}}_{\pm}=e^{V}\bar{D}_{\pm}e^{-V}.
\end{equation}
The field strength of $V$ is given by
\begin{equation}
\Sigma=\frac{1}{2}\left\{ \bar{\mathbb{D}}_{+},\mathbb{D}_{-}\right\} ,
\end{equation}
which is a twisted chiral superfield $\bar{\mathbb{D}}_{+}\Sigma=\mathbb{D}_{-}\Sigma=0$.

The supersymmetric Lagrangian can be written as 
\begin{eqnarray}
\mathcal{L} & = & \int d^{4}\theta\mathcal{K}\left(\mathcal{F},\bar{\mathcal{F}}\right)+\frac{1}{2}\left(\int d\theta^{-}d\theta^{+}\left.\mathcal{W}(\Phi)\right|_{\bar{\theta}^{\pm}=0}+c.c.\right)+\nonumber \\
 &  & +\frac{1}{2}\left(\int d\bar{\theta}^{-}d\theta^{+}\left.\widetilde{\mathcal{W}}(\Lambda)\right|_{\theta^{-}=\bar{\theta}^{+}=0}+c.c.\right),\label{eq:L}
\end{eqnarray}
where the first term involving an arbitrary real function $\mathcal{K}\left(\mathcal{F},\bar{\mathcal{F}}\right)$
is the D-term contribution, the second term involving a superpotential
$\mathcal{W}$ is the F-term contribution, and the third term involving
a twisted superpotential $\widetilde{\mathcal{W}}$ is the twisted
F-term contribution. Here $\mathcal{W}(\Phi)$ and $\widetilde{\mathcal{W}}(\Lambda)$
are required to be holomorphic functions of chiral superfields and
twisted chiral superfields, respectively. 

We are mainly interested in the gauged linear sigma model which describes
a vector superfield $V$ with gauge group $\mathrm{U}(N)$ and field
strength $\Sigma$, coupled with charged chiral multiplets $\Phi_{i}$.
The Lagrangian is given by (\ref{eq:L}), with 
\begin{equation}
\mathcal{K}=-\frac{1}{2e^{2}}\mathrm{Tr}\bar{\Sigma}\Sigma+\mathrm{Tr}\left(\sum_{i}\bar{\Phi}_{i}\Phi_{i}\right),\quad\mathcal{W}=0,\quad\widetilde{\mathcal{W}}=-t\Sigma,
\end{equation}
where $e$ is the gauge coupling constant, and $t=r-\mathrm{i}\vartheta$
is the complex combination of the Fayet-Iliopoulos parameter $r$
and the theta angle $\vartheta$. 

\subsection{$\mathcal{N}=\left(0,2\right)$ supersymmetry}

We can get $\mathcal{N}=\left(0,2\right)$ supersymmetry from $\mathcal{N}=\left(2,2\right)$
supersymmetry by dropping $Q_{-}$ and $\bar{Q}_{-}$. There is only
one $\mathrm{U}(1)_{\mathcal{R}}$ R-symmetry $\mathcal{R}$ satisfying
\begin{equation}
\left[\mathcal{R},Q_{+}\right]=-Q_{+},\quad\left[\mathcal{R},\bar{Q}_{+}\right]=+\bar{Q}_{+}.
\end{equation}
The $\mathcal{N}=\left(0,2\right)$ superspace with coordinates $\left(x^{\mu},\theta^{+},\bar{\theta}^{+}\right)$
is the subspace of $\mathcal{N}=\left(2,2\right)$ superspace with
$\theta^{-}=\bar{\theta}^{-}=0$. 

There are three basic types of $\mathcal{N}=\left(0,2\right)$ superfields.
An $\mathcal{N}=\left(0,2\right)$ chiral superfield $\varPhi$ is
a complex-valued Lorentz scalar obeying
\begin{equation}
\bar{\mathrm{D}}_{+}\varPhi=0,
\end{equation}
which can be expanded as
\begin{equation}
\varPhi=\phi+\sqrt{2}\theta^{+}\psi_{+}-2\mathrm{i}\theta^{+}\bar{\theta}^{+}\partial_{+}\phi,
\end{equation}
where $\phi$ is a complex scalar and $\psi_{+}$ is a right-moving
fermion. 

An $\mathcal{N}=\left(0,2\right)$ Fermi superfield $\varPsi_{-}$
is a left-moving spinor satisfying 
\begin{equation}
\bar{\mathrm{D}}_{+}\varPsi_{-}=\sqrt{2}E\left(\varPhi_{i}\right),
\end{equation}
which can be expanded as
\begin{equation}
\varPsi_{-}=\psi_{-}-\sqrt{2}\theta^{+}G-2\mathrm{i}\theta^{+}\bar{\theta}^{+}\partial_{+}\psi_{-}-\sqrt{2}\bar{\theta}^{+}E\left(\phi_{i}\right)+2\theta^{+}\bar{\theta}^{+}\frac{\partial E}{\partial\phi_{i}}\psi_{+,i},
\end{equation}
where $\psi_{-}$ is a left-moving fermion and $G$ is an auxiliary
field.

The $\mathcal{N}=\left(0,2\right)$ vector superfield $U$ is a real
superfield with the expansion
\begin{equation}
U=A_{0}-A_{1}-2\mathrm{i}\theta^{+}\bar{\lambda}_{-}-2\mathrm{i}\bar{\theta}^{+}\lambda_{-}+2\theta^{+}\bar{\theta}^{+}D,
\end{equation}
where $A_{\mu}$ is the gauge field, $\lambda_{-},\bar{\lambda}_{-}$
are left-moving fermions, and $D$ is a real auxiliary field. All
the fields are in the adjoint representation of the gauge group. The
gauge-covariant super-derivatives $\mathbb{D}_{+}$ and $\bar{\mathbb{D}}_{+}$
are given by 
\begin{eqnarray}
\mathbb{D}_{+} & = & \frac{\partial}{\partial\theta^{+}}-\mathrm{i}\bar{\theta}^{+}\left(\mathcal{D}_{0}+\mathcal{D}_{1}\right),\nonumber \\
\bar{\mathbb{D}}_{+} & = & -\frac{\partial}{\partial\bar{\theta}^{+}}+\mathrm{i}\theta^{+}\left(\mathcal{D}_{0}+\mathcal{D}_{1}\right),
\end{eqnarray}
where
\begin{eqnarray}
\mathcal{D}_{0} & = & \partial_{0}+\mathrm{i}A_{0}+\theta^{+}\bar{\lambda}_{-}+\bar{\theta}^{+}\lambda_{-}+\mathrm{i}\theta^{+}\bar{\theta}^{+}D,\nonumber \\
\mathcal{D}_{1} & = & \partial_{1}+\mathrm{i}A_{1}-\theta^{+}\bar{\lambda}_{-}-\bar{\theta}^{+}\lambda_{-}-\mathrm{i}\theta^{+}\bar{\theta}^{+}D,
\end{eqnarray}
are the gauge-covariant derivatives. We can organize $U$ in terms
of the gauge-invariant field strength $\varUpsilon=\frac{1}{2}\left[\bar{\mathbb{D}}_{+},\mathcal{D}_{0}-\mathcal{D}_{1}\right]$,
which is a Fermi superfield.

We can write down the supersymmetric Lagrangian of an $\mathcal{N}=\left(0,2\right)$
gauged linear sigma model with a vector multiplet $V$ whose field
strength is $\varUpsilon$ coupled to chiral multiplets $\varPhi_{i}$
and the Fermi multiplets $\Psi_{a}$,
\begin{eqnarray}
\mathcal{L} & = & \int d\theta^{+}d\bar{\theta}^{+}\left(\frac{1}{2e^{2}}\mathrm{Tr}\bar{\varUpsilon}\varUpsilon-\frac{\mathrm{i}}{2}\mathrm{Tr}\sum_{i}\bar{\varPhi}_{i}\mathcal{D}_{-}\varPhi_{i}-\frac{1}{2}\mathrm{Tr}\sum_{a}\bar{\varPsi}_{-,a}\varPsi_{-,a}\right)+\nonumber \\
 &  & +\left(\frac{\mathrm{i}t}{2}\int d\theta^{+}\left.\varUpsilon\right|_{\bar{\theta}^{+}=0}+\mathrm{c.c.}\right)-\frac{1}{\sqrt{2}}\left(\int d\theta^{+}\left.\mathrm{Tr}\sum_{a}\varPsi_{-,a}J^{a}\right|_{\bar{\theta}^{+}=0}+c.c.\right),
\end{eqnarray}
where $J^{a}\left(\varPhi_{i}\right)$ are holomorphic functions obeying
\begin{equation}
\sum_{a}E_{a}\left(\varPhi_{i}\right)J^{a}\left(\varPhi_{i}\right)=0.\label{eq:EJ}
\end{equation}

It is sometimes useful to write a theory with $\mathcal{N}=\left(2,2\right)$
supersymmetry in the language of the $\mathcal{N}=\left(0,2\right)$
superspace. An $\mathcal{N}=\left(2,2\right)$ vector multiplet $V$
decomposes into an $\mathcal{N}=\left(0,2\right)$ vector multiplet
$U$ and an $\mathcal{N}=\left(0,2\right)$ chiral multiplet $\Sigma^{\prime}=\left.\Sigma\right|_{\theta^{-}=\bar{\theta}^{-}=0}$.
An $\mathcal{N}=\left(2,2\right)$ chiral multiplet $\Phi$ decomposes
into an $\mathcal{N}=\left(0,2\right)$ chiral multiplet $\varPhi=\left.\Phi\right|_{\theta^{-}=\bar{\theta}^{-}=0}$
and an $\mathcal{N}=\left(0,2\right)$ Fermi superfield $\varPsi_{-}=\frac{1}{\sqrt{2}}\left.\mathbb{D}_{-}\Phi\right|_{\theta^{-}=\bar{\theta}^{-}=0}$,
with
\begin{equation}
E=\frac{1}{2}\left.\bar{\mathbb{D}}_{+}\mathbb{D}_{-}\Phi\right|_{\theta^{-}=\bar{\theta}^{-}=0}=\frac{1}{2}\left.\left\{ \bar{\mathbb{D}}_{+},\mathbb{D}_{-}\right\} \Phi\right|_{\theta^{-}=\bar{\theta}^{-}=0}=\Sigma^{\prime}\varPhi.
\end{equation}
The kinetic terms decompose naturally, while the F-term contribution
specified by the superpotential $\mathcal{W}(\Phi)$ is reduced to
a collection of functions $J^{a}$, one for each $\Phi_{a}=\left(\varPhi_{a},\varPsi_{-,a}\right)$,
with
\begin{equation}
J^{a}=\frac{\partial\mathcal{W}}{\partial\varPhi_{a}}.
\end{equation}
The condition (\ref{eq:EJ}) is satisfied automatically.

\section{Elliptic genus of $\mathcal{N}=\left(0,2\right)$ theories}

We consider the Euclidean path-integral of a two-dimensional $\mathcal{N}=\left(0,2\right)$
supersymmetric theory on a torus $\mathbb{T}^{2}$, in the presence
of a background flat connection for the flavor symmetry. Let $T_{a}$
be the Cartan generators of the flavor symmetry group $G_{f}$. In
the Hamiltonian formalism, the elliptic genus can be defined by \cite{Gadde:2013wq,Gadde:2013ftv,Benini:2013nda,Benini:2013xpa}
\begin{equation}
Z\left(x;q\right)=\mathrm{Tr}_{\mathrm{R}}(-1)^{F}q^{H_{L}}\bar{q}^{H_{R}}\prod_{a}e^{2\pi\mathrm{i}x_{a}T_{a}},
\end{equation}
where the trace is over the Hilbert space of the theory on the spatial
circle, with periodic boundary conditions for fermions. $F$ is the
fermion number. $q=e^{2\pi\mathrm{i}\tau}$ specifies the complex
structure $\tau$ of $T^{2}$. $H_{L}$ and $H_{R}$ are the left-
and right-moving Hamiltonians, respectively. Based on the standard
argument in \cite{Witten:1982df}, the elliptic genus is independent
of $\bar{q}$ if the theory has a discrete spectrum. \footnote{Notice that the elliptic genus can suffer from a holomorphic anomaly
for noncompact models. See \cite{Murthy:2013mya,Ashok:2013pya,Nian:2014fma}
for examples with $\mathcal{N}=\left(2,2\right)$ supersymmetry.} 

We consider a two-dimensional $\mathcal{N}=\left(0,2\right)$ supersymmetric
gauged linear sigma model which is described by a vector multiplet
$V$ with gauge group $G$ of rank $r$, chiral multiplets $\Phi_{i}$
transforming in the representation $\mathfrak{R}\left(\Phi_{i}\right)$
of $G\times G_{f}$, and Fermi multiplets $\Psi_{a}$ transforming
in the representation $\mathfrak{R}\left(\Psi_{a}\right)$ of $G\times G_{f}$.
The elliptic genus has been rigorously derived using the technique
of path integral localization \cite{Benini:2013nda,Benini:2013xpa},
\begin{equation}
Z\left(x;q\right)=\frac{1}{\left|W_{G}\right|}\oint_{\mathrm{JK}}Z_{V}\prod_{i}Z_{\Phi_{i}}\prod_{a}Z_{\Psi_{a}},\label{eq:ZT2}
\end{equation}
where $\left|W_{G}\right|$ is the order of the Weyl group of $G$,
$Z_{V}$, $Z_{\Phi_{i}}$, and $Z_{\Psi_{a}}$ are the contributions
from $V$ without zero-modes of the Cartan generators, $\Phi_{i}$,
and $\Psi_{a}$, respectively. The contour integral is evaluated using
the Jeffrey-Kirwan residue prescription \cite{jeffrey1995localization}.
In terms of the Dedekind eta function $\eta(\tau)$ and the Jacobi
theta function $\theta_{1}\left(\left.z\right|\tau\right)$,
\begin{eqnarray}
\eta(\tau) & = & q^{\frac{1}{24}}\prod_{n=1}^{\infty}\left(1-q^{n}\right),\\
\theta_{1}\left(\left.z\right|\tau\right) & = & \mathrm{i}\sum_{n\in\mathbb{Z}}\left(-1\right)^{n}e^{(2n+1)\pi\mathrm{i}z}q^{\frac{1}{2}\left(n+\frac{1}{2}\right)^{2}},
\end{eqnarray}
the explicit expressions of $Z_{V}$, $Z_{\Phi_{i}}$, and $Z_{\Psi_{a}}$
are given by
\begin{eqnarray}
Z_{V} & = & \left(\frac{2\pi\eta(\tau)^{2}}{\mathrm{i}}\right)^{r}\prod_{I=1}^{r}d\varphi_{I}\prod_{\alpha\in G}\frac{\mathrm{i}\theta_{1}\left(\left.\alpha\cdot\varphi\right|\tau\right)}{\eta(\tau)},\\
Z_{\Phi_{i}} & = & \prod_{\rho\in\mathfrak{R}\left(\Phi_{i}\right)}\frac{\mathrm{i}\eta(\tau)}{\theta_{1}\left(\left.\rho\cdot\zeta\right|\tau\right)},\\
Z_{\Psi_{a}} & = & \prod_{\rho\in\mathfrak{R}\left(\Psi_{a}\right)}\frac{\mathrm{i}\theta_{1}\left(\left.\rho\cdot\zeta\right|\tau\right)}{\eta(\tau)},
\end{eqnarray}
where $\varphi$ parametrizes a Cartan subalgebra of $G$, and $\zeta$
includes both $\varphi$ and $x$. The function $\theta_{1}\left(\left.z\right|\tau\right)$
has no poles, but there are simple zeros at $z\in\mathbb{Z}+\tau\mathbb{Z}$,
with residues of its inverse
\begin{equation}
\frac{1}{2\pi\mathrm{i}}\oint_{z=a+b\tau}\frac{dz}{\theta_{1}\left(\left.z\right|\tau\right)}=\frac{(-1)^{a+b}e^{\mathrm{i}\pi\tau b^{2}}}{2\pi\eta(\tau)^{3}},
\end{equation}
where we have used the identity
\begin{equation}
2\pi\eta(\tau)^{3}=\partial_{z}\theta_{1}\left(\left.0\right|\tau\right).
\end{equation}
In this paper, we often use the abbreviation
\begin{equation}
\theta\left(z\right)\equiv\theta_{1}\left(\left.z\right|\tau\right).
\end{equation}

By taking the degenerate limit $q\to1$ and neglecting an overall
$x$-independent factor, we can reduce the elliptic genus of a two-dimensional
supersymmetric gauge theory to the Witten index of the one-dimensional
supersymmetric quantum mechanics obtained by the standard dimensional
reduction. The contributions of $Z_{V}$, $Z_{\Phi_{i}}$, and $Z_{\Psi_{a}}$
become
\begin{eqnarray}
Z_{V} & = & \prod_{I=1}^{r}d\varphi_{I}\prod_{\alpha\in G}2\sinh\left(\frac{\beta\alpha\cdot\varphi}{2}\right)\\
Z_{\Phi_{i}} & = & \prod_{\rho\in\mathfrak{R}\left(\Phi_{i}\right)}\frac{1}{2\sinh\left(\frac{\beta\rho\cdot\zeta}{2}\right)},\\
Z_{\Psi_{a}} & = & \prod_{\rho\in\mathfrak{R}\left(\Psi_{a}\right)}2\sinh\left(\frac{\beta\rho\cdot\zeta}{2}\right),
\end{eqnarray}
where $\beta$ is the circumference of $\mathbb{S}^{1}$. If we further
reduce to zero dimension, the partition function of the corresponding
supersymmetric matrix model can be obtained by replacing $2\sinh\left(\frac{\beta z}{2}\right)\to z$.

\section{Jeffrey-Kirwan residue formula}

The Jeffrey-Kirwan residue formula introduced in \cite{jeffrey1995localization}
gives a prescription for expressing multiple contour integrals as
a sum of iterated residues. 

Let $\omega$ be a meromorphic $\left(k,0\right)$-form on a $k$-dimensional
complex manifold,
\begin{equation}
\omega=\frac{A(u)}{B(u)}du_{1}\wedge\cdots\wedge du_{k},
\end{equation}
where $A(u)$ and $B(u)$ are two holomorphic functions of $k$ complex
variables $u=\left(u_{1},\cdots,u_{k}\right)$. We assume that $B(u)$
is a product of linear factors,
\begin{equation}
B(u)=\prod_{i}\left(\mathbf{Q}_{i}\cdot u+b_{i}\right),
\end{equation}
where $\mathbf{Q}_{i}$ is the charge vector associated with the singular
hyperplane $H_{i}$,
\begin{equation}
H_{i}=\left\{ \left.u\in\mathbb{C}^{n}\right|\mathbf{Q}_{i}\cdot u+b_{i}=0\right\} .
\end{equation}
Using the standard basis $\left\{ \mathbf{e}_{j}\right\} _{j=1,\cdots,k}$
of $\mathbb{R}^{k}$,
\begin{equation}
\mathbf{e}_{j}=\left(0,\cdots,0,\overset{j}{1},0,\cdots,\overset{k}{0}\right),
\end{equation}
we can write $\mathbf{Q}_{i}$ as
\begin{equation}
\mathbf{Q}_{i}=\sum_{j=1}^{k}\mathbf{Q}_{i,j}\mathbf{e}_{j},
\end{equation}
and 
\begin{equation}
\mathbf{Q}_{i}\cdot u=\sum_{j=1}^{k}\mathbf{Q}_{i,j}u_{j}.
\end{equation}
Clearly, $\omega$ is holomorphic on the complement of $\mathscr{M}^{\mathrm{sing}}=\bigcup_{i}H_{i}$.
Let $\mathscr{M}_{\ast}^{\mathrm{sing}}\subset\mathscr{M}^{\mathrm{sing}}$
be the set of isolated points where $n\geq k$ linearly independent
singular hyperplanes meet. For $u_{\ast}\in\mathscr{M}_{\ast}^{\mathrm{sing}}$,
we denote by $\mathbf{Q}\left(u_{\ast}\right)$ the set of charge
vectors of the singular hyperplanes meeting at $u_{\ast}$,
\begin{equation}
\mathbf{Q}\left(u_{\ast}\right)=\left\{ \left.\mathbf{Q}_{i}\right|u_{\ast}\in H_{i},i=1,\cdots,n\right\} .
\end{equation}
We assume that for each $u_{\ast}\in\mathscr{M}_{\ast}^{\mathrm{sing}}$,
the hyperplane arrangement is projective, which requires that the
set $\mathbf{Q}\left(u_{\ast}\right)$ is contained in a half-space
of $\mathbb{R}^{k}$. This assumption is automatically obeyed when
the hyperplane arrangement is nondegenerate, which means that the
number of hyperplanes meeting at every $u_{\ast}\in\mathscr{M}_{\ast}^{\mathrm{sing}}$
is exactly $k$. Then the residue of $\omega$ at $u_{\ast}$ is given
by its integral over $\prod_{i=1}^{k}\mathcal{C}_{i}$, where $\mathcal{C}_{i}$
is a small circle around $H_{i}$. 

We denote the cone spanned by $\mathbf{Q}_{1},\cdots,\mathbf{Q}_{k}$
by
\begin{equation}
\mathrm{Cone}\left(\mathbf{Q}_{1},\cdots,\mathbf{Q}_{k}\right)=\left\{ \left.\sum_{i=1}^{k}\lambda_{i}\mathbf{Q}_{i}=\boldsymbol{\eta}\right|\lambda_{i}>0\right\} .
\end{equation}
Let $\mathrm{Cone}_{\mathrm{sing}}\left(\mathbf{Q}\right)$ be the
union of the cones generated by all subsets of $\mathbf{Q}$ with
$k-1$ elements. The space $\mathbb{R}^{k}\setminus\mathrm{Cone}_{\mathrm{sing}}\left(\mathbf{Q}\right)$
is a union of connected components, and we call each connected component
a chamber. We can specify a chamber by a generic nonzero vector $\boldsymbol{\eta}\in\mathbb{R}^{k}\setminus\mathrm{Cone}_{\mathrm{sing}}\left(\mathbf{Q}\right)$.
Then the Jeffrey-Kirwan residue formula states that
\begin{equation}
\int\omega\rightsquigarrow\sum_{u_{\ast}\in\mathscr{M}_{\ast}^{\mathrm{sing}}}\underset{u=u_{*}}{\mathrm{JKRes}}\left(\mathbf{Q}\left(u_{*}\right),\boldsymbol{\eta}\right)\omega,
\end{equation}
where the JK-residue operator is defined by the condition
\begin{equation}
\underset{u=u_{*}}{\mathrm{JKRes}}\left(\mathbf{Q}\left(u_{*}\right),\boldsymbol{\eta}\right)\frac{du_{1}\wedge\cdots\wedge du_{k}}{\prod_{i=1}^{k}\left(\mathbf{Q}_{i}\cdot\left(u-u_{\ast}\right)\right)}=\begin{cases}
\frac{1}{\left|\det\left(\mathbf{Q}_{1},\cdots,\mathbf{Q}_{k}\right)\right|}, & \boldsymbol{\eta}\in\mathrm{Cone}\left(\mathbf{Q}_{1},\cdots,\mathbf{Q}_{k}\right)\\
0, & \mathrm{Otherwise}
\end{cases},
\end{equation}
As $\boldsymbol{\eta}$ is varied, the JK-residue is locally constant
but can jump when $\boldsymbol{\eta}$ crosses the boundary of a chamber.
In the simplest case of $k=1$, we have
\begin{equation}
\underset{u=u_{\ast}}{\mathrm{JKRes}}\left(\left\{ q\right\} ,\eta\right)\frac{du}{u-u_{\ast}}=\begin{cases}
\mathrm{sign}(q), & \eta q>0\\
0, & \eta q<0
\end{cases}.
\end{equation}


\providecommand{\href}[2]{#2}\begingroup\raggedright
\endgroup

\end{document}